\documentclass[acmsmall]{acmart}

\setcopyright{cc}
\setcctype{by}
\acmDOI{10.1145/3776681}
\acmYear{2026}
\acmJournal{PACMPL}
\acmVolume{10}
\acmNumber{POPL}
\acmArticle{39}
\acmMonth{1}
\received{2025-07-10}
\received[accepted]{2025-11-06}

\usepackage{scope}
\makeatletter
\patchcmd{\@@addmarginpar}{\ifodd\c@@page}{\ifodd\c@@page\@@tempcnta\m@@ne}{}{}
\makeatother
\reversemarginpar

\begin{document}

\title{Handling Scope Checks \extendedOnly{(Extended Version)}}
\subtitle{A Comparative Framework for Dynamic Scope Extrusion Checks}

\author{Michael Lee}
\email{michael.lee@cl.cam.ac.uk}
\orcid{}
\affiliation{%
  \institution{University of Cambridge}
  \country{United Kingdom}
}
\author{Ningning Xie}
\email{ningningxie@cs.toronto.edu}
\orcid{}
\affiliation{%
  \institution{University of Toronto}
  \country{Canada}
}
\author{Oleg Kiselyov}
\email{oleg@okmij.org}
\orcid{}
\affiliation{%
  \institution{Tohoku University}
  \country{Japan}
}
\author{Jeremy Yallop}
\email{jeremy.yallop@cl.cam.ac.uk}
\orcid{}
\affiliation{%
  \institution{University of Cambridge}
  \country{United Kingdom}
}

\begin{abstract}
  Metaprogramming and effect handlers interact in unexpected, and sometimes undesirable, ways. One example is scope extrusion: the generation of ill-scoped code. Scope extrusion can either be preemptively prevented, via static type systems, or retroactively detected, via dynamic checks. Static type systems exist in theory, but struggle with a range of implementation and usability problems in practice. In contrast, dynamic checks exist in practice (e.g.\ in MetaOCaml), but are understudied in theory. Designers of metaprogramming languages are thus given little guidance regarding the design and implementation of checks. We present the first formal study of dynamic scope extrusion checks, introducing a calculus (\calculusName{}) for describing and evaluating checks. Further, we introduce a novel dynamic check --- the ``Cause-for-Concern'' check --- which we prove correct, characterise without reference to its implementation, and argue combines the advantages of existing dynamic checks. Finally, we extend our framework with refined environment classifiers, which statically prevent scope extrusion, and compare their expressivity with the dynamic checks.
\end{abstract}

\begin{CCSXML}
<ccs2012>
   <concept>
       <concept_id>10011007.10011006.10011008.10011024.10011027</concept_id>
       <concept_desc>Software and its engineering~Control structures</concept_desc>
       <concept_significance>300</concept_significance>
       </concept>
 </ccs2012>
\end{CCSXML}

\ccsdesc[300]{Software and its engineering~Control structures}

\keywords{effect handlers, code generation, metaprogramming, scope extrusion}

\maketitle

\section{Introduction}
Multi-stage programming languages have been used to write code
generators for a wide variety of domains,
from database queries and stream processing to geometry, parsing, and differentiable programming
\citep{DBLP:conf/icfp/RompfA15,DBLP:conf/popl/KiselyovBPS17,DBLP:conf/pepm/CaretteES11,yallop-2023,wang-2019}.
Language constructs for code generation often come with strong
guarantees.
For example, a well-typed code generator written in the MetaML
language~\cite{taha-1999} is guaranteed never to generate ill-typed
code.
However, these guarantees are weakened when code generation constructs
are combined with
effects~\cite{calcagno-00,kameyama-shifting-jfp,kiselyov-14,kameyama-2015,kiselyov-16,parreaux-2020,isoda-24}.

In particular, the combination of code generation constructs and
effects can lead to \textit{scope extrusion}: the inadvertent
generation of code with unbound variables.
For example, in the MacoCaml
program \Cref{code:scope-extrusion-toy-example}, the use of effect
handlers in code generation extrudes the variable \lstinline{x} beyond
its scope:

\begin{code}
\begin{macocaml}
try << let x = 3 in $(perform (Extrude <<x>>)) >> 
with effect Extrude y, k -> << $y + 1 >>
\end{macocaml}
\captionof{listing}{An example of scope extrusion}%
\label{code:scope-extrusion-toy-example}
\end{code}

\noindent
Line 1 installs a handler whose body \lstinline!<< let x = 3 in $(...) >>!
uses \emph{code quotation} to construct
code for a function application.
The expression within quotation marks \lstinline!<< >>! is not
evaluated immediately, but constructs a piece of code that may be
evaluated in the future.
However, the sub-expression prefixed by \lstinline!$! is evaluated
immediately, and performs an effect \lstinline!Extrude!,
transferring control to the most recently installed
handler.
Line 2 shows the handler, which binds the argument \lstinline!<<x>>!
to \lstinline!y! and the continuation delimited by \lstinline!try!
and \lstinline!perform! to \lstinline!k!.
The handler discards the continuation and uses the argument to
construct the code \lstinline!<< x + 1 >>!, in which \lstinline!x! is
unbound.

In \Cref{code:scope-extrusion-toy-example} the extrusion is
simple: \lstinline!<<x>>! leaves the scope of its binder
\lstinline[mathescape]!<<let x = 3 in ...>>! and never returns.
However, handlers that invoke \lstinline!k! might cause control to
re-enter the scope:

\begin{code}
\begingroup\lstset{firstnumber=2}
\begin{macocaml}
with effect Extrude y, k -> continue k << $y + 1 >>
\end{macocaml} 
\endgroup
\captionof{listing}{Revising the handler of \Cref{code:scope-extrusion-toy-example} to bring \lstinline!x! back into scope}%
\label{code:scope-extrusion-toy-example-revised}
\end{code}

\noindent
Here \lstinline!continue! resumes the continuation, returning control
to the point where \lstinline!perform! was invoked, so that the
program ultimately evaluates to a well-scoped code value
\lstinline!<< let x = 3 in x + 1 >>!.

Scope extrusion is a problem in practice as well as in theory.
The strong guarantees attached to multi-stage languages relieve
programmers of the burden of debugging type errors in generated code,
but scope extrusion reintroduces the burden.
\citet{RandIR} report an example: in refactoring the LMS system
to address performance issues, effects were used to perform code motion
optimizations. These effects inadvertently led to scope extrusion errors.
While these errors had simple causes, they were time-consuming to fix due to
the large number of variables involved, and the difficulty of
determining which part of the code generator produced the offending
code.

To avoid the need for programmers to debug generated code, multi-stage
languages with effects often provide help in identifying scope
extrusion.
The key question is \emph{when} to check for problems.
One approach is to track potential extrusion in the type system,
rejecting programs that cannot be shown to be
safe~\citep{calcagno-00,kiselyov-16,isoda-24,parreaux-2020,mint}.
In practice, however, it is difficult to combine the expressiveness that
allows virtuous interactions between effects and code generation
(e.g.~code motion optimizations) with the strictness that excludes all
potential extrusion.
Given the choice between such sophisticated type systems and
simpler but less safe systems, users tend to prefer the latter
\citep{parreaux-2020}.

The other approach is to check dynamically during code generation,
allowing potentially unsafe code generators to run, and identifying
extrusion as it occurs.
This more liberal approach does not have an existing theory, but it is
more common in practice, in part because it can be incorporated into
existing multi-stage languages --- such as MetaOCaml,
Scala~\citep{DBLP:conf/gpce/StuckiBO18} and Typed Template
Haskell~\cite{DBLP:journals/pacmpl/XiePLWYW22} --- without disruption
to their type systems.

There is a range of possible designs for dynamic checks.
At one extreme, scope is checked lazily, once code generation is
complete.  The original MetaOCaml
language used lazy checking, since its static type system,
\emph{environment classifiers}~\citep{DBLP:conf/popl/TahaN03}, prevented some
forms of extrusion, but could not prevent every case.
At the other extreme, BER MetaOCaml checks scope eagerly each time a quotation is
constructed~\citep{kiselyov-14,kiselyov-2024}.
Neither approach is optimal. Lazy checking is \textit{uninformative},  
producing hard-to-debug errors~\citep{RandIR}, \textit{inefficient}, reporting errors much
later than eager checking~\citep{kiselyov-14}, and in some systems can bind variables in \textit{unintended ways}~\citep{kameyama-2015}. On the other hand, eager checking is \textit{incorrect} in a sense that we explicate in \Cref{subsection:eager-dynamic-correctness}, failing to detect occurrences of free variables in certain pathological cases (\Cref{listing:eager-scope-extrusion-unsafe-no-use,listing:eager-scope-extrusion-unsafe-continue}). Further, eager checking is \textit{not continuation-aware}: for example, it incorrectly rejects the safe code generator
in \Cref{code:scope-extrusion-toy-example-revised}~\citep{kiselyov-14}.  

To establish a theory of dynamic checks, we introduce
the \sourceLang{} and \coreLang{} calculi that support multi-stage
programming with effects and handlers, and show how they can be used
to describe and compare eager and lazy checking. We also describe a
new check, the Cause-for-Concern (C4C) check, implemented in MacoCaml,
that combines the advantages of eager and lazy checking.

\paragraph{Contributions}

\Cref{section:overview} presents the eager and lazy approaches informally
using a larger example, and introduces our novel C4C check. 
The subsequent sections present technical contributions:

\begin{itemize}
\item

Two novel calculi, \sourceLang{} and \coreLang{}, designed for the
study of typed multi-stage programming with effects and handlers
(\Cref{chapter:calculus}).
\sourceLang{} is a type safe two-stage calculus, and is the first calculus
to support effect handlers at both compile-time and run-time stages.

\item
A framework based on \sourceLang{} and \coreLang{} that facilitates
formalization and evaluation of different scope extrusion checks as a
family of elaborations from \sourceLang{} to \coreLang{}~(\Cref{chapter:scope-extrusion}).
We use the framework to study a variety of designs: a lazy check
(\Cref{section:lazy-dynamic-check-formal}), an eager check
(\Cref{section:eager-dynamic-check-formal}), and our novel \novelCheck{}
check~(\Cref{section:best-effort-check}).

\item
An extension of \sourceLang{} and \coreLang{} with \citeauthor{kiselyov-16}'s [\citeyear{kiselyov-16}] \textit{refined environment
classifiers} (\Cref{section:refined-environment-classifier-calculus}), with a proof of correctness via
a logical relation (\Cref{subsection:rec-formal-correctness}), and an
evaluation of its expressiveness compared to the dynamic checks
(\Cref{subsection:rec-formal-expressiveness}).

\item
Implementations of the three dynamic checks in the MacoCaml language (\Cref{chapter:implementation}). An implementation with the C4C check is available as an artifact \citep{artifact}.
\end{itemize}

Finally, \Cref{chapter:related} presents related work
and \Cref{chapter:conclusions} concludes.

\section{Overview}\label{section:overview}
While there are many metaprogramming languages, our discussion will be grounded in MacoCaml~\citep{xie-2023,chiang-2024}. The MacoCaml project extends the OCaml programming language with metaprogramming facilities for compile-time program generation: a type constructor \lstinline{'a expr} for code of type \lstinline{'a}, and quote \lstinline{<< >>} and splice \lstinline{$} forms for constructing \lstinline{expr} values.

At a high level, elements of \lstinline{'a expr} correspond to ASTs of type \lstinline{'a}. Quotation converts expressions to ASTs, and splices stop the conversion, allowing evaluation during AST creation. As an example, the metaprogram \lstinline{<<$(print_int (1+2); <<1+2>>) + 0>>} can be thought of as \lstinline{Plus((print_int (1+2); Plus(Int(1), Int(2))), Int(0))}. This conceptual model will be made precise in \Cref{chapter:calculus}.

\Cref{code:staged-matrix-multiplication} shows our running example, adapted from~\citet{kiselyov-14}, which generates code for matrix multiplication. The parameters \lstinline{a}, \lstinline{b}, \lstinline{c} are two-dimensional arrays; \lstinline{a.(0)} accesses the array representing first row of the matrix. Realistic implementations of matrix multiplication typically employ various sophisticated optimizations, but this simple code will be sufficient to highlight the interaction between code generation and effects that is the focus of this paper.

\begin{code}
\begin{macocaml}
macro iter a body = << for i = 0 to length $a - 1 do $(body <<i>>) done >>
macro mmul a b c =
  iter a @@ fun i ->
  iter <<$a.(0)>> @@ fun k ->
  iter <<$b.(0)>> @@ fun j ->
     << $c.($i).($j) <- $c.($i).($j)
                      + $a.($i).($k)
                      * $b.($k).($j) >>
  \end{macocaml}
  \captionof{listing}{Staged code that generates the familiar matrix multiplication code}%
\label{code:staged-matrix-multiplication}
\end{code}

\noindent
Line 1 defines a macro (i.e.~a compile-time function) that generates code for a \lstinline!for! loop from the code for a term of array type \lstinline!a! and the result of calling the \lstinline!body! function with the loop variable \lstinline!i!.  Lines 2--8 define a second macro \lstinline!mmul! that uses \lstinline!iter! to construct a triply-nested loop.  The \lstinline!@@! operator denotes function application, and is used to avoid proliferation of parentheses.

\subsection{Effect Handlers in Staging}
Effect handlers are a composable and customisable mechanism for simulating effects~\citep{pretnar-15}. As with metaprogramming, there are many variants of effect handlers, and we ground our discussion in deep, unnamed effect handlers that permit multi-shot continuations (a calculus is presented in \Cref{section:source-lang}). In this section, our examples use OCaml's deep, unnamed effect handlers, which permit only single-shot continuations~\citep{sivaramakrishnan-21}.

Given the utility of metaprogramming and effect handlers, it is wise to consider how a language that offers both might mediate their interaction. Complete separation may be undesirable, since effects are very useful for relaxing the stack discipline that would otherwise tightly couple the structure of the generated and generating code. Concretely, effects allow the programmer to easily perform \textit{let-insertion}~\citep{kameyama-shifting-jfp,yallop-2019} (\Cref{code:let-insertion-with-quotes-and-effects}):

\begin{code}
\begin{macocaml}
type _ Effect.t += Genlet : int expr * int expr -> int expr t
macro genlet x e = perform (Genlet (x, e))
macro handle_genlet body i = try body i
                             with effect Genlet (x, e), k ->
                               if x == i then <<let y = $e in $(continue k <<y>>)>>
                               else continue k (genlet x e)
\end{macocaml}
\captionof{listing}{Effect handlers and quotes and splices combine to perform let-insertion}%
\label{code:let-insertion-with-quotes-and-effects}
  \end{code}

\noindent
Lines 1--2 define
a new effect constructor \lstinline!Genlet! and a compile-time
function that performs the \lstinline!Genlet! effect.
The \lstinline!Genlet! effect takes two arguments: the first
identifies the insertion point for the new binding, and second the
expression to be bound.
Lines 3--6 define a handler for \lstinline!Genlet! that either
installs a \lstinline!let! binding on the stack (line 5) or forwards
the effect to an outer handler (line 6).
The invocation \lstinline!handle_genlet body i!
wraps the code generated by \lstinline!body i! with
a \lstinline!let! binding for each call to \lstinline!genlet i e!
that takes place during the execution of \lstinline!body!.

In the \lstinline!mmul! example (\Cref{code:staged-matrix-multiplication}), the expression \lstinline{$a.($i).($k)} does not depend on \lstinline{j}, and therefore can be lifted out of the loop, an optimisation known as \textit{loop-invariant code motion}. Effects are a convenient way to perform these types of optimisations in staged programs while maintaining the structure of the generating code.
Without effects, the stack discipline couples the structures of the generating and generated code, so that code motion requires updating the generator to lift the expression \lstinline{$a.($i).($k)} above the \lstinline{iter <<$b.(0)>>} expression.

\Cref{code:let-inserted-matrix-multiplication} shows the
example updated to use let-insertion.
On Line 1, \lstinline!handle_genlet! now installs a \lstinline!Genlet! handler while generating each loop, and
Lines~7 and~8 now perform the \lstinline!genlet! effect.

\begin{code}
\begin{macocaml}
macro iter a body = << for i = 0 to length $a - 1 do $(handle_genlet body <<i>>) done >>
macro mmul a b c =
  iter a @@ fun i ->
  iter <<$a.(0)>> @@ fun k ->
  iter <<$b.(0)>> @@ fun j ->
     << $c.$i.$j <- $c.($i).($j)
                  + genlet k <<$a.($i).($k)>> (*instead of $a.($i).($k)*)
                  * genlet j <<$b.($k).($j)>> (*instead of $b.($k).($j)*) >>
  \end{macocaml}
  \captionof{listing}{Effect handlers and quotes and splices combine to perform let-insertion}%
\label{code:let-inserted-matrix-multiplication}
\end{code}

Unfortunately, it is easy to make mistakes when performing an optimisation of this kind, leading to \textit{scope extrusion}. Assume that, given how arrays are laid out in memory, and the specific design of the cache prefetcher, it is more efficient to interchange the \texttt{j} and \texttt{k} loops:

\begingroup\lstset{firstnumber=3}
\centering
\begin{macocamllst}
iter a @@ fun i ->
iter <<$b.(0)>> @@ fun j ->
iter <<$a.(0)>> @@ fun k -> ...
\end{macocamllst}
\endgroup

\noindent
Should the programmer realise this, they may perform this interchange without changing the let-insertion code. But the use of \lstinline{genlet} in \Cref{code:let-inserted-matrix-multiplication} assumes that the \texttt{k} loop is above the \texttt{j} loop, so this change would result in scope extrusion.

Alternatively, the programmer may identify the wrong let insertion point:

\begingroup\lstset{firstnumber=6}
\centering
\begin{macocamllst}
<< $c.$i.$j <- $c.($i).($j)
             + genlet i <<$a.($i).($k)>> (*should be k, not i*)
             * genlet k <<$b.($k).($j)>> (*should be j, not k*) >>
\end{macocamllst}
\endgroup

\noindent
This mistake also results in scope extrusion. Detecting such errors requires a scope extrusion check.

\subsection{Checking for Scope Extrusion}

In theory, it is possible to adopt a \textbf{lazy check} (\Cref{section:lazy-dynamic-check-formal}), which waits until the end of the program generation process to check that the generated program contains no free variables \citep{kiselyov-14}.
Lazy checking amounts to type checking generated code, an approach used in the original MetaOCaml implementation~\citep{DBLP:conf/popl/TahaN03}, LMS~\citep{RandIR}, and other systems.
The lazy approach has three drawbacks: first, it is inefficient, since it allows a code generator to run to completion after an error has occurred. Second, it produces uninformative error messages that refer to the generated code rather than the code generator. Third, in some systems, it 
can bind variables in unintended ways.

Therefore, BER MetaOCaml instead adopts an \textbf{eager check} \citep{kiselyov-14,kiselyov-2024}, which we describe formally in \Cref{section:eager-dynamic-check-formal}. By checking at various points during the code generation process, the eager check identifies the error early and raises an informative error message \citep[\S5.1]{kiselyov-14}.

While the eager check provides better error reporting than the lazy check, it does not allow effect handlers and metaprogramming to interact as freely as one might desire. For example, a common use-case for effect handlers is \textit{parameterisation}: by choosing different handlers for the same effect, the same piece of code can be specialised in various contexts \citep{wang-2019}.  Parameterisation uses effect handlers in a very simple way; we use it in our example to show that even straightforward uses of effect handlers interact poorly with the eager check.

To extend the matrix multiplication generator with parameterisation, suppose that we wish to generate \lstinline{e1 + e2 * e3} by default, and generate a call to a fused multiply-add instruction \lstinline{__fma(e1, e2, e3)} in contexts where performance takes priority over preserving exactly the expected floating point behaviour. \Cref{code:effect-parameterisation} shows one way to parameterise over these alternatives, first abstracting the choice as an effect, \lstinline!FMA! (lines 1-2), then modifying the body of \lstinline{mmul} to perform the effect (lines 6--8), then defining handlers that can be used to tune the generation process (lines 10--13), and finally installing a handler around the call to \lstinline!mmul! (line 15).  Line 15 makes use of a \emph{top-level splice}, which is used in MacoCaml to insert the code generated by a macro into a larger program.

Unfortunately, the use of \lstinline{FMA} in \Cref{code:effect-parameterisation} is \textit{not} allowed by the eager check, which throws a scope extrusion error when Line 11 or Line 13 is executed. For example, in the body of the \lstinline{hdl_fma_def} handler, the code template \lstinline{<<$x + $y * $z>>} is evaluated in a scope where \lstinline{i}, \lstinline{j}, and \lstinline{k} are free. However, as \lstinline{x} is bound to \lstinline{<<c.($i).($j)>>}, the eager check reports scope extrusion.

\begin{code}
\begin{macocaml}
type _ Effect.t += FMA : int expr * int expr * int expr -> int expr t
macro fma x y z = perform (FMA (x, y, z))

macro mmul a b c =
  ...
     << $c.($i).($j) <- $(fma <<$c.($i).($j)>>
                           (genlet k <<$a.($i).($k)>>)
                           (genlet j <<$b.($k).($j)>>))>>

macro hdl_fma_def body = try body ()
                         with effect FMA (x, y, z), k -> continue k << $x + $y * $z >>
macro hdl_fma_opt body = try body ()
                         with effect FMA (x, y, z), k -> continue k << __fma($x, $y, $z) >>

let code = $(hdl_fma_def @@ fun () -> mmul <<a>> <<b>> <<c>>)
\end{macocaml}
\captionof{listing}{Handlers for selecting a multiply-and-add instruction}
  \label{code:effect-parameterisation}
\end{code}

To support more flexible interaction between effect handlers and metaprogramming, we introduce a novel continuation-aware \textbf{\novelCheck{}} check, explained in detail in \Cref{section:best-effort-check}, which allows code like \Cref{code:effect-parameterisation} to run to completion without reporting scope extrusion.

\section{Calculus}\label{chapter:calculus}
\newcommand{\compilemode}{\textbf{\textsf{\textcolor{compile}{c}}}}
\newcommand{\splicemode}{\textbf{\textsf{\textcolor{splice}{s}}}}
\newcommand{\quotemode}{\textbf{\textsf{\textcolor{quote}{q}}}}

\begin{figure}[ht]
  \centering
  \begin{tikzpicture}
    \begin{scope}[every node/.style={inner sep=0.3cm}, font=\ttfamily\scriptsize, anchor = east, rounded corners, line width = 0.4mm]
  \node[align=left, fill=sourceBackground, draw=sourceHighlight] (source-program) {\,\$(\textbf{do}$\, x \leftarrow \equote[\return{\texttt{0}}]$\\
  \,\quad \textbf{in} $(\lambda z. \return{z})(x))$};
  \node[align=left, fill=coreBackground, draw=coreHighlight] (core-program) at ($(source-program.east) + (5.6cm, 0cm)$){\textbf{tls}(\textbf{do}$\, x \leftarrow \return{\texttt{Ret}(\texttt{Nat}(\texttt{0}))}$ \\
  \quad\quad\textbf{in} $(\lambda z. \return{z})(x)$$)$};

  \node[align=left, fill=effBackground, draw=effHighlight] (normal-form) at ($(core-program.east) + (3.5cm, 0cm)$) {Ret$($Nat$(\texttt{0}))$\\ };
  \end{scope}

  \begin{scope}[every node/.style={font = \scriptsize}]
    \node[fill=sourceHighlight] (source-lang) at ($(source-program.south) + (0cm, 0cm)$ ) {\textcolor{white}\sourceLang{}};
    \node[fill=coreHighlight] (core-lang) at ($(core-program.south) + (0cm, 0cm)$ ) {\textcolor{white}\coreLang{}};
    \node[fill=effHighlight] (eff-lang) at ($(normal-form.south) + (0cm, 0cm)$ ) {\textcolor{white}\efflang{}};
  \end{scope}

  \begin{scope}[scale=8]
  \node[] (elaboration) at ($(source-program.east) + (0.095cm, -0.004cm)$) {\Huge$\leadsto$};
  \node[] (execution) at ($(core-program.east) + (0.095cm, 0.005cm)$) {\Huge$\rightarrow^{\text{\normalsize{$*$}}}$};
  \end{scope}

  \begin{scope}[every node/.style={font = \sffamily\tiny, anchor = south}]
  \node[] (elaboration-caption) at ($(elaboration) + (0cm, 0.3cm)$) {\textbf{Elaboration}};
  \node[align=center] (execution-caption) at ($(execution) + (0cm, 0.3cm)$) {\textbf{Compile-Time} \\ \textbf{Execution}};
  \end{scope}

  \end{tikzpicture}

  \caption{\calculusName{} is first elaborated into \coreLang{}, which is then executed \textbf{at compile-time} to obtain the AST of a run-time \efflang{} program. $\textbf{\texttt{tls}}$ is a marker which tracks the position of the \textbf{t}op-\textbf{l}evel \textbf{s}plice (\Cref{section:core-lang})}%
  \label{fig:elaboration-then-execution}
\end{figure}

To ground the discussion of dynamic scope extrusion checks, we introduce two novel calculi for studying the interaction between typed multi-stage programming and effects and handlers: \sourceLang{} and \coreLang{}.\ \sourceLang{} (\Cref{section:source-lang}) offers metaprogramming in the form of quotes and splices, and effect handlers.\ \coreLang{} offers metaprogramming in the form of AST constructors, and effect handlers. Following~\citet{calcagno-2003}, \sourceLang{} has no operational semantics; programs in \sourceLang{} are instead elaborated into \coreLang{} (\Cref{section:elaboration}), where they may then be executed, to obtain the AST of a run-time program that has no quotes and splices. This process is summarised in \Cref{fig:elaboration-then-execution}. Elaboration simplifies the operational semantics, and is a convenient mechanism for inserting dynamic checks (\Cref{chapter:scope-extrusion}).\ \sourceLang{} and \coreLang{} are both type safe (\Cref{section:metatheory}).

\subsection{The Source Language: \texorpdfstring{\sourceLang{}}{Lambda-Op-Quote-Splice}}\label{section:source-lang}

\sourceLang{} (\Cref{fig:source-syntax-types}) is a language which offers both metaprogramming, in the form of quotes $\equote$ and splices $\splice$, as well as effect handlers~\citep{pretnar-15}. Syntactic \sourceLang{} terms are divided into values, expressions, and handlers, similar to a fine-grained call-by-value approach~\citep{levy-2003}. Ignoring quotes and splices, and adding a continuation term former, $\kappa x. e$ that cannot be written explicitly but may be generated during reduction, one obtains the syntax of a standard base calculus of effects and handlers~\citep{pretnar-15,biernacki-2017,isoda-24}, which we refer to as \efflang{} \extendedOnly{(and which is described in \Cref{appendix:full-efflang-rules})}\unextendedOnly{(and which is described in an appendix in the extended version)}. Briefly, $\return{v}$ lifts a value into an expression, and $\bind{x}{e_1}{e_2}$ sequences expressions. $\op{v}$ performs an effect, suspending the current computation and throwing a value $v$ to be caught by some handler $h$ that was installed using $\handleWith{e}{h}$. Within the body of the handler, $\continue{k}{v}$ can be used to resume the suspended program ($k$), inserting the value $v$ in place of the performed effect.

Metaprogramming systems differ along several key dimensions: they can be homogeneous (where the generating and generated languages coincide) or heterogeneous, two-stage or multi-stage, compile-time or run-time \citep{metaprogramming-survey}.\ \sourceLang{} offers \textit{homogeneous, two-stage, compile-time} metaprogramming. Many practical systems, like MacoCaml and MetaOCaml, are homogeneous. Many practical use cases of MSP involve only two stages~\citep{inoue-2012}, and scope extrusion is often studied in two stage systems~\citep{isoda-24,kiselyov-16}. Similarly, \sourceLang{} offers \textit{deep, unnamed handlers} that \textit{permit multi-shot continuations}, modelling OCaml effect handlers, though generalised to multi-shot continuations. Multi-shot continuations, though not supported by OCaml, are useful for if/case insertion~\citep{yallop-2017}. Other effect systems also allow for shallow or sheep handlers, allow named handlers, or permit only one-shot continuations \citep{effects-bibliography}: we do not study these systems.

Following~\citet{calcagno-2003}, \calculusName{} has no operational semantics, but is instead elaborated into \coreLang{}.\ \coreLang{} programs may then be executed, to obtain the AST of a run-time \efflang{} program that has no quotes and splices. This process is summarised in \Cref{fig:elaboration-then-execution}. Elaboration simplifies the operational semantics and is a convenient mechanism for inserting dynamic checks (\Cref{chapter:scope-extrusion}).

\begin{figure}
\begin{source-desc}
  \footnotesize
  {\normalsize \textbf{Syntax}} \\
  $\begin{array}{@{}llll}
  \text{Values} & v & := & x \mid m \in \mathbb{N} \mid \lambda x. e \\

  \text{Expressions} & e & := & v_1\;v_2 \mid \return{v}  \mid \bind{x}{e_1}{e_2} \mid \op{v} \mid \handleWith{e}{h} \mid \continue{v_1}{v_2} \\
                             &&& \mid \equote \mid \splice \\
  \text{Handlers} & h & := &\returnHandler{x}{e} \mid h;\opHandler{x}{k}{e}
  \end{array}$

  \vspace{3mm}

\begin{minipage}[t]{0.5\textwidth}
  {\normalsize \textbf{Effect sets}} \\
    $\begin{array}{@{}ll}
    \begin{array}{@{}lllr}\textbf{Run-Time} & \xi ::= \emptyset \mid \xi \cup \{ \textsf{op}_i^{0}\}\\
    \textbf{Compile-Time} & \Delta ::= \emptyset \mid \Delta \cup \{ \textsf{op}_i^{-1}\} \end{array}
    \end{array}$
  \end{minipage}%
\begin{minipage}[t]{0.5\textwidth}
   {\textbf{\normalsize {Typing contexts}}}\\
  $\Gamma ::= \cdot \mid \Gamma, x:T^0 \mid \Gamma, x: T^{-1}$
\end{minipage}

\vspace{3mm}

    {\textbf{\normalsize {Types}}}\\
    \begin{minipage}[t]{0.5\textwidth}
      \textbf{Level 0}\\
  $\begin{array}{@{}ll}
     \text{Values }  S^0, T^0 ::= & \mathbb{N}^0 \mid {(\functionType[\xi]{S}{T})}^{0} \\
     & \mid {(\continuationType[\xi]{S}{T})}^{0}\\\vspace{0.4mm}
    \text{Computations} & T^0 \, ! \, \xi \mid T^0 \, ! \, \Delta  \mid T^0 \, ! \,  \Delta;\xi \\
    & \mid (\handlerType{S \, ! \, \xi_1}{T \, ! \, \xi_2})^0\, !\, \Delta\\\vspace{0.4mm}
    \text{Handlers} & (\handlerType{S \, ! \, \xi_1}{T \, ! \, \xi_2})^0
  \end{array}$
\end{minipage}%
  \begin{minipage}[t]{0.5\textwidth}
     \textbf{Level $-$1}\\
    $\begin{array}{@{}ll}
    \text{Values } S^{-1}, T^{-1} ::= & \mathbb{N}^{-1} \mid {(\functionType{S}{T})}^{-1} \\
    & \mid {(\continuationType{S}{T})}^{-1} \mid {\textsf{Code}({T^{0} \, ! \, \xi})}^{-1} \\\vspace{0.4mm}
    \text{Computations} & T^{-1} \, ! \, \Delta \\\\\vspace{0.4mm}
    \text{Handlers} & (\handlerType{S \, ! \, \Delta_1}{T \, ! \, \Delta_2})^{-1}
  \end{array}$
\end{minipage}

\end{source-desc}
\caption{\sourceLang{} syntax and types. For clarity, inferable levels are omitted: e.g.\ ${(\functionType[\xi]{S}{T})}^{0}$ means ${(\functionType[\xi]{S^0}{T^0})}^{0}$.}
\label{fig:source-syntax-types}
\end{figure}

Only expressions can be quoted (values and handlers cannot be): thus, quotes must generate run-time computations. For example, $\equote[\texttt{1}]$ is not valid syntax, instead, one must write $\equote[\return{\texttt{1}}]$. Similarly, $\equote[\return{\texttt{1}}]$ is an expression, not a value, so one must write $\bind{a}{\equote[\return{\texttt{1}}]}{\op{a}}$ rather than $\op{\, \equote[\return{\texttt{1}}] \,}$. However, we will abuse notation and write $\op{\equote[\texttt{1}]}$ in place of $\bind{a}{\equote[\return{\texttt{1}}]}{\op{a}}$.

\subsubsection{Type System}\label{subsection:sourcelang-type-system}
\Cref{fig:source-syntax-types} summarises the \sourceLang{} types. To motivate the type system, consider the following running example $e$ in \sourceLang{} extended with arithmetic:
\[e \triangleq {(\lambda x. \textbf{\texttt{get}}(); \equote[{\bind{y}{\splice[x]}{\textbf{\texttt{readInt}}() + y}}])}\]
Here $e$ is (1) a compile-time function that (2) takes the AST of a run-time computation of type $\mathbb{N} \, ! \, \{ \texttt{print} \}$, (3) performs a compile-time effect (\textbf{\texttt{get}}), and (4) returns the AST of a run-time computation of a different type. The program has the following type:
\begin{center}
\vspace{-3mm}
\begin{tikzpicture}
  \begin{scope}[local bounding box=group 1]
  \node[text width=\linewidth, align = center] (example) at (0, 0) {$(\functionType[]{{\textsf{Code}(\mathbb{N}^0 ! \{ \texttt{print} \})}^{-1}}{\textsf{Code}(\mathbb{N}^0 \, ! \, \{ \texttt{print}, \texttt{readInt} \})^{-1}})^{-1}$};
  '\end{scope}

  \node[align=center, font=\footnotesize](effects) at (-0.9cm,0.3cm){$\{ \texttt{get} \}$};

  \begin{scope}[on background layer, font=\scriptsize\bfseries, text=comment, align=left]
    \draw[draw=comment, line width = 0.3mm] ($(example.north east) + (-3cm, 0.4cm)$) circle[radius=1.4mm] node (annote-1) {\tiny{1}};
    \node[anchor = east, align=right] (annote-1-text) at ($(annote-1.west) + (0.1cm, 0cm)$) {Compile-time function} ;
    \draw[draw=comment, line width =0.3mm] ($(example.north east) + (-2.5cm, 0cm)$) |- (annote-1.east);

    \draw[draw=comment, line width = 0.3mm] ($(effects.north) + (-0.5cm, 0.2cm)$) circle[radius=1.4mm] node (annote-2) {\tiny{3}};
    \node[anchor = east, align=right] (annote-2-text) at ($(annote-2.west) + (0.1cm, 0cm)$) {Compile-time effects} ;
    \draw[draw=comment, line width =0.3mm] (effects.north) |- (annote-2.east);

    \draw[draw=comment, line width = 0.3mm] ($(example.south west) + (3.4cm, -0.4cm)$) circle[radius=1.4mm] node (annote-3) {\tiny{2}};
    \node[anchor = north west, align=left] (annote-3-text) at ($(annote-3.east) + (-0.1cm, 0.27cm)$) {Input: AST of a run-time \\ computation of type $\mathbb{N} \, ! \, \{\texttt{print}\}$} ;
    \draw[draw=comment, line width =0.3mm] ($(example.south west) + (2.9cm, 0cm)$) |- (annote-3.west);

    \draw[draw=comment, line width = 0.3mm] ($(example.south east) + (-5.5cm, -0.4cm)$) circle[radius=1.4mm] node (annote-4) {\tiny{4}};
    \node[anchor = north west, align=left] (annote-4-text) at ($(annote-4.east) + (-0.1cm, 0.27cm)$) {Output: AST of a run-time \\ computation of type $\mathbb{N} \, ! \, \{\texttt{print},\texttt{readInt}\}$} ;
    \draw[draw=comment, line width =0.3mm]  ($(example.south east) + (-6cm, 0cm)$) |- (annote-4.west);
  \end{scope}

\end{tikzpicture}
\end{center}

The type system stratifies types into compile-time ($T^{-1}$) and run-time ($T^0$) levels. The function $e$ has a compile-time type $(S \to T)^{-1}$, and cannot be used as a run-time function of type $(S \to T)^{0}$.

To support compile-time manipulation of run-time programs, the $\textsf{Code}(\effectType[\xi]{T^0})^{-1}$ type makes ASTs of level $0$ computations available at level $-1$. Only \textit{computations}, not values or handlers, can be turned into ASTs.

Effect sets are stratified into $\Delta$ (compile-time) and $\xi$ (run-time). In our running example, suppose $e$ is applied at compile-time to some term $e'$ of the right type. The application has a compile-time effect \texttt{get}, and returns an AST with two run-time effects, \texttt{print} and \texttt{readInt}.
\[\Gamma \vdash^{-1}_{\splicemode{}} e\; e' : \textsf{Code}(\mathbb{N}^0 \, ! \, \{ \texttt{print}, \texttt{readInt} \})^{-1} \, ! \, \{ \texttt{get} \}\]

Splicing the result of application lifts the compile-time AST into a run-time type that has unhandled effects at \textbf{both} compile-time and run-time.
\[\Gamma \vdash^{0}_{\quotemode{}} \splice[(e\;e')] : \mathbb{N}^0 \, ! \, \{ \texttt{get} \} ; \{ \texttt{print}, \texttt{readInt} \} \]

We track compile-time and run-time effects in separate sets. Compile-time effects are tracked in $\Delta (= \{ {\texttt{get}} \})$ and run-time effects in $\xi (= \{ {\texttt{print}}, {\texttt{readInt}} \})$. Distinguishing compile-time and run-time effects stratifies types (\Cref{table:type-strat}): what is a computation at run-time could have been a value at compile-time, and vice versa.

\begin{table}
  \centering
  \newcommand\T{\rule{0pt}{2.6ex}}
\newcommand\B{\rule[-1.2ex]{0pt}{0pt}}
  \caption{Stratification of level $0$ types}%
  \vspace{-4mm}
  \begin{tikzpicture}
    \node (table) {
      \begin{tabular}{ll|l|l|l}
        && \multicolumn{3}{c}{\textbf{Run-Time}} \T\B\\
        && \textbf{Value} & \textbf{Computation} & \textbf{Handler} \T\B \\ \hline
        \multirow{2}{*}{\textbf{Compile-Time}} & \textbf{Value} & $T^0$ & $\effectType[\xi]{T^0}$ & $(\handlerType{S^0 \, ! \, \xi_1}{T^0 \, ! \, \xi_2})^0$ \T\B\\ \cline{2-5}
        & \textbf{Computation} & $\effectType{T^0}$ & $\effectType[\Delta; \xi]{T^0}$ & $\effectType{(\handlerType{S^0 \, ! \, \xi_1}{T^0 \, ! \, \xi_2})^0}$ \T\B\\ \hline
      \end{tabular}
    };
    \begin{scope}[every node/.style={font=\bfseries\tiny , text=comment}]
      \node[anchor = north, align = right] (value-syntax) at ($(table.south) + (-1.6cm, -0.15cm)$) {Type of syntactic \\ level $0$ values $v$};

      \node[anchor = north, align = left] (expr-syntax) at ($(table.south) + (2.3cm, -0.15cm)$) {Type of syntactic \\ level $0$ expressions $e$};

      \node[anchor = north, align = left] (hdlr-syntax) at ($(table.south) + (4.9cm, -0.15cm)$) {Type of syntactic \\ level $0$ handlers $h$};

      \node[anchor = north, align = right] (cmnt) at ($(table.north west) + (2.1cm, -0.5cm)$) {Cannot be \\ directly created};
    \end{scope}

    \begin{scope}[->,>=stealth']
      \draw[comment, dashed] (value-syntax.east) -| ($(value-syntax.east) + (0.14cm, 0.5cm)$);
      \draw[comment, dashed] (expr-syntax.west) -| ($(expr-syntax.west) + (-0.14cm, 0.5cm)$);
      \draw[comment, dashed] (hdlr-syntax.west) -| ($(hdlr-syntax.west) + (-0.14cm, 0.5cm)$);
      \draw[comment, dashed] (cmnt.east) -| ($(cmnt.east) + (0.6cm, -0.6cm)$);
    \end{scope}
  \end{tikzpicture}
  \label{table:type-strat}
\end{table}

In \sourceLang{}, the use of a term typed at level $0$ \textit{always} results in compile-time computation (the second row of \Cref{table:type-strat}). For example, level $0$ values are elaborated into compile-time computations ($\effectType{T^0}$) that evaluate to ASTs of run-time values (\Cref{section:elaboration}):

\[\begin{array}{ll}
  T^0 \, ! \, \Delta  & \text{Compile-time computation, run-time value} \\
  & \textit{Inhabitants: }\text{Level $0$ values $v$, e.g. } \lambda x. \return{x} \\[1mm]
  T^0 \, ! \, \Delta; \xi & \text{Compile-time computation, run-time computation} \\
   & \textit{Inhabitants: } \text{Level $0$ expressions $e$, e.g. } \return{\texttt{1}} \\[1mm]
  (\handlerType{S^0 \, ! \, \xi_1}{T^0 \, ! \, \xi_2})^0\, !\, \Delta & \text{Compile-time computation, run-time handler} \\
   & \textit{Inhabitants: }\text{Level $0$ handlers $h$, e.g. } \{ \returnHandler{x}{\return{x}} \}
\end{array}
\]

Consequently, \textit{syntactic} values ($v$) at level $0$ do not have value \textit{type} ($T^0$). The relationship between syntax and types is more complicated than in \efflang{}. In contrast, level $0$ compile-time value types (the first row in \Cref{table:type-strat}) have no inhabitants in \sourceLang{} (but $T^0$ is used to type \textit{formal parameters} of functions at level $0$, like $x$ in $\lambda x. \return{0}$).

As the stratification is subtle, it is best revisited after covering the typing rules (\Cref{subsection:sourcelang-type-system}), core language (\Cref{section:core-lang}), and elaboration (\Cref{section:elaboration}).
\begin{table}
  \newcommand\T{\rule{0pt}{2.6ex}}
\newcommand\B{\rule[-1.2ex]{0pt}{0pt}}
  \centering
    \caption{The nine \sourceLang{} typing judgements}
    \vspace{-2mm}
  \begin{tabular}{l|l|l|l}
    & \textbf{Value } ($v$) & \textbf{Expression} ($e$) & \textbf{Handler} ($h$) \B \\ \hline
    \textbf{Compile} (\compilemode{}) & $\Gamma \vdash^{0}_{\compilemode{}} v: \effectType{T^{0}}$ & $\Gamma \vdash^{0}_{\compilemode{}} e: \effectType[\Delta ; \xi ]{T^{0}}$ & $\Gamma \vdash^{0}_{\compilemode{}} h: (\handlerType{\effectType[\xi_1]{S^0}}{\effectType[\xi_2]{T^{0}}})^{0} \, ! \, \Delta$ \T\B \\ \hline
    \textbf{Quote} (\quotemode{}) & $\Gamma \vdash^{0}_{\quotemode{}} v: \effectType{T^{0}}$ & $\Gamma \vdash^{0}_{\quotemode{}} e: \effectType[\Delta ; \xi ]{T^{0}}$ & $\Gamma \vdash^{0}_{\quotemode{}} h: (\handlerType{\effectType[\xi_1]{S^0}}{\effectType[\xi_2]{T^{0}}})^{0} \, ! \, \Delta$ \T\B \\ \hline
    \textbf{Splice} (\splicemode) & $\Gamma \vdash^{-1}_{\splicemode{}} v: {T^{-1}}$ & $\Gamma \vdash^{-1}_{\splicemode{}} e: \effectType{T^{-1}}$ & $\Gamma \vdash^{-1}_{\splicemode{}} h: (\handlerType{\effectType[\Delta_1]{S^{-1}}}{\effectType[\Delta_2]{T^{-1}}})^{-1}$ \T\B \\
  \end{tabular}
  \label{table:typing-judgements}
\end{table}

Selected \sourceLang{} typing rules are collated in \Cref{fig:source-cq-typing-rules}. Similar to~\citet{xie-2023}, typing judgements are indexed by one of three compiler modes: \textbf{Compile} (\compilemode{}), \textbf{Quote} (\quotemode{}), or \textbf{Splice} (\splicemode{}). However, unlike~\citet{xie-2023}, typing judgements do not need to be indexed by a level: since \sourceLang{} is a two-level system, each compiler mode uniquely determines a level (\compilemode{}$\mid$\quotemode{} $\mapsto$ $0$, \splicemode{} $\mapsto$ $-1$). For each mode, there are three typing judgements: one for each syntactic category (\Cref{table:typing-judgements}).

Modes are useful for elaboration.\ \compilemode{} identifies code that is \textcolor{compile}{\textbf{ambient}} and \textcolor{compile}{\textbf{inert}} (no surrounding quotes or splices).\ \splicemode{} identifies code that \textcolor{splice}{\textbf{manipulates ASTs}} at compile-time (last surrounding annotation is a splice).\ \quotemode{} identifies code whose \textcolor{quote}{\textbf{ASTs are manipulated}} at compile time (last surrounding annotation is a quote). Accordingly, top-level splices transition from \compilemode{} to \splicemode{}. Quotes transition from \splicemode{} to \quotemode{}. Splices ($\splice$) transition from \quotemode{} to \splicemode{}.\ \Cref{fig:the-need-for-modes} annotates a metaprogram (that evaluates to the AST of $\lambda x. \texttt{1}+\texttt{2}+\texttt{3}$) with modes.
\begin{figure}
\begin{center}
  \begin{tikzpicture}
  \node (example) {$\lambda x. \, \splice[(\bind{f}{(\lambda y. \equote[{\splice[(y)] + \texttt{2}}])}{\bind{a}{\equote[\texttt{1}]}{f a}})] + \texttt{3}$};
  \draw[fill = compile, draw=compile, anchor = north west] ($(example.south west) + (0.05cm, 0.06cm)$) rectangle ($(example.south west) + (0.55cm, -0cm)$);
  \draw[fill = splice, draw=splice, anchor = north west] ($(example.south west) + (0.9cm, 0.06cm)$) rectangle ($(example.south west) + (2.7cm, -0cm)$);
  \draw[fill = splice, draw=splice, anchor = north west] ($(example.south west) + (3.35cm, 0.06cm)$) rectangle ($(example.south west) + (3.55cm, -0cm)$);
  \draw[fill = quote, draw=quote, anchor = north west] ($(example.south west) + (3.75cm, 0.06cm)$) rectangle ($(example.south west) + (4.25cm, -0cm)$);
  \draw[fill = splice, draw=splice, anchor = north west] ($(example.south west) + (4.7cm, 0.06cm)$) rectangle ($(example.south west) + (6.3cm, -0cm)$);
  \draw[fill = quote, draw=quote, anchor = north west] ($(example.south west) + (6.67cm, 0.06cm)$) rectangle ($(example.south west) + (6.87cm, -0cm)$);
  \draw[fill = splice, draw=splice, anchor = north west] ($(example.south west) + (7.2cm, 0.06cm)$) rectangle ($(example.south west) + (8.1cm, -0cm)$);
  \draw[fill = compile, draw=compile, anchor = north west] ($(example.south west) + (8.3cm, 0.06cm)$) rectangle ($(example.south west) + (8.7cm, -0cm)$);

  \begin{scope}[every node/.style={anchor=north, font=\scriptsize}]
  \node at ($(example.south west) + (0.3cm, -0.1cm)$) {\compilemode{}};
  \node at ($(example.south west) + (1.8cm, -0.1cm)$) {\splicemode{}};
  \node at ($(example.south west) + (3.45cm, -0.1cm)$) {\splicemode{}};
  \node at ($(example.south west) + (4cm, -0.1cm)$) {\quotemode{}};
  \node at ($(example.south west) + (5.5cm, -0.1cm)$) {\splicemode{}};
  \node at ($(example.south west) + (6.77cm, -0.1cm)$) {\quotemode{}};
  \node at ($(example.south west) + (7.65cm, -0.1cm)$) {\splicemode{}};
  \node at ($(example.south west) + (8.5cm, -0.1cm)$) {\compilemode{}};
  \end{scope}
  \end{tikzpicture}
  \vspace{-6mm}
  \end{center}
  \caption{A metaprogram annotated with compiler modes}
  \label{fig:the-need-for-modes}
\end{figure}

\newcommand{\cqtypejudge}[3][\Gamma]{{#1} \vdash_{\compilemode \mid \quotemode} {#2} : {#3}}
\newcommand{\ctypejudge}[3][\Gamma]{{#1} \vdash_{\compilemode} {#2} : {#3}}
\newcommand{\qtypejudge}[3][\Gamma]{{#1} \vdash_{\quotemode} {#2} : {#3}}
\newcommand{\stypejudge}[3][\Gamma]{{#1} \vdash_{\splicemode} {#2} : {#3}}

\newcommand{\runtimecomptype}[2]{{#1} \, ! \, {#2}}
\newcommand{\compiletimetype}[1]{{#1}}
\newcommand{\compiletimecomptype}[2]{{#1} \, !  \,{#2}}

The typing judgements for \compilemode{} and \quotemode{} are identical in almost all cases. To avoid repetition, we introduce the notation $\cqtypejudge{e}{T}$ to stand for the two judgements $\ctypejudge{e}{T}$ and $\qtypejudge{e}{T}$. The mode of the conclusion will match the modes of the assumption, unless otherwise stated.

The \sourceLang{} typing rules are mostly standard for a calculus with effect handlers \extendedOnly{(see \Cref{appendix:full-quote-op-rules})}. In \compilemode{} and \quotemode{}, compile-time effects $\Delta$ are threaded through typing judgements, and only level $0$ variables in the context can be accessed. In \splicemode{}, only level $-1$ variables can be accessed. As the levels of types can, in most cases, be inferred: for readability, they too are mostly omitted. The three key rules are \textsc{\splicemode{}-Quote}, \textsc{\quotemode{}-Splice}, and \textsc{\compilemode{}-Splice}, which switch between modes and levels.

\newcommand{\cqmode}{\compilemode{}$\mid$\quotemode{}}
\begin{figure}
\begin{source-desc}
  {\large\textbf{Selected Typing Rules}}
  \\ \textit{Level annotations on types mostly omitted} \\
  \vspace{1mm}\\
  \scriptsize
  \begin{minipage}[t]{0.35\textwidth}
    \fbox{$\Gamma \vdash_{\compilemode{} \mid \quotemode{}} v: \effectType{T^{0}}$}\\[1.05mm]
    \begin{minipage}[t]{0.5\textwidth}
      \begin{center}
    $\inferrule[(\cqmode{}-Nat)]{ \\ }{\cqtypejudge{m}{\runtimecomptype{\mathbb{N}}{\Delta}}}$
    \end{center}
    \end{minipage}%
    \begin{minipage}[t]{0.5\textwidth}
      \begin{center}
    $\inferrule[(\cqmode{}-Var)]{\Gamma(x) = T^0}{\cqtypejudge{x}{\runtimecomptype{T^0}{\Delta}}}$
    \end{center}
    \end{minipage}

  \end{minipage}%
\begin{minipage}[t]{0.65\textwidth}
\fbox{$\Gamma \vdash_{\compilemode{} \mid \quotemode{}} e: \effectType[\Delta; \xi]{T^{0}}$}\\[1mm]
\begin{minipage}[t]{0.6\textwidth}
\centering
\begin{center}
$\inferrule[(\cqmode{}-App)]{\cqtypejudge[\Gamma]{v_1}{\runtimecomptype{(\functionType[\xi]{S}{T})}{\Delta}} \\ \cqtypejudge[\Gamma]{v_2}{\runtimecomptype{S}{\Delta}}}{\cqtypejudge{v_1 v_2}{\runtimecomptype{T}{\Delta;\xi}}}$
\end{center}
\end{minipage}%
\begin{minipage}[t]{0.4\textwidth}
  \begin{center}
  $\inferrule[(\cqmode{}-Splice)]{\stypejudge[\Gamma]{e}{\textsf{Code}(T^0 \, ! \, \xi)^{-1} \, ! \, \Delta}}{\cqtypejudge{\splice}{\runtimecomptype{T^0}{\Delta ; \xi}}}$
  \end{center}
\end{minipage}
\end{minipage}
\vspace{3mm}
\begin{center}

  \begin{minipage}[t]{0.35\textwidth}
   \fbox{$\Gamma \vdash_{\splicemode{}} v: {T^{-1}}$}\\[1.1mm]
    \begin{minipage}[t]{0.5\textwidth}
      \begin{center}
    $\inferrule[(\splicemode{}-Nat)]{ \\ }{\stypejudge{m}{\mathbb{N}}}$
    \end{center}
    \end{minipage}%
    \begin{minipage}[t]{0.5\textwidth}
      \begin{center}
    $\inferrule[(\splicemode{}-Var)]{\Gamma(x) = T^{-1}}{\stypejudge{x}{T^{-1}}}$
    \end{center}
    \end{minipage}
  \end{minipage}%
  \begin{minipage}[t]{0.65\textwidth}
    \fbox{$\Gamma \vdash_{\splicemode{}} e: \effectType{T^{-1}}$}\\[1mm]
    \begin{minipage}[t]{0.6\textwidth}
\begin{center}
$\inferrule[(\splicemode{}-App)]{\stypejudge[\Gamma]{v_1}{\functionType[\Delta]{S}{T}} \\ \stypejudge[\Gamma]{v_2}{S}}{\stypejudge{v_1 v_2}{\effectType{T}}}$
\end{center}
\end{minipage}%
\begin{minipage}[t]{0.4\textwidth}
  \begin{center}
    $\inferrule[(\splicemode{}-Quote)]{ \qtypejudge{e}{\runtimecomptype{T^0}{\Delta ; \xi}}}{\stypejudge{\equote}{\effectType{\compiletimetype{\textsf{Code}(\runtimecomptype{T^0}{\xi})}^{-1}}}}$
  \end{center}
\end{minipage}
 \end{minipage}\\

\end{center}
\end{source-desc}
\caption{Selected typing rules for \sourceLang{}.}%
\label{fig:source-cq-typing-rules}
\end{figure}

A closed \sourceLang{} expression is well-typed if, in \compilemode{}-mode, it can be typed with empty compile-time and run-time effect sets: all effects are provably handled, both at compile-time and run-time.
\begin{definition}[Well-Typed Closed Expression]
  A closed expression $e$ is well-typed if $\ctypejudge[\cdot]{e}{T^0 \, ! \, \emptyset;\emptyset}$
\end{definition}

\subsection{The Core Language: \texorpdfstring{\coreLang{}}{Lambda-Op-AST}}\label{section:core-lang}
\newcommand{\coreConfiguration}[5]{\langle {#1}; {#2}; {#3}; {#4}; {#5} \rangle}
  \renewcommand{\transition}[2]{#1 & \rightarrow & #2}
  \newcommand{\astRule}[1]{\rulename{Ast}{#1}}
  \newcommand{\secRule}[1]{\rulename{Sec}{#1}}

\coreLang{} (\Cref{fig:core}) is a language which offers AST constructors and effect handlers. Syntax is divided into normal forms, terms, and handlers. The syntax of \coreLang{} combines a standard calculus of effect handlers (\efflang{}) with machinery for AST construction, and primitives for scope extrusion checking.

\coreLang{}'s machinery for AST construction comprises one \textbf{AST} node for each \efflang{} term former that can be written by the user (e.g.\ \texttt{Var} for variables, $\texttt{H}_{\texttt{op}}$ for \_;$\opHandler{\_}{\_}{\_}$), as well as \textit{type-annotated} \textbf{formal parameters} ($\alpha_R$, where $R$ is some run-time value pre-type (\Cref{fig:core}), henceforth simply ``type''). Formal parameters represent binding sites, e.g. $x$ in $\lambda x. \return{0}$. Separating ASTs and formal parameters mirrors the approach by~\citet{calcagno-2003}, though they use untyped formal parameters. Additionally, \coreLang{} adds $\gensym{R}$, a primitive for generating fresh formal parameters of type $R$, $\Binder{\alpha}{R}$, where separate calls to \textbf{\texttt{mkvar}} return distinct formal parameters~\citep{taha-1999}.

\coreLang{}'s machinery for scope extrusion checking comprises:
\begin{itemize}
  \item \err{}, an error state for indicating the presence of scope extrusion,
  \item \textbf{\texttt{check}} and $\textbf{\texttt{check}}_\textsf{M}$, guarded \textbf{\texttt{return}}s that either report scope extrusion or \textbf{\texttt{return}} normally,
  \item \textbf{\texttt{dlet}}, a primitive for tracking which variables are well-scoped and which have extruded their scope, and
  \item \textbf{\texttt{tls}}, a marker representing an occurrence of a top-level splice in the source program: at this point, remaining stack frames either introduce a new top-level splice, or construct an AST in an entirely straightforward way, with standard (and thus safe) control flow.
\end{itemize}

\begin{figure}
\begin{core-desc}
  \footnotesize
  {\normalsize\textbf{Syntax}} \\
  $\begin{array}{@{}llll}
    \textbf{Formal Params} & \alpha_R \\
    \textbf{Normal Forms} & n & ::= & x \mid m \in \mathbb{N} \mid \lambda x. t \mid \kappa x. t \mid  \Nat{m} \mid \alpha_R \mid \Var{\alpha}{R} \mid \Lam{n_1}{n_2} \mid \App{n_1}{n_2} \\
    &&& \mid \Continue{n_1}{n_2} \mid \Ret{n} \mid \Do{n_1}{n_2}{n_3} \mid \Op{n} \mid \Hwith{n_1}{n_2}   \\
  &&&\mid \Hret{n_1}{n_2}(n_1, n_2) \mid \Hop{n_1}{n_2}{n_3}{n_4} \\
  \textbf{Terms} & t & := & n_1\;n_2 \mid \return{n}  \mid \bind{x}{t_1}{t_2} \mid \op{t} \mid \handleWith{n}{h} \mid \continue{n_1}{n_2} \\
  &&& \mid \checkfv{n} \mid \checkm{n} \mid \gensym{R} \mid \dlet{n}{t} \mid \tls{t} \mid \err \\
  \textbf{Handlers} & h & := & \returnHandler{x}{t} \mid \opHandler{x}{k}{t} \\
  \end{array}$

\vspace{3mm}

{{\textbf{\normalsize {Typing contexts}}}\\
  $\Gamma ::= \cdot \mid \Gamma, x:T$}

\vspace{3mm}

    {{\normalsize\textbf{Types}}}\\
    \begin{minipage}[t]{0.4\textwidth}
    \textbf{Run-time Pre-types}\\
    $\begin{array}{@{}lllr}
    \text{Effects set } \xi ::= & \emptyset \mid \xi \cup \{ \texttt{op}_i \} \\
    \text{Value type } Q,R ::= & \mathbb{N} \mid  \functionType[\xi]{Q}{R}  \\ \\
    & \mid \continuationType[\xi]{Q}{R} \\
    \text{Computation type} & \effectType[\xi]{R} \\
    \text{Handler type} & \handlerType{\effectType[\xi_1]{Q}}{\effectType[\xi_2]{R}}
    \end{array}$
  \end{minipage}%
  \begin{minipage}[t]{0.6\textwidth}
    \textbf{Types}\\
  $\begin{array}{@{}lllr}
    \text{Effects set } & \Delta ::= \emptyset \mid \Delta \cup \{ \texttt{op}_i \}  \\
    \text{Value type }  S, T ::= & \mathbb{N} \mid  \functionType[\Delta]{S}{T}  \mid \continuationType[\Delta]{S}{T} \\
    & \mid \textsf{FParam}(R) \mid \textsf{AST}(R) \\
    & \mid \textsf{AST}(\effectType[\xi]{R})  \mid \textsf{AST}(\handlerType{\effectType[\xi_1]{Q}}{\effectType[\xi_2]{R}})\\
    \text{Computation type} & \effectType{T} \\
    \text{Handler type} & \handlerType{\effectType{S}}{\effectType[\Delta_2]{T}}
  \end{array}$
\end{minipage}

\vspace{3mm}

  {\normalsize \textbf{Operational Semantics}}\\
  {\textit{Selected Rules}}\\
  {
    \scriptsize

{\textbf{Auxiliary Definitions}}\\
  {\[\begin{array}{lrcl}
    \text{Evaluation Frame } & F & ::= & \bind{x}{[-]\,}{t_2} \mid \handleWith{[-]}{h} \mid \dlet{\Binder{\alpha}{R}}{[-]} \mid \tls{[-]} \\
    \text{Evaluation Context } & E & ::= & [-] \mid E[F] \\ \vspace{1mm} \\
    \text{Domain of Handler} & \textsf{dom}(h) & \triangleq & \textsf{dom}(\returnHandler{x}{t}) = \emptyset, \\
    &&&\textsf{dom}(h;\opHandler{x}{k}{t}) = \textsf{dom}(h) \cup \{ \textbf{\textsf{op}} \} \\     \vspace{1mm}
    \text{Handled Effects} & \textsf{handled}(E) & \triangleq & \textsf{handled}([-]) = \emptyset, \\
    &&& \textsf{handled}(E[\bind{x}{[-]\,}{t_2}]) = \textsf{handled}(E), \\
    &&& \textsf{handled}(E[\handleWith{[-]}{h}]) = \textsf{handled}(E) \cup \textsf{dom}(h), \\
    &&& \textsf{handled}(E[\dlet{\Binder{\alpha}{R}}{[-]}]) = \textsf{handled}(E), \\
    &&& \textsf{handled}(E[\tls{[-]}]) = \textsf{handled}(E)
  \end{array}
  \]}

 {\textbf{Reduction Rules}}\\
 \textit{Mechanisms related to muting and unmuting are \textbf{\textcolor{coreHighlight}{highlighted}}}\\

\begin{minipage}[t]{0.5\textwidth}
  $\begin{array}{@{}lcl}
    \astRule{Gen}\\
  \transition{\coreConfiguration{\gensym{R}}{E}{U}{M}{I}}{\coreConfiguration{\return{\Binder{\alpha}{R}}}{E}{U\cup\{\alpha\}}{M}{I}}
  \end{array}$\\
  \textit{\textcolor{comment}{where $\alpha = \textsf{next}(U), \textsf{next}(U) \notin U, \text{\textsf{next} deterministic}$}}\\[1mm]
  $\begin{array}{@{}lcl}
  \secRule{Chs}\\
   \transition{\coreConfiguration{\checkfv{n}}{E}{U}{M}{I}}{\coreConfiguration{\return{n}}{E}{U}{M}{I}}
  \end{array}$\\
  \textit{\textcolor{comment}{if $\freevars{n} \subseteq \projfvs{E}$}}

  $\begin{array}{@{}lcl}
  \secRule{Chf}\\
   \transition{\coreConfiguration{\checkfv{n}}{E}{U}{M}{I}}{\coreConfiguration{\err}{E}{U}{M}{I}}
  \end{array}$\\
  \textit{\textcolor{comment}{if $\freevars{n} \not\subseteq \projfvs{E}$}}
\end{minipage}%
\begin{minipage}[t]{0.5\textwidth}%
    $\begin{array}{@{}lcl}
      \secRule{Tls}\\
       \transition{\coreConfiguration{\tls{\return{n}}}{E}{U}{M}{I}}{\coreConfiguration{\return{n}}{E}{U}{\textcolor{coreHighlight}{\emptyset}}{\textcolor{coreHighlight}{\top}}}
    \end{array}$\\[\lineskip]
    \\[2.5mm]
  $\begin{array}{@{}lcl}
    \secRule{Cms}\\
   \transition{\coreConfiguration{\checkm{n}}{E}{U}{M}{I}}{\coreConfiguration{\return{n}}{E}{U}{M}{I}}\end{array}$\\
    \textit{\textcolor{comment}{if $\freevars{n} \setminus M \subseteq \projfvs{E}$}}\\
  $\begin{array}{@{}lcl}
    \secRule{Cmf}\\
   \transition{\coreConfiguration{\checkm{n}}{E}{U}{M}{I}}{\coreConfiguration{\err}{E}{U}{M}{I}}
  \end{array}$\\
  \textit{\textcolor{comment}{if $\freevars{n} \setminus M \not\subseteq \projfvs{E}$}}
\end{minipage}
\\[1mm]
  $\begin{array}{@{}lcl}
  \secRule{Dlt} \\
  \transition{\coreConfiguration{\dlet{\Binder{\alpha}{R}}{\return{n}}}{E}{U}{M}{I}}{\coreConfiguration{\return{n}}{E}{U}{\textcolor{coreHighlight}{M'}}{\textcolor{coreHighlight}{I'}}}\end{array}$\\
\textit{\textcolor{coreHighlight}{if $\textsf{len}(E) > I$ then $M' = M, I' = I$, else $M' = \emptyset, I' = \top$}}\\[1.4mm]
$\begin{array}{@{}lcl}
   \effectRule{Op}\\
   \transition{\coreConfiguration{\op{v}}
                                                   {E_1[\handleWith{E_2}{h}]}
                                                   {U}
                                                   {M}
                                                   {I}}
                                {\coreConfiguration{c[v/x, \text{cont}/ k]}
                                                   {E_1}
                                                   {U}
                                                   {\textcolor{coreHighlight}{M \cup \projfvs{E_2}}}
                                                   {\textcolor{coreHighlight}{I'}}}\end{array}$\\
  \textit{\textcolor{comment}{where cont $=\kappa x. \, \handleWith{E_2[\return{x}]}{h}$ and $\opHandler{x}{k}{c} \in h$ and $\textbf{\textsf{op}} \notin \textsf{handled}(E_2)$ and \textcolor{coreHighlight}{$I' = \textsf{min}(\textsf{len}(E_1), I)$}}}
}

\end{core-desc}
\caption{\coreLang{}: syntax, types, and operational semantics.}
\label{fig:core}
\end{figure}

Notice that, while the calculus provides the \textit{machinery} for scope extrusion checking, it does not demand that one \textit{use} it, or use it \textit{properly}. Scope extrusion checking is not a language feature, but an algorithm one builds on top of the calculus.
\subsubsection{Operational Semantics}

The operational semantics of \coreLang{} is defined over configurations $ \langle t; E; U; M; I \rangle $. At a high level, $t$ are terms and $E$ are evaluation contexts, defined as a stack of evaluation frames, à la \citet{DBLP:conf/lfp/FelleisenWFD88}. $U$ acts as a source of fresh names. $M$ is a set of \emph{muted} variables, i.e.\ those that do not trigger a scope extrusion error, even if they have extruded their scope. $I$ indicates the point at which variables in $M$ should be \textit{unmuted}, by setting $M$ to $\emptyset$. Collectively, $M$ and $I$ determine whether to perform the check immediately ($M = \emptyset$), or defer it to a later point (marked by $I$). Deferring checking in the presence of a continuation that could later be used to recover from scope extrusion is used by the C4C check, making it ``continuation-aware''. The semantics for the lazy and eager checks can be more simply given as 3-tuple transition systems, which are straightforward projections of the 5-tuple system used to compare the three checks.

The operational semantics is mostly as expected for a calculus with effect handlers. Interesting rules are collated in \Cref{fig:core}, \unextendedOnly{and full rules in an appendix in the extended version}\extendedOnly{and full rules in \Cref{appendix:full-corelang-rules}}.

In the \textsc{Ast-Gen} rule, $U$ ensures freshness by recording previously generated names. To ensure determinacy of the semantics, fresh names are chosen by some (unspecified) deterministic process.

The \textbf{\texttt{check}} primitive acts like a guarded \textbf{\texttt{return}}. For some arbitrary normal form $n$ of AST type, either all the free variables of $n$ are properly scoped, so $\checkfv{n}$ reduces to $\return{n}$ (\textsc{Sec-Chs}), or some  free variables of $n$ are not properly scoped, so $\checkfv{n}$ reduces to $\err$ (\textsc{Sec-Chf}).
Following~\citet{kiselyov-2024}, \textbf{\texttt{dlet}}s declare that variables are properly scoped, by placing a frame of the form $\dlet{\Binder{\alpha}{R}}{[-]}$ on the evaluation context $E$. The notation $\projfvs{E}$ filters out the variables declared in this manner from $E$. For example, $\projfvs{\dlet{\Binder{\alpha}{R}}{\bind{x}{[-]}{t}}} = \{ \Var{\alpha}{R} \}$. Given a term $E[t]$, $\Var{\alpha}{R}$ in $t$ is ``declared safe'' in $E$ if $\Var{\alpha}{R} \in \projfvs{E}$ (\Cref{dfn:declared-safe}).
\begin{definition}[Declared Safe]\label[definition]{dfn:declared-safe}
  Given a term $E[t]$, $\Var{\alpha}{R}$ in $t$ is \emph{\textbf{declared safe}} in $E$ if $\Var{\alpha}{R} \in \projfvs{E}$
\end{definition}
Given a normal form ${n}$ in some evaluation context $E$, where $n$ is an AST, $n$ is properly scoped in $E$ (that is, $\textbf{\texttt{check}} \, n$ succeeds) if and only if the free \texttt{Var}s of $n$, written $\freevars{n}$, have all been declared safe in $E$, i.e.~$\freevars{n} \subseteq \projfvs{E}$. As \coreLang{} is an elaboration target for \sourceLang{}, it is up to the elaboration to use \textbf{\texttt{dlet}} and \textbf{\texttt{check}} appropriately.

The \textbf{\texttt{check}}$_\textsf{M}$ construct is a variant of \textbf{\texttt{check}}. As \Cref{section:best-effort-check} explains, \textbf{\texttt{check}}$_\textsf{M}$ additionally ignores some \textit{muted} variables, treating them as properly scoped ($\checkm{n}$ succeeds if $\freevars{n} \text{\colorbox{yellow}{$ \, \setminus \, M$}} \subseteq \projfvs{E}$).

\textsc{Sec-Tls}, \textsc{Sec-Dlt}, and \textsc{Eff-Op} mute or unmute variables.\ \Cref{section:best-effort-check} explains muting and unmuting. Ignoring muting and unmuting, \textsc{Sec-Tls} and \textsc{Sec-Dlt} silently remove a $\tls{[-]}$ and $\dlet{\Binder{\alpha}{R}}{[-]}$ frame respectively, and \textsc{Eff-Op} gives handlers the expected, standard behaviour.

\subsubsection{Type System}
\coreLang{} types are mostly standard. The key additions are an \textsf{FParam} type for formal parameters and an \textsf{AST} type for abstract syntax trees (\Cref{fig:core}).

The \coreLang{} typing rules (\Cref{fig:core-typing-rules}) are extremely straightforward. Under the typing rules, a well-typed AST can be \textit{ill-scoped}; for example, $\cdot \vdash {\Var{\alpha}{R}}: {\textsf{AST}(R)}$ is a valid typing judgement. Scope extrusion checks are effectively invisible to the type system. The only complex case is \textbf{\texttt{err}}, which can be assigned any type in any context, similarly to \textbf{\texttt{abort}}~\citep{scherer-2017}.

A closed \coreLang{} term is well-typed if it can be typed with an empty effects set.
\begin{definition}[Well-Typed Closed Term]
A closed term $t$ is well-typed if $\cdot \vdash t: T \, ! \,  \emptyset$
\end{definition}
\begin{figure}
  \begin{core-desc}
    {\large \textbf{Typing Rules}}\\
     \footnotesize
    \textit{Selected Rules}
    \begin{center}
    \begin{minipage}[t]{0.24\textwidth}
      \centering
    $\inferrule[(FParam)]{ \\ }{\type{\Binder{\alpha}{R}}{\textsf{FParam}(R)}}$
    \end{minipage}%
    \begin{minipage}[t]{0.3\textwidth}
      \centering
    $\inferrule[(Var-AST)]{ \type{n}{\textsf{FParam}(R)} }{\type{\texttt{Var}{(n)}}{\textsf{AST}(R)}}$
    \end{minipage}%
    \begin{minipage}[t]{0.26\textwidth}
      \centering
    $\inferrule[(Mkvar)]{ \\ }{\type{\gensym{R}}{\effectType{\textsf{FParam}(R)}}}$
    \end{minipage}%
    \begin{minipage}[t]{0.2\textwidth}
      \centering
    $\inferrule[(Err)]{  \\  }{\type{\err}{\effectType{T}}}$
    \end{minipage}

    \vspace{3mm}

    \begin{minipage}[t]{0.35\textwidth}
      \centering
    $\inferrule[(Lambda-AST)]{\type{n_1}{\textsf{FParam}(Q)}\\{\type{n_2}{\textsf{AST}(\effectType[\xi]{R})}}}{\type{\Lam{n_1}{n_2}}{\textsf{AST}(\functionType[\xi]{Q}{R})}}$
    \end{minipage}%
    \begin{minipage}[t]{0.2\textwidth}
      \centering
    $\inferrule[(Tls)]{ \\\\ \type{t}{\effectType{T}}}{\type{\tls{t}}{\effectType{T}}}$
    \end{minipage}%
    \begin{minipage}[t]{0.25\textwidth}
      \centering
    $\inferrule[(DLet)]{\type{n}{\textsf{FParam}(R)} \\ \type{t}{\effectType{T}}}{\type{\dlet{n}{t}}{\effectType{T}}}$
    \end{minipage}%
    \begin{minipage}[t]{0.2\textwidth}
      \centering
    $\inferrule[(Check)]{\type{n}{T} \\ T \text{ of } \textsf{AST } \text{type} }{\type{\checkfv{n}}{\effectType{T}}}$
    \end{minipage}

  \end{center}
  \end{core-desc}

\caption{Selected \coreLang{} typing rules}
\label{fig:core-typing-rules}
\end{figure}

\subsection{Elaboration from \texorpdfstring{\sourceLang{}}{Lambda-Op-Quote-Splice} to \texorpdfstring{\coreLang{}}{Lambda-Op-AST}}\label{section:elaboration}
This section describes an elaboration ($\llbracket - \rrbracket$) from \sourceLang{} to \coreLang{}. This elaboration is simple: it does not insert any dynamic scope extrusion checks. Other elaborations in \Cref{chapter:scope-extrusion}, which do insert checks, extend this elaboration.

\newcommand{\elaborate}[1]{\llbracket #1 \rrbracket}
\newcommand{\erase}[1]{\textsf{erase}(#1)}
\newcommand{\AST}[1]{\textsf{AST}(#1)}
\newcommand{\Code}[1]{\textsf{Code}(#1)}

The elaboration is defined on typing judgements: \sourceLang{} judgements elaborate to \coreLang{} judgements. This decomposes into four elaborations: on effect sets, types, contexts, and terms.

\subsubsection{Elaborating Effect Sets and Types}

\renewcommand{\cqmode}{\renewcommand{\cqmode}{\compilemode{} \mid \quotemode{}}
}
\begin{figure}
  \begin{source-desc}
    \footnotesize

    {\normalsize\textbf{Type Elaboration}}\\
        \textit{Selected Rules}\\[1mm]
    \begin{minipage}[t]{0.5\textwidth}
      $\begin{array}{@{}lll}
  \elaborate{T^0} & = & \AST{\erase{T^0}}\\
  \elaborate{\effectType[\xi]{T^0}} & = & \AST{\erase{\effectType[\xi]{T^0}}}\\
  \elaborate{\effectType[\Delta]{T^0}} & = & \effectType[\elaborate{\Delta}]{\AST{\erase{T^0}}}\\
  \elaborate{\effectType[\Delta ; \xi]{T^0}} & = & \effectType[\elaborate{\Delta}]{\AST{\erase{\effectType[\xi]{T^0}}}}
      \end{array}$\end{minipage}%
      \begin{minipage}[t]{0.5\textwidth}
      $\begin{array}{@{}lll}
        \elaborate{\mathbb{N}^{-1}} & = & \mathbb{N} \\
  \elaborate{(\functionType{S^{-1}}{T^{-1}})^{-1}} & = & \functionType[\elaborate{\Delta}]{\elaborate{S^{-1}}}{\elaborate{T^{-1}}} \\
  \elaborate{(\continuationType{S^{-1}}{T^{-1}})^{-1}} & = & \continuationType[\elaborate{\Delta}]{\elaborate{S^{-1}}}{\elaborate{T^{-1}}} \\
  \elaborate{\textsf{Code}({\effectType[\xi]{T^{0}}})^{{-1}}} & = & {\textsf{AST}({\erase{{\effectType[\xi]{T^{0}}}}})}
      \end{array}$
    \end{minipage}

    \vspace{3mm}

    {\normalsize\textbf{Context Entry Elaboration}}\\
      \begin{minipage}[t]{0.2\textwidth}$\elaborate{\cdot} = \cdot$
      \end{minipage}%
      \begin{minipage}[t]{0.5\textwidth}$ \elaborate{\Gamma, x:T^0}  = \elaborate{\Gamma}, x: \textsf{FParam}(\erase{T^0})$
      \end{minipage}%
      \begin{minipage}[t]{0.3\textwidth}$\elaborate{\Gamma, x:T^{-1}} = \elaborate{\Gamma}, x: \elaborate{T^{-1}}$
      \end{minipage}

  \vspace{3mm}

{
  \renewcommand{\cqmode}{\compilemode{} \mid \quotemode{}}
    {\normalsize\textbf{Term Elaboration}}\\
    \textit{Selected Rules (AST)}\\[1mm]
    \begin{minipage}[t]{0.65\textwidth}
    $\begin{array}{@{}lll}
      \elaborate{x}_{\cqmode{}} & = & \varToAST{x}\\
      \elaborate{\lambda x: T^0. \, e}_{\cqmode} & = & \bind{x}{\gensym{\erase{T^0}}}{}\\
      &&{\bind{\texttt{body}}{\elaborate{e}_{\cqmode}}{\return{\Lam{x}{\texttt{body}}}}}\\
    \end{array}$
  \end{minipage}%
    \begin{minipage}[t]{0.35\textwidth}
      $\begin{array}{@{}lll}
        \elaborate{x}_{\splicemode{}} & = & x\\
        \elaborate{\lambda x: T^0. \, e}_{\splicemode{}} & = & \lambda x. \elaborate{e}_{\splicemode{}}\\[8pt]
      \end{array}$
    \end{minipage}}

    \vspace{2mm}

    \textit{Selected Rules (Quote/Splice)}\\[1mm]
    \begin{minipage}[t]{0.33\textwidth}
      \centering
      $\elaborate{\splice}_{\quotemode{}} = {\elaborate{e}_{\splicemode{}}}$
    \end{minipage}%
    \begin{minipage}[t]{0.33\textwidth}
      \centering
      $\elaborate{\splice}_{\compilemode{}} = \tls{\elaborate{e}_{\splicemode{}}}$
    \end{minipage}%
    \begin{minipage}[t]{0.33\textwidth}
      \centering
      $\elaborate{\equote}_{\splicemode{}} = \elaborate{e}_{\quotemode{}}$
    \end{minipage}
  \end{source-desc}
  \caption{Selected elaboration rules from \sourceLang{} to \coreLang{}.}%
  \label{fig:elaboration}
\end{figure}

Elaboration of effect sets is the identity.
To define the elaboration of types (\Cref{fig:elaboration}), it is convenient to refer to a helper function, \textsf{erase}\extendedOnly{ (\Cref{appendix:auxiliary-erase})}. Given a level $0$ type, \textsf{erase} \textit{erases} all the level annotations (and elaborates effect sets), e.g.~$\textsf{erase}((\functionType[\xi]{S^0}{T^0})^0) = \functionType[\elaborate{\xi}]{S}{T}$. In a nutshell, level $0$ types elaborate into \textsf{AST} types, and level $-1$ types elaborate into themselves (sans level annotations), except for \textsf{Code} types, which elaborate into \textsf{AST} types.
\subsubsection{Elaborating Contexts}
Elaboration of contexts is subtle (\Cref{fig:elaboration}). Level $0$ types in the context elaborate into \textsf{FParam}, rather than \textsf{AST} types. Elaboration of contexts thus requires a separate elaboration for context entries, and cannot rely naïvely on the elaboration on types. To see why level $0$ types elaborate into \textsf{FParam} types, notice that the only cases where the context $\Gamma$ is extended with a level $0$ variable occur in \compilemode{} or \quotemode{}. These modes build ASTs, and thus $x$ must be an \textsf{FParam}.

\subsubsection{Elaborating Terms}
Elaboration of terms (\Cref{fig:elaboration}) assumes that all formal parameters have been annotated with their types, for example $\lambda x: \mathbb{N}^0. \; e$. The elaboration for terms is moderated by the \textbf{mode}: \compilemode{}, \quotemode{}, or \splicemode{}. At a high level, in \compilemode{} and \quotemode{}-mode, one builds ASTs. To ensure formal parameters are appropriately renamed,
the elaboration must use \textbf{\texttt{mkvar}}.

Elaboration does not differ significantly between  \compilemode{} and \quotemode{}-modes, except in the rule for splice, where \textbf{\texttt{tls}} is inserted in \compilemode{}-mode, but not in \quotemode{}-mode. The \compilemode{} and \quotemode{}-modes become important when building scope extrusion checks. Elaboration in \splicemode{}-mode is effectively the identity.

\subsubsection{Elaborating Typing Judgements}\label{subsection:typing-judgement-elaboration}
Elaboration of typing judgements can now be defined compositionally. For example, the typing judgement for lambdas in \compilemode{}-mode is elaborated by applying the elaboration component-wise:
\newcommand{\typejudge}[3][\Gamma]{#1 \vdash #2 : #3}
\[
{\inferrule{\typejudge[\elaborate{\Gamma, x: S}]{\elaborate{e}_{\compilemode{}}}{\elaborate{\runtimecomptype{T}{\Delta;\xi}}}}{\typejudge[\elaborate{\Gamma}]{\elaborate{\lambda x.e}_{\compilemode{}}}{\elaborate{\runtimecomptype{(\functionType[\xi]{S}{T})}{\Delta}}}}}
\]

Letting $Q = \erase{S}$, $R = \erase{T}$, and $\elaborate{e}_{\compilemode{}} = t$, and applying the elaboration functions defined above, we obtain \Cref{derivation:elaborated}, which, assuming that the premise is a valid typing derivation, corresponds to a valid \coreLang{} typing derivation.
\begin{typederivation}[H]
  \vspace{-3.4mm}
  \small
\[
{\inferrule{\typejudge[\elaborate{\Gamma}, x: \textsf{FParam}(Q)]{t}{\effectType{\AST{\effectType[\xi]{R}}}}}{\typejudge[\elaborate{\Gamma}]{\bind{x}{\gensym{\erase{T^0}}}{\bind{\texttt{body}}{t}{\return{\Lam{x}{\texttt{body}}}}}}{\effectType{\AST{\functionType[\xi]{Q}{R}}}}}}
\]
\vspace{-4mm}
\caption{The elaborated derivation of $\ctypejudge{\lambda x. e}{\functionType[\xi]{S}{T}}$}
\label{derivation:elaborated}
\end{typederivation}

\subsection{Metatheory}\label{section:metatheory}
Well-typed \sourceLang{} programs elaborate into well-typed \coreLang{} programs:

\begin{theorem}[Elaboration Preservation]
  If $\Gamma \vdash_{\star} e: \tau$ then $\elaborate{\Gamma} \vdash \elaborate{e}_{\star}: \elaborate{\tau}$, where $\star = \compilemode{} \mid \quotemode{} \mid \splicemode{}$ and $\tau$ is a level $0$ or level $-1$ value, computation, or handler type.
\end{theorem}

The proof is by induction on the typing rules, e.g.\ \Cref{derivation:elaborated} in \Cref{subsection:typing-judgement-elaboration}.

Additionally, the core language \coreLang{} has progress and preservation properties.

\begin{theorem}[Progress]
If $\cdot \vdash {E[t]}: {\effectType{T}}$ then for all $U, M, I$ either
\begin{enumerate}
\item $t$ is of the form $\return{n}$ and $E = [-]$,
\item $t$ is of the form $\op{v}$ for some $\textsf{op} \in \Delta$, and $\texttt{op} \notin \textsf{handled}(E)$
\item $t$ is of the form $\err$
\item $\exists \, t', E', U', M', I'$ such that $\langle t; E; U; M; I \rangle \rightarrow \langle t';E';U';M';I'\rangle$
\end{enumerate}
\end{theorem}
Note the third clause, which may be used by the calculus to report scope extrusion.

The proof of progress is by induction over the typing derivation. Most cases are standard, and have been shown by~\citet{bauer-2014}. The proof need only consider the typing rules for AST construction and scope extrusion checking, all of which are straightforward.

\begin{theorem}[Reduction Preservation]
If $\cdot \vdash E[t]: \effectType{T}$ and $\langle t; E; U; M; I \rangle \to \langle t'; E'; U'; M'; I' \rangle$
then $\cdot \vdash E'[t']: \effectType{T}$
\end{theorem}
The proof is by induction over the operational semantics. Once again, one need only consider the rules for AST construction and scope extrusion checking, which are simple.

As a corollary, we obtain a notion of type safety.

\begin{corollary}[Type Safety]\label[corollary]{cor:core-type-safety}
  If $\ctypejudge[\cdot]{e}{T^0 \, ! \, \emptyset;\emptyset}$ then either
  \begin{enumerate}
    \item $\langle \elaborate{e}_{\compilemode{}}; [-]; \emptyset
    ; \emptyset; \top \rangle \to^{\omega}$,
    \item $\langle \elaborate{e}_{\compilemode{}}; [-]; \emptyset; \emptyset; \top \rangle \to^{*} \langle \err; E; U; M; I \rangle$ for some $E$, $U$, $M$, $I$, or
    \item $\langle \elaborate{e}_{\compilemode{}}; [-]; \emptyset; \emptyset; \top \rangle \to^{*} \langle \return{n}; [-]; U; M; I \rangle$ for some $U$, $M$, $I$
  \end{enumerate}
where the initial configuration comprises an elaborated term, the empty evaluation context, an empty set indicating that no variables have been previously generated, another empty set indicating no variables have been muted, and $\top$, indicating that there is (currently) no plan to unmute variables.
\end{corollary}

Importantly, this notion of type safety is weak. A semantics which always reports a scope extrusion error (\textbf{\texttt{err}}) would be type safe under this definition, as would a semantics which never reports scope extrusion. Due to the potential presence of scope extrusion, the third case of \Cref{cor:core-type-safety} cannot additionally claim that the normal form $n$ represents a well-typed \efflang{} program.

Finally, underneath a top-level splice, quotation and splice are duals.

\begin{theorem}[Quote-Splice Duality]\label{thm:quote-splice-duality}
  Under a top-level splice, quotation and splice are duals:\\
  \begin{minipage}[t]{0.5\textwidth}
  \centering
$\splice[{\equote}] =_{\quotemode{}} e$
\end{minipage}%
\begin{minipage}[t]{0.5\textwidth}
  \centering
$\equote[{\splice}] =_{\splicemode{}} e$\end{minipage}
\end{theorem}
\noindent
where $=_{\star}$ means ``elaborates to contextually equivalent \coreLang{} programs in $\star$ mode''. Parameterising by the mode is necessary, since it affects the result of elaboration. It is possible to prove something stronger: they elaborate to the same syntactic \coreLang{} program (contextual equivalence follows from reflexivity). The proof of \Cref{thm:quote-splice-duality} is by inspection of the definition of elaboration, where:
\begin{center}
\begin{minipage}[t]{0.5\textwidth}
  \centering
  $  \elaborate{\splice[{\equote}]}_{\quotemode{}} = t \iff \elaborate{e}_{\quotemode{}} = t$
\end{minipage}%
\begin{minipage}[t]{0.5\textwidth}
  \centering
  $\elaborate{\equote[{\splice}]}_{\splicemode{}} = t \iff \elaborate{e}_{\splicemode{}} = t$
\end{minipage}
\end{center}

\section{Dynamic Scope Extrusion Checks}\label{chapter:scope-extrusion}
\newcommand{\scoped}[2][\Theta]{\textsf{Scoped}_{{#1}, {#2}}} This section uses \calculusName{} to formulate precise definitions of scope extrusion (including existing approaches \citep{kiselyov-14,isoda-24}), and properties of scope extrusion checks.
\newcommand{\indentone}{\quad\quad\quad\quad\quad\quad\quad\quad
    \quad\quad\quad\quad\quad\quad\;\;\;}
    \newcommand{\indenttwo}{\quad\quad\quad\quad\quad\;\,\,}
    \newcommand{\indentthree}{\quad\quad\quad\quad}

\begin{figure}
  \scriptsize
\begin{subfigure}[t]{0.33\textwidth}
  {
    \vspace{0pt}
$\def\arraystretch{1.8} \begin{array}{@{}l}
  \bind{x}{\gensym{\mathbb{N}}}{} \\
  \hl{\textbf{\texttt{dlet}}}(\, {x}, \textbf{\texttt{do}} \; {\texttt{body}_1} \leftarrow \\
  \quad \hl{\textbf{\texttt{check}}}(\,\textbf{\texttt{tls}}({\bind{y}{\gensym{\mathbb{N}}}{}} \\
    \quad\quad
    \textbf{\texttt{do}}\;{\texttt{body}_2} \leftarrow \\
    \quad\quad\quad {(\bind{a}{{\varToAST{x}}}{}}\\
    \quad\quad\quad\;\,
    \bind{b}{{\varToAST{y}}}{}\\
    \quad\quad\quad\;\,
    \return{\texttt{Plus}(a, b)}) \, \textbf{\texttt{in}}\\
    \quad\quad
    \return{\Lam{y}{\texttt{body}_2}})\,)\,\textbf{\texttt{in}}\\
  \quad {\return{\Lam{x}{\texttt{body}_1}}}\,)
\end{array}$
}
\caption{Lazy}
\label{fig:lazy-example}
\end{subfigure}%
\begin{subfigure}[t]{0.33\textwidth}
  {
    \vspace{0pt}
  \renewcommand{\varToAST}[1]{\hl{\textbf{\texttt{check}}}\, \texttt{Var}(#1)}
$\def\arraystretch{1.8} \begin{array}{@{}l}
  \bind{x}{\gensym{\mathbb{N}}}{} \\
  \fadedHL{\textbf{\texttt{dlet}}}({x}, \textbf{\texttt{do}} \; {\texttt{body}_1} \leftarrow \\
  \quad \textbf{\texttt{\fadedHL{check}}}( \textbf{\texttt{tls}}(\bind{y}{\gensym{\mathbb{N}}}{}\\
  \quad\quad {\hl{\textbf{\texttt{check}}}(\hl{\textbf{\texttt{dlet}}} (y, }
    \textbf{\texttt{do}}\;{\texttt{body}_2} \leftarrow \\
    \quad\quad\quad {(\bind{a}{{\varToAST{x}}}{}}\\
    \quad\quad\quad\;\,
    \bind{b}{{\varToAST{y}}}{}\\
    \quad\quad\quad\;\,
    \textbf{\texttt{\hl{check}}}\;{\texttt{Plus}(a, b)}) \, \textbf{\texttt{in}}\\
    \quad\quad
    \return{\Lam{y}{\texttt{body}_2}})))) \, \textbf{\texttt{in}}\\
  \quad {\return{\Lam{x}{\texttt{body}_1}}})
\end{array}$
}
\caption{Eager}
\label{fig:eager-example}
\end{subfigure}%
\begin{subfigure}[t]{0.33\textwidth}
  {
    \vspace{0pt}
  \renewcommand{\varToAST}[1]{\hl{$\textbf{\texttt{check}}_{\textsf{M}}$}\, \texttt{Var}(#1)}
$\def\arraystretch{1.8} \begin{array}{@{}l}
  \bind{x}{\gensym{\mathbb{N}}}{} \\
  \fadedHL{\textbf{\texttt{dlet}}}({x}, \textbf{\texttt{do}} \; {\texttt{body}_1} \leftarrow \\
  \quad \hl{$\textbf{\texttt{check}}_\textsf{M}$}( \textbf{\texttt{tls}}(\bind{y}{\gensym{\mathbb{N}}}{}\\
  \quad\quad \hl{$\textbf{\texttt{check}}_\textsf{M}$} ( \fadedHL{\textbf{\texttt{dlet}}} (y,
    \textbf{\texttt{do}}\;{\texttt{body}_2} \leftarrow \\
    \quad\quad\quad {(\bind{a}{{\varToAST{x}}}{}}\\
    \quad\quad\quad\;\,
    \bind{b}{{\varToAST{y}}}{}\\
    \quad\quad\quad\;\,
    \hl{$\textbf{\texttt{check}}_\textsf{M}$}\;{\texttt{Plus}(a, b)}) \, \textbf{\texttt{in}}\\
    \quad\quad
    \return{\Lam{y}{\texttt{body}_2}})))) \, \textbf{\texttt{in}}\\
  \quad {\return{\Lam{x}{\texttt{body}_1}}})
\end{array}$
}
\caption{\NovelCheck{}}
\label{fig:best-effort-example}
\end{subfigure}

\caption{Elaboration of $\lambda x:\mathbb{N}. \; \splice[{\equote[\lambda{y}: \mathbb{N}. \; x+y]}]$ under different checks}
\end{figure}

\subsection{Properties of Dynamic Scope Extrusion Checks}\label{section:properties}
Since checks are defined as term elaborations, we use $\elaborate{-}^{\textbf{Check}}$ to indicate an arbitrary check. We refer to the term elaboration in \Cref{section:elaboration} as naïve elaboration.

Given a definition of scope extrusion as a predicate $\Phi$ on configurations, a check is \textbf{correct} if, whenever the naïve elaboration of a well-typed \sourceLang{} expression $e$ reduces to a configuration exhibiting scope extrusion ($\Phi(\langle t;E;U;M;I\rangle)$), the elaboration of $e$ with the check reduces to $\err{}$. The \textbf{permissiveness} of a scope extrusion check refers to the set of well-typed \sourceLang{} expressions whose elaborations do not reduce to \textbf{\texttt{err}}, even if they exhibit scope extrusion.

\begin{definition}[Correctness of a Dynamic Scope Extrusion Check]
 Given a predicate on configurations $\Phi$, a dynamic scope extrusion check $\elaborate{-}^{\textbf{Check}}$ is \textbf{\emph{correct}} with respect to $\Phi$ if for all closed, well-typed \sourceLang{} expressions $e$,
 $\langle \elaborate{e} ; [-]; \emptyset;\emptyset; \top \rangle \to^{*} \langle t;E;U;M;I \rangle \land \Phi(\langle t;E;U;M;I \rangle) \implies \langle \elaborate{e}^{\textbf{Check}} ; [-]; \emptyset;\emptyset; \top \rangle  \to^{*} \; \langle \textbf{\texttt{err}};E';U';M';I' \rangle$ for some $E', U', M', I'$.
\end{definition}

\begin{definition}[Permissiveness of a Dynamic Scope Extrusion Check]
 Let \emph{\textsf{WellTyped}} be the set of closed, well-typed \sourceLang{} expressions.
 The \emph{\textbf{permissiveness}} of a dynamic scope extrusion check is defined as
 $\{ e \in \textsf{WellTyped}\, \mid \langle \elaborate{e}^{\textbf{Check}};[-];\emptyset;\emptyset;\top \rangle \not\to^{*} \langle \err ; E ; U ; M ; I \rangle \}$
\end{definition}

\subsection{Lazy Check}\label{section:lazy-dynamic-check-formal}
A \coreLang{} configuration exhibits \textbf{lazy scope extrusion} if it is the \textit{result} of compile-time execution and is improperly scoped. This formalises the definition by \citet{kiselyov-14}.

\begin{definition}[Lazy Scope Extrusion]\label[definition]{def:lazy-scope-extrusion} A \coreLang{} configuration of the form $\langle t;E;U;M;I\rangle$ exhibits \textbf{\emph{lazy scope extrusion}} if $t = \return{n}$ for some $n$ of \textsf{AST} type, $E = E'[\tls{[-]}]$ for some $E'$, and $\freevars{n} \not\subseteq \projfvs{E}$.
\end{definition}

The lazy check, ${\elaborate{-}}^{\textbf{Lazy}}$, augments the naïve elaboration in two ways (\Cref{fig:lazy-example}). First, \textbf{\texttt{check}}s are performed after top-level splices: ($\elaborate{\splice[e]}_{\compilemode{}}^{\textbf{Lazy}} \triangleq {\checkfv{(\tls{\elaborate{e}_{\splicemode{}}^{\textbf{Lazy}}})}}$). Second, \textbf{\texttt{dlet}}s are inserted to ensure variables bound outside top-level splices (in \compilemode{}-mode) are declared safe (\Cref{dfn:declared-safe}) in the context surrounding the top-level splice. Elaboration of formal parameters in \compilemode{}-mode (but not \quotemode{}-mode) should insert \textbf{\texttt{dlet}}s:
{
  \footnotesize
\[\elaborate{\lambda x: T^0. \, e}_{\compilemode{}}^{\textbf{Lazy}} = \bind{x}{\gensym{\erase{T^0}}}{ \dlet{x}{\bind{\texttt{body}}{\elaborate{e}_{\compilemode{}}^\textbf{Lazy}}{\return{\Lam{x}{\texttt{body}}}}}}\]
}

Due to the simplicity of the algorithm, verifying the correctness (with respect to lazy scope extrusion) and permissiveness of the check is trivial: the lazy check detects scope extrusion if, and only if, naïve elaboration would exhibit lazy scope extrusion after reduction.

\begin{theorem}[Correctness and Permissiveness of the Lazy Check]
For all closed, well-typed \sourceLang{} programs $e$,$\langle \elaborate{e}^{\textbf{Lazy}}; [-]; \emptyset; \emptyset; \top \rangle \to^{*} \langle \err; E; U; M; I \rangle \iff \text{For some $E', U', M', I'$, }\\
\langle \elaborate{e}; [-]; \emptyset; \emptyset; \top \rangle \to^{*} \langle \return{n}; E'; U'; M'; I' \rangle$, and $\langle \return{n}; E'; U'; M'; I' \rangle$ exhibits lazy scope extrusion
\end{theorem}

The lazy check thus characterises the set of \sourceLang{} programs that it is safe to permit.
This set is used to define the \textit{expressiveness} of a check, where the lazy check is \textit{maximally} expressive:

\begin{definition}[Expressiveness of a Dynamic Scope Extrusion Check]
Define the set $\textsf{\emph{Safe}} \triangleq \{ e \in \textsf{WellTyped} \mid \langle \elaborate{e}^{\textbf{Lazy}}; [-]; \emptyset ; \emptyset ; \top \rangle \not\to^{*} \langle \err ; E ; U ; M ; I \rangle \}$. Then the \emph{\textbf{expressiveness}} of a dynamic scope extrusion check is defined as
$\{ e \in \textsf{{Safe}}\, \mid \langle \elaborate{e}^{\textbf{Check}};[-];\emptyset;\emptyset;\top \rangle \not\to^{*} \langle \err ; E ; U ; M ; I \rangle \}$
\end{definition}

Given a scope extrusion check, every rejected program that would be permitted by the lazy check is considered a false positive:
\begin{definition}[False Positives of a Dynamic Scope Extrusion Check]
The \emph{\textbf{false positives}} of a dynamic scope extrusion check are defined as $\{ e \in \textsf{Safe}\, \mid \langle \elaborate{e}^{\textbf{Check}};[-];\emptyset;\emptyset;\top \rangle \to^{*} \langle \err ; E ; U ; M ; I \rangle \}$
\end{definition}

However, due again to its simplicity, the lazy check is considered unsuitable for practical use. \citet{RandIR}, who use the lazy check, report the following:

\begin{leftbar}{lightgrey}
  \it
\noindent Bugs in our implementation \ldots would manifest in errors such as:
\begin{lstlisting}[language=plain,style=plain]
forward reference extends over definition of value x1620
[error] val x1343 = x1232(x1123, x1124, x1180, x1181,
x1223, x1224, x1223, x1229, x1216, x1120, x1122, x1121)
\end{lstlisting}
\ldots [A] large piece of code is processed before we hit
this error \ldots The root cause of bugs such as this one often proved to
be very simple but heavily obfuscated in the code it manifested in.
\end{leftbar}

The lazy check is uninformative: since it waits until the end of evaluation, errors refer to the generated code rather than the generating program \citep{kiselyov-14}. This obfuscation makes debugging difficult for all programs. In addition, the lazy check has to wait for the end of evaluation before reporting errors. This creates an inefficiency in debugging large staged programs, like the ones generated by \citeauthor{RandIR}. Additionally, \citet[\S4.1]{kameyama-2015} note that in some systems, the
lazy check can result in unintendedly bound variables.

\subsection{Eager Check}\label{section:eager-dynamic-check-formal}
A configuration exhibits \textbf{eager scope extrusion} if it $\textbf{\texttt{return}}$s an improperly scoped AST at any point in the execution. \Cref{dfn:eager-scope-extrusion} thus generalises \Cref{def:lazy-scope-extrusion}.
\begin{definition}[Eager Scope Extrusion]\label[definition]{dfn:eager-scope-extrusion} A \coreLang{} configuration of the form  $\langle t;E;U;M;I\rangle$
exhibits \emph{\textbf{eager scope extrusion}} if $t = \return{n}$ for some $n$ of \textsf{AST} type, and $\freevars{n} \not\subseteq \projfvs{E}$
\end{definition}

It is possible to define an eager check by extending the lazy check (\Cref{fig:eager-example}). In addition to the top-level splice $\textbf{\texttt{check}}$, and the \compilemode{}-mode \textbf{\texttt{dlet}}s, the eager check adds \textbf{\texttt{check}}s for ASTs constructed in \quotemode{}-mode, for example:
\[\elaborate{v_1 v_2}_{\quotemode{}}^{\textbf{Eager}} = \bind{f}{\elaborate{v_1}_{\quotemode{}}^{\textbf{Eager}}}{\bind{a}{\elaborate{v_2}_{\quotemode{}}^{\textbf{Eager}}}{\checkfv{\App{f}{a}}}}\]
Unlike both the naïve elaboration and the lazy check, the eager check produces $\checkfv{\App{f}{a}}$ instead of $\return{\App{f}{a}}$. Consequently, to prevent false positives, variables bound in \quotemode{}-mode must generate \textbf{\texttt{dlet}}s:
{
  \footnotesize
\[
\begin{array}{rcl}
\elaborate{\lambda x: T^0. \, e}_{\quotemode{}}^{\textbf{Eager}} = \bind{x}{\gensym{\erase{T^0}}}{\checkfv{(\dlet{x}{\bind{\texttt{body}}{\elaborate{e}_{\quotemode{}}^\textbf{Eager}}{\return{\Lam{x}{\texttt{body}}}}})}}
\end{array}\]
}

Intuitively, the eager check performs a check whenever an AST is built. Hence, assume that evaluation reduces to a configuration that exhibits eager scope extrusion. Let the offending AST be $n$. The error is detected and reported when, in some evaluation context $E$, $n$ is used to build a bigger AST $n'$, and not all free variables in $n'$ are declared safe in $E$ (\Cref{listing:eager-scope-extrusion-check-eg}). \citet{kiselyov-14} observes that the overhead of checking on AST construction is negligible.

\begin{code}
 \begin{source}
  $\begin{array}{l}
      \$(\textbf{\texttt{do}} \, z \leftarrow (\textbf{\texttt{handle}} \; \equote[{\, \lambda x. \,\splice[({{\, \textbf{\texttt{op}}(\equote[x]) \,}})]\,}] \\
      \quad\quad\quad\quad\,\; \textbf{\texttt{with}} \, \{ \textbf{\texttt{return}}(u) \mapsto {\equote[\texttt{0}]}; \textbf{\texttt{op}}(y, k) \mapsto {\return{y}}\})\\
      \quad \textbf{\texttt{in}} \, \equote[{\splice[z] + 1}])
    \end{array}$
 \end{source}
 \captionof{listing}{Extrusion is reported when $z$ is used $\equote[{\splice[z]} + 1]$ in a context where $\Var{x}{\mathbb{N}}$ is not declared safe}%
 \label{listing:eager-scope-extrusion-check-eg}
\end{code}

The eager check models the BER MetaOCaml check described by \citet{kiselyov-24}. The model can be verified by executing the BER MetaOCaml N153 translations of \Cref{listing:eager-scope-extrusion-check-eg,listing:eager-scope-extrusion-unsafe-no-use,listing:eager-scope-extrusion-unsafe-continue,listing:eager-scope-extrusion-looks-unsafe} in the accompanying artifact \citep{artifact}.

\subsubsection{Correctness of the Eager Check}\label{subsection:eager-dynamic-correctness}
The eager check is \emph{not} correct with respect to eager scope extrusion. Evaluation may result in eager scope extrusion that is never detected. For example, the offending AST could be discarded (\Cref{listing:eager-scope-extrusion-unsafe-no-use}).

\begin{code}
 \begin{source}
  $\begin{array}{l}
      \$(\textbf{\texttt{handle}} \; \equote[\,{\lambda x. \,\splice[(\,{{\textbf{\texttt{op}}(\equote[x])}}\,)]}\,] \\
      \;\;\,\, \textbf{\texttt{with}} \, \{ \textbf{\texttt{return}}(u) \mapsto {\equote[\texttt{0}]}; \textbf{\texttt{op}}(y, k) \mapsto {\bind{w}{\return{y}}{\equote[\texttt{0}]}}\})
    \end{array}$
 \end{source}
 \captionof{listing}{The eager check does not report eager scope extrusion when the offending AST is discarded.}%
 \label{listing:eager-scope-extrusion-unsafe-no-use}
\end{code}

A notable property of the eager check is that it allows a program to recover from scope extrusion by \textit{resuming} a continuation. In \Cref{listing:eager-scope-extrusion-unsafe-continue}, the program restores the captured evaluation context, which declares $\Var{x}{\mathbb{N}}$ safe. Only then is $\Var{x}{\mathbb{N}}$ used to build an AST, so the checks pass.

\begin{code}
 \begin{source}
  $\begin{array}{l}
      \$ (\textbf{\texttt{handle}} \; \equote[\, {\lambda x. \,\return{\splice[( \, {{\textbf{\texttt{op}}(\equote[x])}} \,)]}} \,] \\
      \;\;\,\, \textbf{\texttt{with}} \, \{ \textbf{\texttt{return}}(u) \mapsto {\return{u}}; \textbf{\texttt{op}}(y, k) \mapsto {\bind{u}{\return{y}}{\continue{k}{u}}}\})
    \end{array}$
 \end{source}
 \captionof{listing}{The eager check does not report cases where the offending AST is used only in safe ways.}%
 \label{listing:eager-scope-extrusion-unsafe-continue}
\end{code}

The incorrectness of the eager check (i.e.~that it does not report all eager
scope extrusion) arises naturally from the definitions. \citet{kiselyov-14}
defines eager scope extrusion as the occurrence of a free variable at any
point in the evaluation. The eager \textit{check}, in contrast, is only
invoked when the free variable is used, e.g.~executed or used to
construct larger pieces of code. The incorrectness of the eager check, however,
can be desirable. Since
\Cref{listing:eager-scope-extrusion-unsafe-no-use,listing:eager-scope-extrusion-unsafe-continue} are in \textsf{Safe},
permissiveness makes the eager check more expressive.

\subsubsection{Expressiveness of the Eager Check}\label{subsection:eager-dynamic-expressiveness}
In the presence of first-class continuations, the eager check is
\emph{not} maximally expressive. It reports false positives, such as \Cref{listing:eager-scope-extrusion-looks-unsafe}.

\begin{code}
 \begin{source}
  $\begin{array}{l}
      \$ (\textbf{\texttt{handle}} \; \equote[\, {\lambda x. \, {\splice[(\, {{\textbf{\texttt{op}}(\equote[x])}} \, )]}}\,] \\
      \;\;\,\,\textbf{\texttt{with}} \, \{ \textbf{\texttt{return}}(u) \mapsto {\return{u}}; \textbf{\texttt{op}}(y, k) \mapsto {\bind{u}{\equote[{\splice[y]} + \texttt{0}]}{\continue{k}{u}}}\})
    \end{array}$
 \end{source}
 \captionof{listing}{A false positive: a safe program that fails the eager check.}%
 \label{listing:eager-scope-extrusion-looks-unsafe}
\end{code}

In \Cref{listing:eager-scope-extrusion-looks-unsafe}, the offending AST ($\Var{x}{\mathbb{N}}$) is used in a context where $\Var{x}{\mathbb{N}}$ is not declared safe, and thus the eager check reports an error. However, if evaluation had been allowed to proceed, the evaluation context binding $\Var{x}{\mathbb{N}}$ and declaring it safe would have been restored, and all variables would have been properly scoped.

Comparing \Cref{listing:eager-scope-extrusion-unsafe-continue}, which passes the eager check, with \Cref{listing:eager-scope-extrusion-looks-unsafe}, which fails the check, shows that the check is unpredictable: it is difficult to characterise its expressiveness without referring to the operational semantics.  Unfortunately, $\equote[\splice[e]] \neq_{\splicemode{}} \equote[{\splice[e] + \texttt{0}}]$. More generally, for program fragments $P$ and $P'$, $P[e] =_{\splicemode{}} P'[e] \centernot\implies \equote[{P[\splice[{\equote[e]}]]}] =_{\splicemode{}} \equote[{P'[\splice[{\equote[e]}]]}]$.

The unpredictability arises from the design of the eager check. \citet[Footnote 10]{kiselyov-14} notes that the eager check can report false positives in the
presence of first class continuations, but has not observed such cases in practice.
We say that the eager check is not \textit{continuation-aware}.

\subsection{Cause-for-Concern (\NovelCheck{}) Check}\label{section:best-effort-check}
If the lazy check is too impractical, and the eager check too unpredictable, might it be possible to find a ``goldilocks'' solution? Such a check should allow the program in \Cref{listing:eager-scope-extrusion-looks-unsafe}, and be permissive in a predictable way. A configuration exhibits \textbf{inevitable scope extrusion} when it \textit{must} cause lazy scope extrusion.
\begin{definition}[Inevitable Scope Extrusion] A \coreLang{} configuration of the form $t;E;U;M;I$
exhibits \textbf{\emph{inevitable scope extrusion}} if $\langle t;E;U;M;I\rangle \to^{*} \langle t';E';U';M';I'\rangle$ and $\langle t';E';U';M';I'\rangle$ exhibits lazy scope extrusion.
\end{definition}
This section describes a Cause-for-Concern (\novelCheck{}) check that approximates inevitable scope extrusion, though with false positives.
Elaboration for the \novelCheck{} check is a slight variation of elaboration for the eager check, with $\textbf{\texttt{check}}_\textsf{M}$ replacing \textbf{\texttt{check}} (\Cref{fig:best-effort-example}). For example,
{
  \footnotesize
\[
\begin{array}{rcl}
\elaborate{\lambda x: T^0. \, e}_{\quotemode{}}^{\textbf{BE}} = \bind{x}{\gensym{\erase{T^0}}}{\checkm{(\dlet{x}{\bind{\texttt{body}}{\elaborate{e}_{\quotemode{}}^\textbf{BE}}{\return{\Lam{x}{\texttt{body}}}}})}}
\end{array}\]
}

To understand the \novelCheck{} check, consider \Cref{fig:core-eager-looks-unsafe}, where \Cref{listing:eager-scope-extrusion-looks-unsafe} is elaborated using the eager check into \coreLang{} and simplified for readability (e.g. $\checkfv{t}$ rather than $\bind{x}{t}{\checkfv{x}}$). The failing check is \colorwave[quote]{underlined}.
\begin{figure}
\begin{center}
\tikzset{snake it/.style={decoration={snake,amplitude=.4mm,segment length=2mm},decorate}}

  \begin{tikzpicture}
  \begin{scope}[every node/.style={font=\ttfamily\footnotesize}]
    \node[align=left] (example) {$\begin{array}{l}
      \textbf{\texttt{handle}}\\
      \quad \textbf{\texttt{do}} \, \Binder{x}{\mathbb{N}} \leftarrow \gensym{\mathbb{N}}  \, \textbf{\texttt{in}} \\
       \quad \textbf{\texttt{check}}(\textbf{\texttt{dlet}}({\Binder{x}{\mathbb{N}}}, \, \textbf{\texttt{do}} \, {\texttt{body}} \leftarrow \, (\bind{a}{{\Var{x}{\mathbb{N}}}}{\textbf{\texttt{op}}}(a)) \; \textbf{\texttt{in}} \, {\return{\Lam{\Binder{x}{\mathbb{N}}}{\texttt{body}} }}) \\
      \textbf{\texttt{with}} \\
      \quad \{ \textbf{\texttt{return}}(u) \mapsto {\return{u}}; \\
      \quad\;\, \textbf{\texttt{op}}(y, k) \mapsto {\bind{w}{\checkfv{\texttt{Plus}(y, \texttt{Nat}(\texttt{0}))}}{\continue{k}{w}}}\}
    \end{array}$};
  \end{scope}

   \path [draw=quote,snake it,thick]
    ($(example.south) + (-2.2, 0.03)$) -- ($(example.south) + (0.5, 0.03)$);

  \end{tikzpicture}

\end{center}
\caption{The result of elaborating \Cref{listing:eager-scope-extrusion-looks-unsafe} using the eager check}
\label{fig:core-eager-looks-unsafe}
\end{figure}

The check fails because when $\textbf{\texttt{op}}$ is performed, the variable $\Var{x}{\mathbb{N}}$ is no longer declared safe in the new evaluation context. Since $y$ is bound to $\Var{x} {\mathbb{N}}$, checking $\texttt{Plus}(y, \texttt{Nat}(\texttt{0}))$ reports an error. The problem is that the continuation $k$ can be used to bind $\Var{x}{\mathbb{N}}$. It is not clear, when the \texttt{Plus} AST is constructed and checked, that eager scope extrusion \textit{must} lead to lazy scope extrusion. To make the check more expressive, therefore, it may be useful to temporarily allow $\Var{x}{\mathbb{N}}$ to extrude its scope, delaying error detection until one \textit{must} have lazy scope extrusion.

The $\textbf{\texttt{check}}_\textsf{M}$ primitive checks for scope extrusion, but allows a set of muted variables $M$ to temporarily extrude their scope. In our example, we may mute $\Var{x}{\mathbb{N}}$, by adding it to $M$. The \coreLang{} operational semantics (\Cref{fig:core}) automates this process, strategically muting and unmuting variables at key points:
\begin{itemize}
\item When effects are performed, the variables which are no longer declared safe in the new evaluation context (like $\Var{x}{\mathbb{N}}$) are added to the set of muted variables (\textsc{Eff-Op}).

\item Variables are unmuted when there are no bound continuations, and thus no way to resume a continuation $k$ that could bind $\Var{x}{\mathbb{N}}$. This point is identified by tracking the maximal length $I$ of the stack $E$ that was never captured by the handling of an effect.
\end{itemize}
Intuitively, $I$ is a stack mark (or continuation mark), which tracks
the point where effects and exceptions are indistinguishable.
The C4C check acts like the lazy check before this point, and like the
eager check after it.  Stack marks are used in the eager check
implementation \citep[Appendix B]{kiselyov-14}, and in the semantics
and implementation of languages with continuations
\citep{DBLP:conf/pldi/FlattD20,kiselyov-2012}.
However, since the eager check is not continuation-aware, stack marks
play only a limited role.

As an example, the program in \Cref{fig:core-unmute-example} builds the AST of $\lambda z. \, (\lambda x. \, x + \texttt{0}) (\texttt{1})$. Let \textcolor{selected}{\texttt{body}} be the program in \Cref{fig:core-eager-looks-unsafe}.
\begin{figure}
\begin{subfigure}[t]{0.4\textwidth}
  \footnotesize
  \centering
$\begin{array}{@{}l}
  \textbf{\texttt{check}}_\textsf{M}(\textbf{\texttt{dlet}}(\Binder{z}{\mathbb{N}}, \textbf{\texttt{do}} \; b \leftarrow \\
  \quad\quad  (\textbf{\texttt{do}} \; {f} \leftarrow \textcolor{selected}{\texttt{body}} \\
  \quad\quad\; \textbf{\texttt{in}} \; \textbf{\texttt{do}} \; {a}\leftarrow {\return{\texttt{Nat}(\texttt{1})}} \\
  \quad\quad\; \textbf{\texttt{in}} \; {\checkm{\texttt{App}(f, a)}})\; \\
  \quad\textbf{\texttt{in}} \; \return{\Lam{\Binder{z}{\mathbb{N}}}{b}}))
\end{array}$
\caption{Initial term}
\label{fig:core-unmute-example}
\end{subfigure}%
\begin{subfigure}[t]{0.6\textwidth}
   \footnotesize
     \centering
$\begin{array}{@{}l}
  \textbf{\texttt{check}}_\textsf{M}(\textbf{\texttt{dlet}}(\Binder{z}{\mathbb{N}}, \textbf{\texttt{do}} \; b \leftarrow \\
  \quad\quad  (\textbf{\texttt{do}} \; {f} \leftarrow \hl{[}\; \return{\Lam{\Binder{x}{\mathbb{N}}}{\texttt{Plus}(\Var{x}{\mathbb{N}}, \Nat{\texttt{0}})}}\; \hl{]} \\
  \quad\quad\; \textbf{\texttt{in}} \; \textbf{\texttt{do}} \; {a}\leftarrow {\return{\texttt{Nat}(\texttt{1})}} \\
  \quad\quad\; \textbf{\texttt{in}} \; {\checkm{\texttt{App}(f, a)}})\; \\
  \quad \textbf{\texttt{in}} \; \return{\Lam{\Binder{z}{\mathbb{N}}}{b}}))
\end{array}$
\caption{Reduced term}
\label{fig:core-unmute-example-result}
  \end{subfigure}
  \caption{(a) A \coreLang{} program that generates the AST of $\lambda z. \, (\lambda x. \, x + \texttt{0}) (\texttt{1})$. (b) The result of reducing the program in (a) to the point where variables may be unmuted.}
\end{figure}
The surrounding context around \textcolor{selected}{\texttt{body}} is identified by $I$: it is never captured by the handling of any effect, and thus must have no references to the captured continuation $k$.

If the stack was never captured by the handling of an effect (for example, no operations were performed), then $I$ is set to $\top$, $\forall n \in \mathbb{N}, \top \geq n$. Performing an effect can thus \textit{decrease} $I$, but never increase it. This is the side condition on \textsc{Eff-Op}.

During reduction, when the length of the stack is less than, or equals to, $I$, there must not be any remaining references to any continuations $k$, and thus $I$ may be reset to $\top$, and all muted variables may be unmuted. The program in \Cref{fig:core-unmute-example} eventually reduces to the term in \Cref{fig:core-unmute-example-result}. $\hl{[}-\hl{]}$ separates the evaluation context (outside) and the term (inside). At this point, the length of the stack is less than or equal to $I$. It is safe to unmute all muted variables. When there are no muted variables, $\textbf{\texttt{check}}_\textsf{M}$ and \textbf{\texttt{check}} have the same behaviour.

However, altering the semantics in such a manner means that any transition could unmute variables. To keep the semantics standard, and to more closely model the implementation of the check, we associate the act of unmuting with \textbf{\texttt{dlet}} and \textbf{\texttt{tls}}. A transition from \textbf{\texttt{dlet}} conditionally unmutes variables (\textsc{Sec-Dlt}, \Cref{fig:core}).
In \Cref{fig:core-unmute-example-result}, the transition from $\dlet{\Binder{z}{\mathbb{N}}}{\return{n}}$ unmutes variables. Hence, $\Var{x}{\mathbb{N}}$ is still muted when the \texttt{App} constructor is checked, but unmuted when the outer \texttt{Lam} constructor is checked.

Additionally, a transition from \textbf{\texttt{tls}} \textit{unconditionally} unmutes variables, since the evaluation context beyond \textbf{\texttt{tls}} must be inert, and thus can never be captured by a handler (\textsc{Sec-Tls}).

As a $\textbf{\texttt{check}}_\textsf{M}$ can never fail where a \textbf{\texttt{check}} succeeds, the \novelCheck{} check is at least as permissive as the eager check.

\subsubsection{Correctness of the \novelCheck{} check}\label{subsection:best-effort-correct}
The \novelCheck{} check is correct with respect to inevitable scope extrusion. The proof is simple: either one of the non-top-level splice $\textbf{\texttt{check}}_\textsf{M}$s reports an error, or none do. The latter case degenerates to the lazy check, where the top-level splice $\textbf{\texttt{check}}_\textsf{M}$ must report an error.

\begin{theorem}[Correctness of the \NovelCheck{} Check] Given a closed, well-typed \sourceLang{} expression $e$, if  $\langle \elaborate{e}; [-]; \emptyset; \emptyset; \top \rangle$ exhibits inevitable scope extrusion
then there exists $E$, $U$, $M$, $I$ such that $\langle \elaborate{e}^{\textbf{BE}}; [-]; \emptyset; \emptyset; \top \rangle \to^{*}\langle \err{}; E; U; M; I \rangle$
\end{theorem}

\subsubsection{Expressiveness of the \NovelCheck{} Check}\label{subsection:best-effort-expressive}
The \novelCheck{} check is not maximally expressive. In particular, it does not allow the program in \Cref{listing:best-effort-imperfect}.

\begin{code}
  \begin{source}
    $
    \begin{array}{l}
      \$(\langle \langle \lambda x. \$ (\textbf{\texttt{handle}} \, \equote[{\lambda y. \$(\textbf{\texttt{op}}(\equote[y]); \return{y})}]\\
      \quad\quad\quad\quad\textbf{\texttt{with}} \, \{ \returnHandler{u}{\return{\equote[\texttt{0}]}}; \opHandler{z}{k}{\return{z}}\})\rangle\rangle;\\
      \;\, \equote[{\texttt{1}}])
    \end{array}
    $
  \end{source}
  \captionof{listing}{The \novelCheck{} check reports false positives.}
  \label{listing:best-effort-imperfect}
\end{code}

\Cref{listing:best-effort-imperfect} attempts to build the AST $\lambda x.\return{y}$, where $y$ has extruded its scope, but then throws it away, returning the AST of \texttt{1}. Critically, the constructor of the outer lambda, $\lambda x. [-]$, is never captured by any effect. Hence, \Cref{listing:best-effort-imperfect} eventually reduces to a configuration:
\[\langle \dlet{\Binder{x}{\mathbb{N}}}{\return{\Lam{\Binder{x}{\mathbb{N}}}{\Var{y}{\mathbb{N}}}}}; E[\textbf{\texttt{check}}[-]]; U; \{\Var{y}{\mathbb{N}}\} ; I \rangle\]
where $\textsf{len}(E[\checkm{[-]}]) \leq I$. The subsequent transition unmutes $\Var{y}{\mathbb{N}}$, and the surrounding $\textbf{\texttt{check}}_\textsf{M}$ fails, as $\Var{y}{\mathbb{N}}$ is free, unmuted, and not declared safe in $E$.

A Cause-for-Concern property characterises the expressiveness of the \novelCheck{} check\footnote{and gives it its name, which, unlike the lazy and eager check, describes its user-facing behaviour, not its operation}. The property is defined informally as follows: assume the check reports an error, and let the offending AST be $n$. Now re-wind to the point of the failing check, and consider an alternative execution where all the $\textbf{\texttt{check}}_\textsf{M}$s are erased (turned into $\textbf{\texttt{return}}$s). In this counter-factual execution, all ASTs $n'$ that are constructed from $n$ have at least one variable that is not declared safe in its evaluation context. Consequently, in \Cref{listing:best-effort-imperfect}, the only way to safely use $\lambda x.\return{y}$ is to throw it away.

\begin{theorem}[Cause-for-Concern Property]\label{thm:best-effort-cause-for-concern} Assuming a closed, well-typed \sourceLang{} expression $e$, if $\exists$ $E$, $U$, $M$, $I$ such that $\langle \elaborate{e}^{\textbf{BE}}; [-];\emptyset; \emptyset; \top \rangle \to^{*}\langle \checkm{n}; E; U; M; I \rangle$, and $\langle \checkm{n}; E; U; M; I \rangle \to \langle \err{}; E; U; M; I \rangle$, then, assuming $\langle \return{n}; \textsf{erase-checks}(E); U; M; I \rangle \to^{*} \langle \return{n'}; E'; U'; M'; I' \rangle$, and $n$ a subtree of $n'$, it must be that $\freevars{n'} \not\subseteq \projfvs{E'}$.
\end{theorem}

The proof of \Cref{thm:best-effort-cause-for-concern} is by contradiction. Informally, if  $\freevars{n'} \subseteq \projfvs{E'}$, then all the variables in $n$ must be declared safe. This implies that when the initial $\checkm{n}$ failed, there was a continuation on the stack which can declare the variables in $n$ safe. But then $M$ cannot be empty, so the check would not have failed.

The expressiveness of the eager check cannot be characterised by the Cause-for-Concern property, with \Cref{listing:eager-scope-extrusion-looks-unsafe} being a counter-example. Hence, the \novelCheck{} check is \textbf{more} expressive, and more \textbf{predictably} expressive, than the eager check. 

Like the eager check, the expressiveness of the \novelCheck{} check can be empirically verified by executing MacoCaml translations of \Cref{listing:eager-scope-extrusion-check-eg,listing:eager-scope-extrusion-unsafe-no-use,listing:eager-scope-extrusion-unsafe-continue,listing:eager-scope-extrusion-looks-unsafe,listing:best-effort-imperfect} in the accompanying artifact \citep{artifact}. 

\newcommand{\yes}{\textbf{\textcolor{splice}{\sffamily Y}}}
\newcommand{\no}{\textbf{\textcolor{quote}{\sffamily N}}}
\newcommand{\maybe}{\textbf{\textcolor{compile}{\sffamily ?}}}

\definecolor{yes}{HTML}{046052}
\definecolor{no}{HTML}{8A0630}
\begin{table}
  \footnotesize
  \captionsetup{width=.45\textwidth}
  \newcommand\T{\rule{0pt}{2.6ex}}       
\newcommand\B{\rule[-1.2ex]{0pt}{0pt}}
\begin{minipage}[t]{0.5\textwidth}
  \centering
  \caption{Correctness comparison}
  \vspace{-2.5mm}
  \begin{tabular}{l|c|c|c|c|c}
    & \multicolumn{5}{c}{\textbf{Listings}}
    \\[1mm] &
    \ref{listing:eager-scope-extrusion-unsafe-no-use} &
    \ref{listing:eager-scope-extrusion-unsafe-continue} &
    \ref{listing:eager-scope-extrusion-looks-unsafe} &
    \ref{listing:best-effort-imperfect} &
    \ref{listing:refined-environment-classifiers-safe} \T\B\\ \hline
    \textbf{Lazy}~(\S\ref{section:lazy-dynamic-check-formal}) & \yes & \yes & \yes & \yes & \yes \T\B\\
    \textbf{Eager}~(\S\ref{section:eager-dynamic-check-formal}) & \no & \no & \yes & \yes & \yes  \T\B\\
    \textbf{\NovelCheck{}}~(\S\ref{section:best-effort-check}) & \yes & \yes & \yes & \yes & \yes \T\B\\
    \textbf{Ref.\ Env.\ Classifiers}~(\S\ref{section:refined-environment-classifiers-formal}) & \yes & \yes & \yes & \yes & \yes \T\B\\
    \hline
  \end{tabular}
  \label{table:correctness-comparison}
  \end{minipage}%
\begin{minipage}[t]{0.5\textwidth}
  \centering
     \caption{Expressiveness comparison}
       \vspace{-2.5mm}
  \begin{tabular}{l|c|c|c|c|c}
    & \multicolumn{5}{c}{\textbf{Listings}}
    \\[1mm] &
    \ref{listing:eager-scope-extrusion-unsafe-no-use} &
    \ref{listing:eager-scope-extrusion-unsafe-continue} &
    \ref{listing:eager-scope-extrusion-looks-unsafe} &
    \ref{listing:best-effort-imperfect} &
    \ref{listing:refined-environment-classifiers-safe} \T\B\\ \hline
    \textbf{Lazy}~(\S\ref{section:lazy-dynamic-check-formal}) & \yes & \yes & \yes & \yes & \yes \T\B\\
    \textbf{Eager}~(\S\ref{section:eager-dynamic-check-formal}) & \yes & \yes & \no & \no  & \yes  \T\B\\
    \textbf{\NovelCheck{}}~(\S\ref{section:best-effort-check}) & \yes & \yes & \yes & \no  & \yes \T\B\\
    \textbf{Ref.\ Env.\ Classifiers}~(\S\ref{section:refined-environment-classifiers-formal})& \no & \no & \no & \no & \yes \T\B\\ \hline
  \end{tabular}
  \label{table:expressiveness-comparison}
  \end{minipage}
  \vspace{-4mm}
\end{table}

\subsection{Evaluation of \texorpdfstring{\sourceLang{}}{Lambda-Op-Quote-Splice}} \label{section:evaluation}

We have demonstrated that \sourceLang{} is an appropriate language for encoding
and evaluating scope extrusion checks.
\Cref{table:correctness-comparison,table:expressiveness-comparison} summarize
the correctness and expressiveness of the checks (with refined
environment classifiers and \Cref{listing:refined-environment-classifiers-safe}
discussed in the next section). The accompanying artifact \citep{artifact} provides translations of \Cref{listing:eager-scope-extrusion-check-eg,listing:eager-scope-extrusion-unsafe-no-use,listing:eager-scope-extrusion-unsafe-continue,listing:eager-scope-extrusion-looks-unsafe,listing:best-effort-imperfect} in both BER MetaOCaml N153 and MacoCaml. The first three rows of \Cref{table:correctness-comparison,table:expressiveness-comparison} can be verified empirically by executing these translations.
Unifying these checks under \sourceLang{} facilitated comparative evaluation
with reference to the same set of programs.
Moreover, formalising scope extrusion in \sourceLang{} aided development of the
novel \novelCheck{} check, which finds a sweet spot between the eager and lazy
checks.

It is worth re-iterating that the focus of \sourceLang{} is on scope
extrusion.
There are additional interesting questions related to bindings in
generated code that this paper does not consider.
In particular, \sourceLang{} does not prevent shadowing: it is possible in
\sourceLang{} to generate programs with multiple binders that use the
same formal parameter.
For example, using multi-shot continuations, it is possible in
\sourceLang{} to generate the code $\Lam{x}{\Lam{x}{\texttt{body}}}$,
since the binder $\Lam{x}{-}$ can be captured in a continuation that
is re-instated in a nested manner.
None of the three dynamic checks is able to detect every instance of
scope shadowing.
Restricting \sourceLang{} to permit only one-shot continuations, as in
systems like \texttt{MetaOCaml} and \texttt{MacoCaml}, would prevent
shadowing, but in practice multi-shot continuations are useful in
multi-staged programming, e.g.\ for case-insertion
\citep[\S4.4]{yallop-2017}, and do not compromise type safety.

\section{Extension: Refined Environment Classifiers}\label{section:refined-environment-classifiers-formal}

\newcommand{\recCoreLang}{\coreLang{}$^\gamma$}

This section presents \recLang{}, an extension to
\sourceLang{} with \textit{refined environment classifiers}~\citep{kiselyov-16}
to statically prevent scope extrusion,
following \citet{isoda-24}. \Cref{chapter:scope-extrusion} illustrates the use of \sourceLang{} to compare dynamic scope extrusion checks, and \recLang{} shows how to extend the framework to describe and evaluate static prevention techniques, too.

\subsection{The Calculus}\label{section:refined-environment-classifier-calculus}

\Cref{fig:refined-env-classifiers-types} presents the types and selected
typing rules of \recLang{}.
The calculus shares its syntax with \sourceLang{}, extending it with a
simplified\footnote{
\citeauthor{isoda-24}'s typing rules for handlers and continuations are
polymorphic over the classifier, to allow for let-insertion.} version of \citeauthor{isoda-24}'s type system.

Intuitively, a \textit{classifier} represents a scope that permits a set of free
variables.
An AST is considered well-scoped at a given scope if it is well-typed
and all its free variables are permitted by the scope.

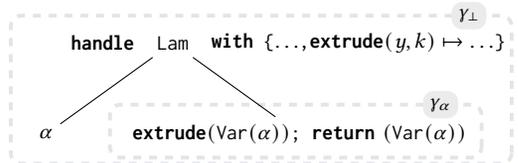
\begin{wrapfigure}{r}{0.48\textwidth}
\vspace{-12pt}
\centering
\footnotesize
\begin{tikzpicture}[level 1/.style={sibling distance=12em},yscale=0.8]

  \begin{scope}[every node/.style={font=\ttfamily}]
  \node (handler) {Lam}
  child {node (var) {$\alpha$}}
  child {node (body) {\textbf{extrude}$($\texttt{Var}$(\alpha)$$)$; $\return(\texttt{Var}(\alpha))$}};
  \end{scope}

  \begin{scope}
    \node[anchor=east] (lam-left) at ($(handler.west) + (-0.1cm, 0cm)$) {\textbf{\texttt{handle}}};
    \node[anchor=west] (lam-right) at ($(handler.east) + (0.1cm, 0cm)$) {\texttt{\textbf{with} \{$\ldots$,\textbf{extrude}$(y, k)\mapsto\ldots$\} }};
  \end{scope}

  \begin{scope}[on background layer]
    \draw[line width=0.5mm, rounded corners,envclassifier1, dashed] ($(body.south east) + (0.5cm,-0.2cm)$) rectangle ($(lam-left.north west) + (-0.7cm,0.2cm)$);

    \draw[line width=0.5mm, rounded corners,envclassifier1, dashed] ($(body.south east) + (0.1cm,-0.1cm)$) rectangle ($(body.north west) + (-0.2cm,0.2cm)$);

    \begin{scope}[every node/.style={rounded corners, fill=envclassifier2, font=\scriptsize}]
      \node (gammabottom) at ($(lam-left.north west) + (5.4cm,0.2cm)$) {$\gamma_{\bot}$};
      \node (gammaalpha) at ($(body.north west) + (4.2cm,0.2cm)$) {$\gamma_{\alpha}$};
    \end{scope}
  \end{scope}

\end{tikzpicture}\vspace{-3mm}
\caption{Refined environment classifiers}
\label{fig:classifier-ast-scope-extrusion}
\vspace{-10pt}
\end{wrapfigure}

As an example, consider \Cref{fig:classifier-ast-scope-extrusion}, where there are
two classifiers: $\gamma_{\alpha}$ is the scope that permits only
\texttt{Var}($\alpha$), and $\gamma_\bot$ the scope that permits no variables
(the ``top-level''). To capture the nesting of scopes, classifiers are related
by a partial order $\gamma \sqsubseteq \gamma'$, with $\gamma$ the outer
scope, and $\gamma'$ the inner scope; in this case, we have $\gamma_\bot
\sqsubseteq \gamma_{\alpha}$.

Classifiers prevent scope extrusion by checking that created ASTs are
well-scoped, and moreover, that manipulating ASTs preserves well-scopedness.
Specifically, they prevent variables being lifted into
scopes where they are not permitted. In
\Cref{fig:classifier-ast-scope-extrusion}, the \textbf{\texttt{extrude}} effect
attempts to lift \texttt{Var$(\alpha)$} to a handler in the $\gamma_{\bot}$
scope (where \texttt{Var}$(\alpha)$ is not permitted).
\Cref{fig:classifier-ast-scope-extrusion} cannot be typed, regardless of the
body of the handler.

\paragraph{Syntax.}

We annotate level $-1$ \textsf{Code} types with a classifier $\gamma$. For level
$0$ types, classifiers are associated in the typing contexts and in the typing judgement.
We define \textit{extended} \recLang{} types, a notion useful for defining the
logical relation in \Cref{subsection:rec-formal-correctness}.

\begin{definition}[Extended \recLang{} type]\label[definition]{dfn:extended-source-type}
An \emph{{extended \recLang{} type}} is either:
  \begin{enumerate}
    \item A level $-1$ type, e.g.~$(\textsf{Code}(\mathbb{N}^{0} \, ! \,  \emptyset)^{\gamma})^{-1}$;
    \item A level $0$ type annotated with a classifier, e.g.~$\mathbb{N}^{0} (\gamma)$; or
    \item A level $0$ formal parameter type, which is a level $0$ value type (e.g.~$\mathbb{N}^{0}$) annotated with a classifier $\gamma$, and an \underline{underline}, to indicate that it is elaborated into an \textsf{FParam} type, \underline{$\mathbb{N}^{0} (\gamma)$}.
  \end{enumerate}
\end{definition}

The typing context $\Gamma$ maps terms to their types, and tracks their
environment classifiers for level $0$ types. Additionally, it tracks
classifiers $\gamma$ and their partial ordering $\gamma \sqsubseteq \gamma'$.
A level $-1$ type $T^{-1}$ is well-formed under a context $\Gamma$, written
$\Gamma \vdash T^{-1}$, if all its classifiers are in $\Gamma$.
A context is well-formed if it contains the \textit{least}
classifier $\gamma_{\bot}$, and if all its types are well-formed.
We will assume all contexts are well-formed.

\paragraph{Typing.}

Most typing rules are straightforwardly adapted, with key rules listed in
\Cref{fig:refined-env-classifiers-types}.
The \compilemode{}$\mid$\quotemode{}\textsc{-Var} rule says that a variable with
classifier $\gamma$ is well-typed under the classifier $\gamma$.
Of particular interest is the \compilemode{}$\mid$\quotemode{}-\textsc{Lambda}
rule. As classifiers formalise the notion of scope, this rule introduces a new
scope, represented as a fresh classifier $\gamma' \notin \Gamma$
and associates the variable $x$  with $\gamma'$.
Moreover,
since $\gamma'$ is created within the scope of $\gamma$, we have $\gamma
\sqsubseteq \gamma'$.

The \compilemode{}$\mid$\quotemode{}\textsc{-Sub-Expr} and
$\splicemode{}\textsc{-Sub}$ rules formalise the nesting of scopes: to show a
term is well-scoped in some nested scope $\gamma$, it suffices to show that it
is well-scoped in any of its parents $\gamma'$, with $\gamma' \sqsubseteq
\gamma$.

\newcommand\handle{\textbf{\texttt{handle}}\xspace}

Following \citet{isoda-24}, operations (\splicemode{}\textsc{-Op}),
continuations (\splicemode{}\textsc{-Continue}), and handlers
(\splicemode{}\textsc{-Handle}) are restricted to \textsf{Code} types, although it
generalises easily to non-Code types. These rules work in concert to prevent
scope extrusion. Specifically, the \handle construct operates on \textsf{Code}
types, and acts at a scope $\gamma$. As a result, handlers cannot change the
scopes in which the result of computation is permitted, e.g. changing the type
from $\textsf{Code}(\mathbb{N})^\gamma$ to a different classifier
$\textsf{Code}(\mathbb{N})^{\gamma'}$. Notably, each handler clause (and thus
handled effect) inherits this scope $\gamma$. As a result, the values passed to
handled effects, should they be code types, must be tagged with a classifier
that may be substituted for $\gamma$. Since binders introduce new classifiers
$\gamma'$, where $\gamma'$ cannot be substituted for $\gamma$ (since $\gamma'
\not\sqsubseteq \gamma$), examples that result in scope extrusion do not type
check.

Lastly, we note that since contexts must contain the least classifier
$\gamma_{\bot}$, an expression $e$ is a closed, well-typed expression if
$\gamma_{\bot} \vdash_{\compilemode}^{\gamma_{\bot}} e : T^{0} \, ! \,
\emptyset; \emptyset$.

\paragraph{Elaboration.}

Like \sourceLang{}, \recLang{} does not have an operational semantics, but is
elaborated into \recCoreLang{} terms, where formal parameters are annotated with
classifiers (e.g.\ $\alpha_R^{\gamma}$). Classifiers show up only in the formal
parameters, and are invisible to the types. Since elaboration does not require
any dynamic scope extrusion checking machinery, \recCoreLang{} does not have
\textbf{\texttt{check}}, $\textbf{\texttt{check}}_\textsf{M}$, \textbf{\texttt{dlet}},
\textbf{\texttt{tls}}, and \textbf{\texttt{err}}. Consequently, \recCoreLang{}
configurations are of the form $\langle t;E;U\rangle$.

Elaboration is similar to \Cref{section:elaboration}, except that elaboration of
types erases classifiers, elaboration of context entries erases proof-theoretic
terms, and elaboration of terms assumes binders have been annotated with an
extended source type, and does \textit{not} erase classifiers. Finally,
elaboration of top-level splice does not insert \textbf{\texttt{tls}}.

\paragraph{Weakening.}
We prove a weakening lemma, which is useful for our later proof of correctness.
As types are stratified into two levels, and into value, computation, and
handler types, there are various sub-lemmas. As an example, we present weakening
for level 0 computations:

\newcommand{\rcqtypejudge}[4][\Gamma]{{#1} \vdash_{\compilemode \mid \quotemode}^{#2} {#3} : {#4}}
\newcommand{\rctypejudge}[4][\Gamma]{{#1} \vdash_{\compilemode}^{#2} {#3} : {#4}}
\newcommand{\rqtypejudge}[4][\Gamma]{{#1} \vdash_{\quotemode}^{#2} {#3}: {#4}}
\newcommand{\rstypejudge}[3][\Gamma]{{#1} \vdash_{\splicemode} {#2} : {#3}}

\begin{lemma}[Weakening for Level 0 Computations]\label[lemma]{lemma:weakening}
  If $\rcqtypejudge{\gamma}{e}{T^0 \, ! \, \Delta}$ then
  \begin{enumerate}
    \item $\rcqtypejudge[\Gamma, (x:S^0)^{\gamma'}]{\gamma}{e}{T^0 \, ! \, \Delta}$, for arbitrary $\gamma' \in \Gamma$, $x \notin \Gamma$;
    \item $\rcqtypejudge[\Gamma, (x:S^{-1})]{\gamma}{e}{T^0 \, ! \, \Delta}$, where $\Gamma \vdash S^{-1}$, $x \notin \Gamma$;
    \item $\rcqtypejudge[\Gamma, \gamma']{\gamma}{e}{T^0 \, ! \, \Delta}$, for arbitrary $\gamma' \notin \Gamma$
    \item $\rcqtypejudge[\Gamma, \gamma' \sqsubseteq \gamma'']{\gamma}{e}{T^0 \, ! \, \Delta}$, for arbitrary $\gamma', \gamma'' \in \Gamma$
  \end{enumerate}
\end{lemma}

\begin{figure}
  \begin{rec-desc}
    \footnotesize

    {{\textbf{\normalsize {Typing contexts}}}\\
    $\Gamma ::= \cdot \mid \Gamma, (x:T^{0})^{\gamma} \mid \Gamma, x:T^{-1} \mid \Gamma, \gamma \mid \Gamma, \gamma \sqsubseteq \gamma' $}

    \vspace{3mm}

    {\normalsize\textbf{Types}}\\
    $\begin{array}{@{}lllr}

    \textbf{Level $-$1} & \text{Values} & T^{-1} ::=
    \ldots \mid {({\textsf{Code}({T^{0} \, ! \, \xi})}^{\textbf{\hl{$\gamma$}}})}^{-1}
    \end{array}$

    \vspace{3mm}
  {\normalsize\textbf{Typing Rules}}\\
  \textit{Selected Rules}
    \scriptsize
  \begin{center}
  \begin{minipage}[t]{0.2\textwidth}
    \centering
    $\inferrule[(\compilemode{}$\mid$\quotemode{}-Var)]{(x: T^0)^\gamma \in \Gamma}{\rcqtypejudge{\gamma}{x}{\runtimecomptype{T^0}{\Delta}}}$
  \end{minipage}%
  \begin{minipage}[t]{0.5\textwidth}
    \centering
$\inferrule[(\compilemode{}$\mid$\quotemode{}-Lambda)]{\rcqtypejudge[\Gamma, \gamma', \gamma \sqsubseteq \gamma', (x: S)^{\gamma'}]{\gamma'}{e}{\runtimecomptype{T}{\Delta;\xi}} \\
\gamma \in \Gamma \\
\gamma' \notin \Gamma
}{\rcqtypejudge{\gamma}{\lambda x.e}{\runtimecomptype{(\functionType[\xi]{S}{T})}{\Delta}}}$
\end{minipage}%
\begin{minipage}[t]{0.3\textwidth}
    \centering
  $\inferrule[(\splicemode{}-Op)]
    {\strut \stypejudge{v}{\compiletimetype{S}} \\ \texttt{op} \in \Delta  \\ \texttt{op}: \compiletimetype{S} \rightarrow \compiletimetype{\textsf{Code}(\effectType[\xi]{T})^{\gamma}} \in \Sigma }
    {\stypejudge{\op{v}}{\effectType{\compiletimetype{\textsf{Code}(\effectType[\xi]{T})^{\gamma}}}}}$
  \end{minipage}

\vspace{3mm}

  \begin{minipage}[t]{0.5\textwidth}
    \centering
  $\inferrule[(\splicemode{}-Continue)]
    { \\\\ \stypejudge{v_1}{\compiletimetype{\continuationType{\compiletimetype{\textsf{Code}(\effectType[\xi_1]{S})^{\gamma}}}{\compiletimetype{\textsf{Code}(\effectType[\xi_2]{T})^{\gamma'}}}}} \\ \stypejudge{v_2}{\compiletimetype{\textsf{Code}(\effectType[\xi_1]{S})^{\gamma}}}}
    {\stypejudge{\continue{v_1}{v_2}}{\effectType{\compiletimetype{\textsf{Code}(\effectType[\xi_2]{T})^{\gamma'}}}}}$
  \end{minipage}%
  \begin{minipage}[t]{0.5\textwidth}
    \centering
  $\inferrule[(\splicemode{}-Handle)]
    {\stypejudge{e}{\effectType{\compiletimetype{\textsf{Code}(\effectType[\xi_1]{S})^{\gamma}}}} \\ \stypejudge{h}{\compiletimetype{\handlerType{\effectType[\Delta_1]{\compiletimetype{(\textsf{Code}(\effectType[\xi_1]{S})^{\gamma})}}}{\effectType[\Delta_2]{\compiletimetype{(\textsf{Code}(\effectType[\xi_2]{T})^{\gamma})}}}}} \\ \forall \textsf{op} \in \Delta_1 \setminus \Delta_2. \, \textsf{op} \in \textsf{dom}(h)}
    {\stypejudge{\handleWith{e}{h}}{\effectType[\Delta_2]{\compiletimetype{\textsf{Code}(\effectType[\xi_2]{T})^{\gamma}}}}}$
  \end{minipage}

  \vspace{3mm}

\begin{minipage}[t]{0.25\textwidth}
  \centering
  $\inferrule[(\compilemode{}$\mid$\quotemode{}-Splice)]{ \\\\ \stypejudge[\Gamma]{e}{\effectType{\textsf{Code}(T \, ! \, \xi)^{\gamma}}}}{\rcqtypejudge{\gamma}{\splice}{\runtimecomptype{T}{\Delta ; \xi}}}$
\end{minipage}%
\begin{minipage}[t]{0.25\textwidth}
  \centering
  $\inferrule[(\splicemode{}-Quote)]{ \\\\ \rqtypejudge{\gamma}{e}{\runtimecomptype{T}{\Delta ; \xi}}}{\stypejudge[\Gamma]{\equote}{\textsf{Code}(T \, ! \, \xi)^{\gamma} \, ! \, \Delta}}$
\end{minipage}%
\begin{minipage}[t]{0.25\textwidth}
  \centering
  $\inferrule[(\compilemode{}$\mid$\quotemode{}-Sub-Expr)]{\strut \Gamma \vDash \gamma' \sqsubseteq \gamma \\ \rcqtypejudge{\gamma'}{\splice}{\runtimecomptype{T}{\Delta ; \xi}}}{\rcqtypejudge{\gamma}{e}{\runtimecomptype{T}{\Delta ; \xi}}}$
\end{minipage}%
\begin{minipage}[t]{0.25\textwidth}
  \centering
  $\inferrule[(\splicemode{}-Sub)]{\Gamma \vDash \gamma' \sqsubseteq \gamma \\ \stypejudge[\Gamma]{e}{\textsf{Code}(T \, ! \, \xi)^{\gamma'} \, ! \, \Delta}}{\stypejudge[\Gamma]{e}{\textsf{Code}(T \, ! \, \xi)^{\gamma} \, ! \, \Delta}}$
\end{minipage}

\end{center}
\end{rec-desc}

\caption{\recLang{}: types and selected typing rules.}
\label{fig:refined-env-classifiers-types}
\end{figure}

\subsection{Correctness of Refined Environment Classifiers}
\label{subsection:rec-formal-correctness}

In this section, we prove the correctness of refined environment classifiers:
every well-typed \recLang{} term produces a well-scoped AST on termination.
However, with an elaboration-based semantics, directly reasoning about
\recLang{} is challenging. As a result, we employ Tait-style logical relations
\citep{tait-67} to demonstrate that typing guarantees are preserved by
elaboration \citep{benton-09}, thereby establishing correctness of refined environment
classifiers.

\begin{figure}
\begin{rec-desc}\arraycolsep=1.4pt
  \footnotesize
  {\normalsize\textbf{The $\scoped{T}$ Logical Relation}}
  \vspace{3mm}

  \textbf{Context of Proof Theoretic Terms}\\[1mm]
  $\Theta := \gamma_\bot \mid \Theta, \gamma \mid \Theta, \gamma' \sqsubseteq \gamma$\\

  \textbf{Normal Forms} \\
    {\scriptsize{\textit{In the following, let $\tau$ be shorthand for any of $T^{0} (\gamma)$, $\effectType[\xi]{T^{0}} (\gamma)$, $(\handlerType{\effectType[\xi_1]{S^0}}{\effectType[\xi_2]{T^0}})^{0}(\gamma)$, or ${(\textsf{Code}(\effectType[\xi]{T^0})^{\gamma})^{-1}}$}}}\\[1mm]
  $
  \begin{array}{@{}lllr}
    n \in \scoped{\mathbb{N}^{-1}} & \triangleq & n \in \mathbb{N} \\
    n \in \scoped{{\tau}} & \triangleq & \cdot \vdash n \in \elaborate{\tau} \text{ and } \Theta \vdash \freevars{n} \subseteq \textsf{permitted}(\gamma) \\
    \vspace{2mm}
    n \in \scoped{\underline{{T}^{0}(\gamma)}} & \triangleq & \texttt{Var}({n}) \in \scoped{{T^{0}} (\gamma)} \\ \vspace{2mm}
    n \in \scoped{(\functionType{{S}^{-1}}{{T}^{-1}})^{-1}} & \triangleq & \forall n' \in \scoped{{S}^{-1}}, n \, n' \in \scoped{\effectType{T^{-1}}} \\ \vspace{2mm}
    n \in \scoped{(\continuationType{{S}^{-1}}{{T}^{-1}})^{-1}} & \triangleq & \forall n' \in \scoped{{S}^{-1}}, \continue{n}{n'} \in \scoped{\effectType{T^{-1}}} \\\\ \vspace{2mm}
    \textbf{Handlers} \\
    h \in \scoped{(\handlerType{\effectType[\Delta_1]{{S}^{-1}}}{\effectType[\Delta_2]{{T}^{-1}}})^{-1}} & \triangleq & \text{if } h = \returnHandler{x}{t_{\text{ret}}}\\ \vspace{2mm}
    && \quad \forall n' \in \scoped{{S}^{-1}}, t_{\text{ret}}[n'/x] \in \scoped{\effectType[\Delta_2]{T^{-1}}} \\
    && \text{else } h = h'; \opHandler{x}{k}{t_{\text{op}}}, \textsf{op}: A^{-1} \to B^{-1} \\
    && \quad h' \in \scoped{(\handlerType{\effectType[\Delta_1]{{S}^{-1}}}{\effectType[\Delta_2]{{T}^{-1}}})^{-1}} \text{ and}\\
    && \quad \forall n \in \scoped{{A}^{-1}}, n' \in \scoped{\continuationType[\Delta_2]{{B}^{-1}}{T^{-1}}}, t_{\text{op}}[n/x, n'/k] \in \scoped{\effectType[\Delta_2]{T^{-1}}}
  \end{array}
$\\

\textbf{Terms}\\
{\scriptsize{\textit{In the following, let $\effectType{\tau}$ be shorthand for any of $\effectType{T^{0}} (\gamma)$, $\effectType[\Delta; \xi]{T^{0}} (\gamma)$, $\effectType{(\handlerType{\effectType[\xi_1]{S^0}}{\effectType[\xi_2]{T^0}})^{0}} (\gamma)$, or $\effectType{T^{-1}}$}}}\\
{\scriptsize{\textit{Given a compile-time computation type $\effectType{\tau}$, let $\tau$ refer to the corresponding value type. e.g.\ if $\effectType{\tau} = \effectType[\Delta; \xi]{T^{0}} (\gamma)$, then $\tau = \effectType[\xi]{T^{0}} (\gamma)$}}}\\[1mm]
$\scoped{\effectType{\tau}} \hspace{2mm} \triangleq \hspace{2mm}$ The smallest property on terms $t$ such that either:
\begin{enumerate}
  \item For arbitrary $U$ consistent with $t$, exists $U'$ such that $\langle t;[-];U \rangle \to^{*} \langle \return{n}; [-]; U' \rangle$, such that $U'$ consistent with $n$, and $n \in \scoped{\tau}$
  \item For arbitrary $U$ consistent with $t$, exists $U'$ such that $\langle t;[-];U \rangle \to^{*} \langle \op{n}; E; U' \rangle$ where $\textbf{\texttt{op}} \not\in \textsf{handled}(E)$, $U'$ consistent with $E[\op{n}]$, and
  \begin{enumerate}
    \item $\textsf{op}: A^{-1} \to B^{-1}$,
    \item $n \in \scoped{A^{-1}}$, and
    \item for all $n' \in \scoped{B^{-1}}$, $E[n'] \in \scoped{\effectType{\tau}}$
  \end{enumerate}
\end{enumerate}
Where, in this context, consistent with $t$ means that for all $\Var{\alpha^\gamma}{R}$ or $\Binder{\alpha^\gamma}{R} \in t$, $\alpha \in U$. This side condition ensures that we use \textbf{\texttt{mkvar}} correctly.
\end{rec-desc}
\caption{The definition of the \textsf{Scoped} logical relation}
\label{fig:logical-relation-defn}
\end{figure}

\Cref{fig:logical-relation-defn} presents the logical relation, \textsf{Scoped},
defined on core language (\recCoreLang{}) terms. The relation is
indexed by a context of proof-theoretic terms $\Theta$ and an \textit{extended}
\recLang{} type (\Cref{dfn:extended-source-type}). Given a context $\Gamma$,
$\pi_{\gamma}({\Gamma})$ projects out only the proof theoretic terms. For
example, given $\Gamma = \gamma_\bot, \gamma_1, \gamma_\bot \sqsubseteq
\gamma_1, \textcolor{comment}{(x:\mathbb{N}^0)^{\gamma_1}}, \gamma_2, \gamma_1
\sqsubseteq \gamma_2, \textcolor{comment}{y:(\textsf{Code}(\mathbb{N}^0 \, ! \,
  \emptyset)^{\gamma_2})^{-1}}$, the proof theoretic part of the context is
$\pi_{\gamma}({\Gamma}) = \gamma_\bot, \gamma_1, \gamma_\bot \sqsubseteq
\gamma_1, \gamma_2, \gamma_1 \sqsubseteq \gamma_2$, which is an instance of
$\Theta$.

The two key definitions are the relation on the $T^{0} (\gamma)$ value type ($\scoped{T^{0} (\gamma)}$), and the relation on terms ($\scoped{\effectType{\tau}}$). For a normal form $n$ to be in $\scoped{T^{0} (\gamma)}$, $n$ must be of type $\textsf{AST}(\erase{T^0})$, and the free variables of $n$ need to be permitted within the scope represented by $\gamma$. Permissibility assumes some known partial order on classifiers, e.g. $\gamma' \sqsubseteq \gamma$, which is carried by the index $\Theta$. $\scoped{\effectType{\tau}}$ is defined as a least fixed point, following similar definitions by \citet{plotkin-2025} and \citet{kuchta-2023}, giving rise to the principle of \textsf{Scoped}-Induction:

\begin{inductionPrinciple}[\textsf{Scoped}-Induction]
  For a property $\Phi$ on closed terms of type $\elaborate{\effectType{\tau}}$,
  \begin{enumerate}
    \item if $\langle t; [-]; U \rangle \to^{*} \langle \return{n};[-];U' \rangle$ implies $\Phi(t)$, and
    \item if $\langle t; [-]; U \rangle \to^{*} \langle \op{n};E;U' \rangle$ where $\textbf{\texttt{op}} \not\in \textsf{handled}(E)$, $\textsf{op}: A^{-1} \to B^{-1}$, $n \in \scoped{A^{-1}}$, and for arbitrary $n' \in \scoped{B^{-1}}$, $\Phi(E[n'])$ implies $\Phi(t)$,
  \end{enumerate}
  then for all $t \in \scoped{\effectType{\tau}}$, $\Phi(t)$
\end{inductionPrinciple}

The proof additionally relies on a closure lemma \citep{kuchta-2023} and a notion of closed substitution $\rho \vDash \Gamma$. Care must be taken with substitution of level $0$ variables, since these should be in the logical relation for \textsf{FParam}s rather than \textsf{AST}s (clause 2 in \Cref{dfn:closed-substitution}).

\begin{lemma}[Closure under Anti-Reduction]\label[lemma]{lemma:closure-reduction}
  Assume $\langle t;E;U \rangle \to^{*} \langle t';E';U' \rangle$. Then $E'[t'] \in \scoped{\effectType{\tau}} \implies E[t] \in \scoped{\effectType{\tau}}$
\end{lemma}
\begin{definition}[Closed Substitution]\label[definition]{dfn:closed-substitution}
  Given a context $\Gamma$, and assuming $\Theta = \pi_{\gamma}(\Gamma)$, the set of closed substitutions $\rho \vDash \Gamma$ are defined inductively as follows:
  \begin{enumerate}
    \item $() \vDash \gamma_{\bot}$
    \item If $\rho \vDash \Gamma$, then for arbitrary $\gamma \in \Gamma$, $n \in \scoped{\underline{T^{0} (\gamma)}}$, $(\rho, n/x) \vDash \Gamma, (x:T^0)^{\gamma}$
    \item If $\rho \vDash \Gamma$, $\Gamma \vdash T^{-1}$, and $n \in \scoped{T^{-1}}$, then $(\rho, n/x) \vDash \Gamma, (x:T^{-1})$
    \item If $\rho \vDash \Gamma$ then $\rho \vDash \Gamma, \gamma$, for arbitrary $\gamma \not\in \Gamma$
    \item If $\rho \vDash \Gamma$ then $\rho \vDash \Gamma, \gamma \sqsubseteq \gamma'$, for arbitrary $\gamma, \gamma' \in \Gamma$
  \end{enumerate}
\end{definition}

Finally, we introduce a $\Theta$-truncation lemma, which allows us to discard proof theoretic terms ($\gamma$, $\gamma \sqsubseteq \gamma'$) should they not be necessary for the proof.

\begin{lemma}[$\Theta$-truncation] \label[lemma]{lemma:theta-truncation}
Assume an AST $n$. If $\textsf{Var}(\alpha^{\gamma'}_S)$ does not occur in $\textsf{FVs}^0(n)$, and $\textsf{Var}(\alpha^{\gamma'}_S)$ is the only variable tagged with classifier $\gamma'$, then
$n\in\mathsf{Scoped}_{(\Theta, \gamma', \gamma \sqsubseteq \gamma'),\tau}$ implies $n\in\mathsf{Scoped}_{\Theta,\tau}$
\end{lemma}

Stratification of types and mode-indexing decomposes the fundamental lemma into many sub-lemmas; here we present one such sub-lemma:

\begin{lemma}[Fundamental Lemma {[\compilemode{}, {$\effectType[\Delta; \xi]{T^0}$}]} of the \textsf{Scoped} Logical Relation] \label[lemma]{lemma:fundamental}
  If $\Gamma \vdash_{\compilemode{}}^{\gamma} e: T^{0} \, ! \, \Delta ; \xi$ then for $\Theta = \pi_\gamma({\Gamma})$, and for all $\rho$ such that $\rho \vDash \Gamma$,$\elaborate{e}_{\compilemode{}}(\rho) \in \scoped[\Theta]{T^{0} \, ! \, \Delta ; \xi (\gamma)}$
\end{lemma}

Proof of \Cref{lemma:fundamental} is by induction on the \recLang{} typing rules. In the \textsc{\compilemode{}-Lambda} case, it suffices to show that for $\rho \vDash \Gamma$, $\bind{x}{\gensym{\erase{S^0 (\gamma')}}}{\bind{\texttt{body}}{\elaborate{e}_{\compilemode}({\rho})}{\return{\Lam{x}{\texttt{body}}}}}$ is
in $\scoped{\effectType{(\functionType[\xi]{S^0}{T^0})^0} (\gamma)}$. This reduces to ${\bind{\texttt{body}}{\elaborate{e}_{\compilemode}({\rho, \Binder{\alpha^{\gamma'}}{S} / x})}{\return{\Lam{\Binder{\alpha^{\gamma'}}{S}}{\texttt{body}}}}}$. By anti-reduction (\Cref{lemma:closure-reduction}) it suffices to show that this term is in the logical relation. By weakening (\Cref{lemma:weakening}), and the induction hypothesis (IH), $\elaborate{e}_{\compilemode}({\rho, \Binder{\alpha^{\gamma'}}{S} / x}) \in \scoped[\Theta']{\effectType[\Delta; \xi]{T^{0}} (\gamma')}$, where $\Theta' =  \Theta, \gamma', \gamma \sqsubseteq \gamma'$. It suffices to show:
\[\forall t \in \scoped[\Theta']{\effectType[\Delta; \xi]{T^{0}} (\gamma')}\text{, } \bind{\texttt{body}}{t}{\return{\Lam{\Binder{\alpha^{\gamma'}}{S}}{\texttt{body}}}} \text{ in } \scoped{\effectType{(\functionType[\xi]{S^0}{T^0})^0} (\gamma)}\]
Applying \textsf{Scoped}-Induction,
 \begin{enumerate}
  \item $t \in \scoped[\Theta']{\effectType[\Delta; \xi]{T^{0}} (\gamma')}$ reduces to some $\return{n}$\\
   $\bind{\texttt{body}}{\return{n}}{\return{\Lam{\Binder{\alpha^{\gamma'}}{S}}{\texttt{body}}}}$ reduces to $\return{\Lam{\Binder{\alpha^{\gamma'}}{S}}{n}}$, where $\Var{\alpha^{\gamma'}}{S}$ is bound. By IH, $n \in \scoped[\Theta']{\effectType[\xi]{T^{0}} (\gamma')}$. Thus, all the free variables in $n$ are permitted by $\gamma'$. By the typing rules, only $\alpha$ is annotated with classifier $\gamma'$. Hence, using \Cref{lemma:theta-truncation}, under $\Theta$, the free variables of $\Lam{\Binder{\alpha^{\gamma'}}{S}}{n}$ are permitted by $\gamma$. The conclusion thus follows from anti-reduction.
  \item $t \in \scoped[\Theta']{\effectType[\Delta; \xi]{T^{0}} (\gamma')}$ reduces to $E[\op{n}]$, $\op{n} \text{ unhandled}$ \\
  As $\bind{\texttt{body}}{[-]}{\return{\Lam{\Binder{\alpha^{\gamma'}}{S}}{\texttt{body}}}}$ introduces no handlers, the conclusion follows immediately from the \textsf{Scoped}-Induction hypothesis and anti-reduction.
\end{enumerate}

Using the logical relation, and a type safety result identical to \citeauthor{bauer-2014}'s [\citeyear{nanevski-contextual}] Corollary 4.2, we prove the correctness of refined environment classifiers:

\begin{theorem}[Correctness of Refined Environment Classifiers]\label{thm:refined-env-classifiers-correct}
  If $\gamma_\bot \vdash ^{\gamma_\bot}_{\compilemode{}} e: T^0 \, ! \, \emptyset ; \emptyset $, and $\elaborate{e}_{\compilemode{}} = t$,
  \noindent{}then for some $U$, $\langle t;[-]; \emptyset \rangle \to^{*} \langle \return{n}; [-] ; U \rangle$, and $\freevars{n} = \emptyset$
\end{theorem}

\subsection{Expressiveness of Refined Environment Classifiers}\label{subsection:rec-formal-expressiveness}

\recLang{} prevents scope extrusion by looking only at the argument to the
effect, not at the handler. In a well-typed \recLang{} program, the only variables
that may be passed to an effect $\textbf{\texttt{op}}$ are those that are in
scope when the handler for $\textbf{\texttt{op}}$ is defined, for example, the
variable $z$ in \Cref{listing:refined-environment-classifiers-safe}:

\begin{code}
\begin{rec}
$\begin{array}{l}
  \lambda z. \$(\textbf{\texttt{handle}} \; \equote[\, {\lambda x. \, \return{\splice[( \, {{\textbf{\texttt{op}}(\equote[z])}} \, )]}} \,] \\
  \quad \quad \,\, \textbf{\texttt{with}} \, \{ \textbf{\texttt{return}}(u) \mapsto {\return{u}}; \textbf{\texttt{op}}(y, k) \mapsto {\continue{k}{()}}\})
\end{array}$
\end{rec}
\captionof{listing}{Refined environment classifiers allow variables to be passed to an effect, so long as the variable can never cause a scope extrusion error (e.g.~$z$ may be passed, since it is bound outside the handler definition).}%
\label{listing:refined-environment-classifiers-safe}
\end{code}

\Cref{table:expressiveness-comparison} summarizes the expressiveness of
refined environment classifiers on our set of programs; as shown, refined
environment classifiers are less expressive than all the dynamic checks.

\section{Implementation}\label{chapter:implementation}
We have implemented the various dynamic checks in the MacoCaml compiler, and
made an implementation with the C4C check available as an artifact \citep{artifact}.
MacoCaml implements quotation via elaboration in a similar manner to
the elaboration of \Cref{section:elaboration}, albeit targeting a
lower-level intermediate language Lambda rather than ASTs.
We have extended the elaboration with extrusion checking similarly to the extended
elaborations of \Cref{section:lazy-dynamic-check-formal,section:eager-dynamic-check-formal,section:best-effort-check}.

The MacoCaml implementation closely follows the description in \Cref{chapter:scope-extrusion}. The implementation realises \textbf{\texttt{check}}, \textbf{\texttt{dlet}}, and \textbf{\texttt{err}} as a \textit{mode of use} of effects and handlers:
\begin{enumerate}
  \item $\checkfv{n}$ is implemented by performing a \texttt{FreeVar} effect, passing it the free variables of $n$.\\
  $\checkm{n}$ is similar, except that it additionally relies on a \texttt{Mute} effect that is performed within the handlers of effects besides \texttt{FreeVar} to implement the \textsc{Eff-Op} rule (\Cref{fig:core}).
  
  \item $\dlet{\Binder{\alpha}{R}}{t}$ is implemented as a handler of the \texttt{FreeVar} effect: it subtracts $\Var{\alpha}{R}$ from the set of free variables, and either:
  \begin{enumerate}
    \item resumes the continuation, if the set of free variables is now empty (i.e.~if all free variables are declared safe), or
    \item performs another \texttt{FreeVar} effect, to check that the remaining free variables are declared safe. If the check returns successfully, the continuation is resumed.
  \end{enumerate}
  \item \textbf{\texttt{err}} is implemented as an unhandled \texttt{FreeVar} effect.
\end{enumerate}

\section{Related Work}\label{chapter:related}
Using mutation and control effects for code generation, particularly
for let-insertion, has a long history \cite{lawall-94,sumii-hybrid};
\citet[\S8]{kameyama-shifting-jfp}
and \citet[\S5.3]{kameyama-2015} give a thorough overview.  The danger
of generating code with unbound variables has
also become apparent. There are two lines of work dealing with the
problem: prevention and detection.
Most of the prevention research focuses on designing an appropriate
type system, such as closed types~\citep{calcagno-00} or environment
classifiers~\citep{DBLP:conf/popl/TahaN03} (although the latter
prevents scope extrusion arising from \texttt{eval} rather than from effects).  The
majority of type systems aimed at preventing scope extrusion are
considerably more complex \citep{kameyama-2015,kiselyov-16,parreaux-2020,isoda-24}, and are essentially variations of \citeauthor{nanevski-contextual}'s [\citeyear{nanevski-contextual}]
Contextual Modal Type Theory.

Besides types, one may also guarantee the absence of scope extrusion
by restricting the scope of mutation (so-called \emph{weak separability}
\cite{mint}) or by restricting the scope of control effects by placing an
effect handler under every future-stage binder
\cite{kameyama-shifting-jfp}. Continuation-passing or monadic
transformations ~\cite{swadi-monadic} amount to the same.

Most of the prevention techniques limit, often severely, the
expressiveness of the language.\ \citet{kameyama-2015} proposed
a set of benchmarks to evaluate expressiveness of program generation systems; at
that time only \citet{kameyama-2015} passed all the benchmarks.

Whereas \emph{prevention} techniques statically reject potentially unsafe code-generating programs when compiling the code generator, \emph{detection} techniques operate when executing the code generator, alerting the metaprogrammer when a code
fragment with a scope-extruded variable has been
generated. Since generated code must eventually be compiled, the
simplest detection technique is to do nothing during code generation, instead
relying on the compiler of the generated code to report extrusion.
This approach is what \Cref{section:overview} calls the \emph{lazy check}; as we stressed,
it suffers from severe usability problems in practice.\ \citet{kiselyov-14} took efforts to implement the \textit{eager check}, detecting scope extrusion as soon as it occurs, before
the complete code is generated, with informative error
messages. However, the design and implementation were not formalised. Hence, it was difficult to evaluate the check (or even
to tell if it detects errors at the earliest possible point).

Very few of the prevention approaches have been implemented in systems
that are used in practice (or, at least, that are used for realistic, larger-scale
examples): examples include Mint~\cite{mint}, Contextual Squid~\cite{parreaux-2020} and StagedHaskell~\cite{kameyama-2015}. Mint is
very restrictive, outright prohibiting let-insertion and
assert-insertion beyond binders. The other, type-based approaches,
permit optimisations such as let-insertion and loop interchange, but they are very
complex.  Contextual Squid, implemented on top of Scala-2 macros,
required access to Scala compiler internals which is no longer available in
Scala-3. StagedHaskell relied on tricky Haskell type class
programming, where type annotations are often required, and where the types can
become quite complex and the error messages
incomprehensible. The complexity of the types was unfortunately necessary:
\S4.1 of \citeauthor{kameyama-2015}'s work showed subtle and serious problems
that could arise with \emph{unintendedly} bound variables in a version
of their system with simpler types.
\citet{parreaux-2020} reports that some users found
dealing with contextual types to be too much of a burden.\
\citet{kiselyov-14} gives more discussion of practical aspects of the
scope extrusion check.

Many practical metaprogramming systems such as Template Haskell~\cite{sheard-02} rely
on generation-time detection of scope extrusion, in particular the
lazy check, i.e.~offshoring all the detection to the compiler
that compiles the generated code. The notable exception is (BER)
MetaOCaml~\cite{kiselyov-14,kiselyov-24}, where \textit{eager} scope
extrusion detection is the principal feature.

\section{Conclusions}\label{chapter:conclusions}
We have presented the first formal framework for comparing scope
extrusion checks, based on the calculus \sourceLang{} and its
elaboration into the \coreLang{} core language.
Using the framework, we have modelled the two main approaches to
checking scope extrusion, lazy and eager checking, and developed a new
check which combines the best properties of both, and which interacts
well with effects and handlers.
We have incorporated the new check into the MacoCaml implementation.
Our framework also extends to modelling the refined environment
classifier system for preventing scope extrusion, and we expect that
it could be similarly extended to model other static systems.

\begin{acks}
We thank the anonymous POPL reviewers and Neel Krishnaswami, whose comments helped to improve the work,
and Alistair O'Brien and Yulong Huang, whose comments helped to improve the artifact.
This work is funded by Jane Street Capital, by Ahrefs, and by the Natural Sciences and Engineering Research Council of Canada.
\end{acks}

\bibliographystyle{ACM-Reference-Format}
\bibliography{scope}

\begin{extendedonly}
\pagebreak
\appendix

\section{Expanded versions of the matrix multiplication examples}\label[appendix]{appendix:matmul}
We illustrate the code generated by the examples
in \Cref{section:overview}, using as the running example the following
MacoCaml code, which defines a function \lstinline{f} whose body is
generated by a call to the \lstinline{mmul} macro:

\begin{macocamllst}
let f a b c = $(mmul <<a>> <<b>> <<c>>)
\end{macocamllst}

For \Cref{code:staged-matrix-multiplication}, \lstinline{mmul} generates the following code:

\begin{macocamllst}
let f a b c = for i = 0 to length a - 1 do
                for k = 0 to length a.(0) - 1 do 
                  for j = 0 to length b.(0) - 1 do 
                    c.i.j <- c.i.j + a.i.k * b.k.j
                  done 
                done 
              done 
\end{macocamllst}

For the let-inserting variant in \Cref{code:staged-matrix-multiplication}, \lstinline{mmul} generates the following code:

\begin{macocamllst}
let f a b c = for i = 0 to length a - 1 do
                for k = 0 to length a.(0) - 1 do 
                  let e1 = a.i.k in 
                    for j = 0 to length b.(0) - 1 do 
                      let e2 = b.k.j in 
                        c.i.j <- c.i.j + a.i.k * b.k.j
                    done 
                  done 
                done 
\end{macocamllst}

Notice how the \lstinline{Genlet} effect has been used to 
perform loop-invariant code motion. 

However, if \lstinline{mmul} is faulty because the programmer 
exchanges the loops without modifying the let insertion code, 
then the generated program is instead as follows: 

\begin{macocamllst}
let f a b c = for i = 0 to length a - 1 do
                for j = 0 to length b.(0) - 1 do 
                  let e2 = b.k.j in (*k not in scope*)
                    for k = 0 to length a.(0) - 1 do 
                      let e1 = a.i.k in 
                        c.i.j <- c.i.j + a.i.k * b.k.j
                    done 
                  done 
                done 
\end{macocamllst}

\noindent
where scope extrusion has occurred on line 3.

Further, if \lstinline{mmul} is faulty because the programmer 
has identified the wrong loop when performing let-insertion, then 
the generated program is instead as follows:

\begin{macocamllst}
let f a b c = for i = 0 to length a - 1 do
                let e1 = a.i.k in (*k not in scope*)
                  for k = 0 to length a.(0) - 1 do 
                    let e2 = b.k.j in (*j not in scope*)
                      for j = 0 to length b.(0) - 1 do  
                        c.i.j <- c.i.j + a.i.k * b.k.j
                      done 
                    done 
                  done 
\end{macocamllst}

\noindent
where scope extrusion has occurred on lines 2 and 4.

\section{\efflang{}: A base calculus of effects and handlers}\label[appendix]{appendix:full-efflang-rules}

\newcommand{\print}[1]{\texttt{\textbf{print}}(\texttt{#1})}
\newcommand{\accum}[1]{\texttt{\textbf{accum}}(\texttt{#1})}
\newcommand{\readInt}[1]{\texttt{\textbf{read\_int}(#1)}}

\efflang{} is a base calculus with deep effect handlers and multi-shot continuations, broadly similar to the calculus of \citet{pretnar-15}, except that \efflang{} treats handlers as a distinct syntactic category, uses generic effects, disambiguates functions from continuations at both the type and syntax level (like \citet{isoda-24}), and has multiple typing rules for handlers (like \citet{biernacki-2017}).
\Cref{listing:efflang-running-example} presents a sample \efflang{} program that evaluates to $22$ (i.e.~$1 + (1 + ((20 * 2) - (10 * 2)))$).

\begin{code}
  \centering
  \begin{efflst}
    $\begin{array}{l}
      \textbf{\texttt{handle}}\;\bind{x}{\accum{20}}{{\bind{y}{\accum{10}}{\return{(x - y)}}}} \\
      \textbf{\texttt{with}}\;\{ \textbf{\texttt{accum}}(v, k) \mapsto {\bind{z}{\continue{k}{(v * 2)}}{\return{1 + z}}}\}
    \end{array}$
  \end{efflst}
  \captionof{listing}{A \efflang{} program that returns $\texttt{22}$.}
  \label{listing:efflang-running-example}
  \end{code}

\begin{figure}[ht]
  \begin{eff-desc}
    \footnotesize
\begin{minipage}[t]{0.5\textwidth}
    {\normalsize{\textbf{Syntax}}}\\
  $\begin{array}{@{}llll}
  \text{Values} & v & := & x \mid m \in \mathbb{N} \mid \lambda x. c \mid \kappa x.c\\

  \text{Computations} & c & := & v_1\;v_2 \mid \return{v} \\
                             &&& \mid \bind{x}{c_1}{c_2} \\
                             &&& \mid \op{v}  \\
                             &&& \mid \handleWith{c}{h} \\
                             &&& \mid \continue{v_1}{v_2} \\ 
  \text{Handlers} & h & := &\returnHandler{x}{c} \\
                             &&& \mid h;\opHandler{x}{k}{c}
  \end{array}$
  
\end{minipage}
\begin{minipage}[t]{0.5\textwidth}
  {\normalsize \textbf{Types}}\\
  $\begin{array}{@{}lllr}
    \text{Effects set} & \Delta ::= \emptyset \mid \Delta \cup \{ \texttt{op}_i \} \\ \\
    \text{Value type} & S, T ::= \mathbb{N} \mid \functionType{S}{T}  \mid \continuationType{S}{T} \\
    \text{Computation type} & \effectType{T} \\
    \text{Handler type} & \handlerType{\effectType[\Delta_1]{S}}{\effectType[\Delta_2]{T}}
  \end{array}$
\end{minipage}

\vspace{3mm}

  \arraycolsep=3pt
  
  {\normalsize\textbf{Operational Semantics}}\\
  \renewcommand{\effconfiguration}[2]{\langle {#1}; {#2} \rangle}
  \renewcommand{\transition}[2]{#1 & \rightarrow & #2}
  \footnotesize
  \textbf{Auxiliary Definitions}
  {\footnotesize
    \[\begin{array}{lrcl}
    \text{Evaluation Frame } & F & ::= & \bind{x}{[-]\,}{c_2} \mid \handleWith{[-]}{h} \\
    \text{Evaluation Context } & E & ::= & [-] \mid E[F] \\ \vspace{1mm} \\
    \text{Domain of Handler} & \textsf{dom}(h) & \triangleq & \textsf{dom}(\returnHandler{x}{c}) = \emptyset, \\
    &&&\textsf{dom}(h;\opHandler{x}{k}{c}) = \textsf{dom}(h) \cup \{ \textbf{\textsf{op}} \} \\  
    \vspace{1mm} \text{Handled Effects} & \textsf{handled}(E) & \triangleq & \textsf{handled}([-]) = \emptyset, \\ 
    &&& \textsf{handled}(E[\bind{x}{[-]\,}{c_2}]) = \textsf{handled}(E), \\
    &&& \textsf{handled}(E[\handleWith{[-]}{h}]) = \textsf{handled}(E) \cup \textsf{dom}(h),
  \end{array}
  \]}

{\noindent\textbf{Reduction Rules}}
  {\footnotesize
\[
  \begin{array}{rrcl}
  \reductionRule{App} & \transition{\effconfiguration{(\function{x}{c})v}{E}}{\effconfiguration{c[v/x]}{E}}\\
  \reductionRule{Seq} & \transition{\effconfiguration{\bind{x}{\return{v}}{c}}{E}}{\effconfiguration{c[v/x]}{E}}\\
  \reductionRule{Hdl} & \transition{\effconfiguration{\handleWith{\return{v}}{h}}{E}}{\effconfiguration{c[v/x]}{E}} \quad (\text{where $\returnHandler{x}{c} \in h$)}\\
  \vspace{1mm} \\ 
  \congruenceRule{Psh} & \transition{\effconfiguration{F[c]}{E}}{\effconfiguration{c}{E[F]}} \\
  \congruenceRule{Pop} & \transition{\effconfiguration{\return{v}}{E[F]}}{\effconfiguration{F[\return{v}]}{E}}\\
  \vspace{1mm} \\
  \effectRule{Op} & \transition{\effconfiguration{\op{v}}{E_1[\handleWith{E_2}{h}]}}\effconfiguration{c[v/x, \kappa x. \, \handleWith{E_2[\return{x}]}{h} / k]}{E_1}\\
  &&& \text{(where $\textbf{\texttt{op}} \in \textsf{dom}(h)$ and $\textbf{\textsf{op}} \notin \textsf{handled}(E_2)$)}\\
  \effectRule{Cnt} & \transition{\effconfiguration{\continue{(\kappa x. E_2[\return{x}])}{v}}{E_1}}{\effconfiguration{\return{v}}{E_1[E_2]}}

\end{array}
\]
  }
  \end{eff-desc}
\caption{\efflang{}: syntax, types, operational semantics. }
\label{fig:efflang}
\end{figure}

\Cref{fig:efflang} shows the syntax of \efflang{} terms and types, and the operational semantics, given on configurations of the form $\effconfiguration{c}{E}$ for a term $c$ and evaluation context  $E$, in the style of \citet{felleisen-87}. Evaluation contexts are represented as a stack of evaluation frames $F$, à la \citet{kiselyov-2012}.  

For clarity, we compress multiple stack frames using nesting. For example, instead of $\bind{x}{-\,}{c_2} :: \bind{y}{-\,}{c_1}$, we write $\bind{x}{(\bind{y}{-\,}{c_1})}{c_2}$. 

Types are divided into value types (for example, $\mathbb{N}$), computation types ($\effectType[\{\textbf{\texttt{print}} \}]{\mathbb{N}}$), and handler types ($\handlerType{\effectType[\{\textbf{\texttt{print}} \}]{\mathbb{N}}}{\effectType[\emptyset]{\mathbb{N}}}$).

Effect signatures may be recursive, and so \efflang{} supports writing non-terminating programs, e.g.:

\begin{center}
\begin{eff}
  $\handleWith{(\lambda \_ . \textbf{\texttt{recursive}} ()) ()}{\textbf{\texttt{recursive}}(\_, k) \mapsto {\continue{k}{(\lambda \_. \textbf{\texttt{recursive}} ())}}}$
\end{eff}
\end{center}

\begin{figure}
  \begin{eff-desc}
    {\large\textbf{Typing Rules}}\\
    \vspace{1mm}\\
    \scriptsize
    \fbox{$\Gamma \vdash v : T$}\\
    \begin{center}
    \begin{minipage}[t]{0.2\textwidth}
      \centering
      $\inferrule[(Nat)]
      { \\ }
      {\type{m}{\mathbb{N}}}$
      \end{minipage}%
  \begin{minipage}[t]{0.2\textwidth}
    \centering
  $\inferrule[(Var)]
  {\Gamma(x) = T}
  {\type{x}{T}}$
  \end{minipage}%
  \begin{minipage}[t]{0.3\textwidth}
    \centering
  $\inferrule[(Lambda)]
    {\type[, x:S]{c}{\effectType{T}}}
    {\type{\function{x}{c}}{\functionType{S}{T}}}$
  \end{minipage}%
  \begin{minipage}[t]{0.3\textwidth}
  \centering
$\inferrule[(Continuation)]
  {\type[, x:S]{c}{\effectType{T}}}
  {\type{\continuation{x}{c}}{\continuationType{S}{T}}}$
\end{minipage}  
  \vspace{3mm}

\end{center}
  
  \fbox{$\Gamma \vdash c : \effectType{T}$}\\
  \begin{center}
    
  \begin{minipage}[t]{0.25\textwidth}
    \centering
  $\inferrule[(App)]
    {\type{v_1}{S \oset{\text{\tiny{$\Delta$}}}\longrightarrow T} \\ \type{v_2}{S}}
    {\type{v_1 \, v_2}{\effectType{T}}}$
  \end{minipage}%
  \begin{minipage}[t]{0.25\textwidth}
    \centering
  $\inferrule[(Continue)]
    {\type{v_1}{\continuationType{S}{T}} \\ \type{v_2}{S}}
    {\type{\continue{v_1}{v_2}}{\effectType{T}}}$
  \end{minipage}%
  \begin{minipage}[t]{0.25\textwidth}
    \centering
  $\inferrule[(Return)]
    { \\\\ \type{v}{T}}
    {\type{\return{v}}{\effectType{T}}}$
  \end{minipage}%
  \begin{minipage}[t]{0.25\textwidth}
    \centering
  $\inferrule[(Do)]
    {\type{c_1}{\effectType[\Delta]{S}} \\ \type[, x: S]{c_2}{\effectType{T}}}
    {\type{\bind{x}{c_1}{c_2}}{\effectType{T}}}$
  \end{minipage}
  
  \vspace{3mm}
  
  \begin{minipage}[t]{0.5\textwidth}
    \centering
  $\inferrule[(Op)]
    {  \\\\ \type{v}{S} \\ \texttt{op}: S \rightarrow T \in \Sigma \\ \texttt{op} \in \Delta}
    {\type{\op{v}}{\effectType{T}}}$
  \end{minipage}%
  \begin{minipage}[t]{0.5\textwidth}
    \centering
  $\inferrule[(Handle)]
    {\type{c}{\effectType[\Delta_1]{S}} \\ \type{h}{\handlerType{\effectType[\Delta_1]{S}}{\effectType[\Delta_2]{T}}} \\ \forall \textsf{op} \in \Delta_1 \setminus \Delta_2. \, \textsf{op} \in \textsf{dom}(h)}
    {\type{\handleWith{c}{h}}{\effectType[\Delta_2]{T}}}$
  \end{minipage}\\

  \vspace{3mm}

\end{center}

\fbox{$\Gamma \vdash h: \handlerType{\effectType[\Delta_1]{S}}{\effectType[\Delta_2]{T}}$}\\
\begin{center}

  \begin{minipage}[t]{0.5\textwidth}
    \centering
  $\inferrule[(Ret-Handler)]
    {  \\\\  \\\\   {\type[, x:S]{c}{\effectType[\Delta_2]{T}}}}
    {\type{\returnHandler{x}{c}}{\handlerType{\effectType[\Delta_1]{S}}{\effectType[\Delta_2]{T}}}}$
  \end{minipage}%
  \begin{minipage}[t]{0.5\textwidth}
    \centering
  $\inferrule[(Op-Handler)]
    { \texttt{op}: A \to B \in \Sigma \\ 
      \type{h}{\handlerType{\effectType[\Delta_1]{S}}{\effectType[\Delta_2]{T}}}\\
      \type[, x:A, k:{\continuationType[\Delta_2]{B}{T}} ]{c}{\effectType[\Delta_2]{T}}\\
      \Delta_1 \subseteq \Delta_2 \cup \{ \texttt{op} \} \\
             \opHandler{x'}{k'}{c'} \notin h}
    {\type{h ; \opHandler{x}{k}{c}}{\handlerType{\effectType[\Delta_1]{S}}{\effectType[\Delta_2]{T}}}}$
  \end{minipage}
\end{center}
  \end{eff-desc}
  \caption{\efflang{} typing rules}
  \label{fig:efflang-type-system}
  \end{figure}

The typing rules for terms are standard (\Cref{fig:efflang-type-system}). Since types are stratified, so are typing judgements: $\Gamma \vdash v: T$ (values), $\Gamma \vdash c: \effectType{T}$ (computations), and $\Gamma \vdash h: \handlerType{\effectType[\Delta_1]{S}}{\effectType[\Delta_2]{T}}$ (handlers).

A closed computation is well-typed if it can be typed with an empty effects set.  
\begin{definition}[Well-Typed Closed Computation]
  $c$ is a closed, well-typed computation if $\cdot \vdash {c}: {T \, ! \, \emptyset}$
\end{definition}

\subsubsection{Metatheory}
Discussion around scope extrusion builds on some metatheoretic properties of \efflang{}, which are proven by \citet{bauer-2014}.

\renewcommand{\effconfiguration}[2]{\langle {#1}; {#2} \rangle}
\renewcommand{\transition}[2]{#1 \rightarrow #2}

\begin{theorem}[Progress]\label{thm:progress}
If $\cdot \vdash {E[c]}: {\effectType{T}}$ then either 
\begin{enumerate}
\item $c$ is of the form $\return{v}$ and $E = [-]$,
\item $c$ is of the form $\op{v}$ for some $\textsf{op} \in \Delta$, and $\texttt{op} \notin \textsf{handled}({E})$
\item $\exists \, c', E'$ such that $\transition{\effconfiguration{c}{E}}{\effconfiguration{c'}{E'}}$
\end{enumerate}
\end{theorem}

\begin{theorem}[Preservation]\label{thm:preservation}
If $\cdot \vdash {E[c]}: {\effectType{T}}$ and $\transition{\effconfiguration{c}{E}}{\effconfiguration{c'}{E'}}$, then $\cdot \vdash {E'[c']}: {\effectType{T}}$
\end{theorem}

\begin{corollary}[Type Safety]\label{thm:type-safety}
  If $\cdot \vdash {c}: {\effectType[\emptyset]{T}}$ then either 
\begin{enumerate}
\item $\langle c; [-] \rangle \to^{\omega}$ (non-termination)
\item $\langle c; [-] \rangle \to^{*} \langle \return{v}; [-] \rangle$
\end{enumerate}
\end{corollary}

\section{Full \calculusName{} rules}\label[appendix]{appendix:full-quote-op-rules}

\begin{figure}
\begin{source-desc}
  \footnotesize
  {\normalsize \textbf{Syntax}} \\
  $\begin{array}{@{}llll}
  \text{Values} & v & := & x \mid m \in \mathbb{N} \mid \lambda x. e \\

  \text{Expressions} & e & := & v_1\;v_2 \mid \return{v}  \mid \bind{x}{e_1}{e_2} \mid \op{e} \mid \handleWith{e}{h} \mid \continue{v_1}{v_2} \\
                             &&& \mid \equote \mid \splice \\
  \text{Handlers} & h & := &\returnHandler{x}{e} \mid h;\opHandler{x}{k}{e}
  \end{array}$

  \vspace{3mm}

\begin{minipage}[t]{0.5\textwidth}
  {\normalsize \textbf{Effect sets}} \\
    $\begin{array}{@{}ll}
    \begin{array}{@{}lllr}\textbf{Run-Time} & \xi ::= \emptyset \mid \xi \cup \{ \textsf{op}_i^{0}\}\\
    \textbf{Compile-Time} & \Delta ::= \emptyset \mid \Delta \cup \{ \textsf{op}_i^{-1}\} \end{array}
    \end{array}$
  \end{minipage}%
\begin{minipage}[t]{0.5\textwidth}
   {\textbf{\normalsize {Typing contexts}}}\\
  $\Gamma ::= \cdot \mid \Gamma, x:T^0 \mid x: T^{-1}$
\end{minipage}

\vspace{3mm}

    {\textbf{\normalsize {Types}}}\\
    \begin{minipage}[t]{0.5\textwidth}
      \textbf{Level 0}\\
  $\begin{array}{@{}ll}
     \text{Values }  S^0, T^0 ::= & \mathbb{N}^0 \mid {(\functionType[\xi]{S^0}{T^0})}^{0} \\
     & \mid {(\continuationType[\xi]{S^0}{T^0})}^{0}\\\vspace{0.4mm}
    \text{Computations} & T^0 \, ! \, \xi \mid T^0 \, ! \, \Delta  \mid T^0 \, ! \,  \Delta;\xi \\
    & \mid (\handlerType{S^0 \, ! \, \xi_1}{T^0 \, ! \, \xi_2})^0\, !\, \Delta\\\vspace{0.4mm}
    \text{Handlers} & (\handlerType{S^0 \, ! \, \xi_1}{T^0 \, ! \, \xi_2})^0
  \end{array}$
\end{minipage}%
  \begin{minipage}[t]{0.5\textwidth}
     \textbf{Level $-$1}\\
    $\begin{array}{@{}ll}
    \text{Values } S^{-1}, T^{-1} ::= & \mathbb{N}^{-1} \mid {(\functionType{S}{T})}^{-1} \\
    & \mid {(\continuationType{S}{T})}^{-1} \mid {\textsf{Code}({T^{0} \, ! \, \xi})}^{-1} \\\vspace{0.4mm}
    \text{Computations} & T^{-1} \, ! \, \Delta \\\\\vspace{0.4mm}
    \text{Handlers} & (\handlerType{S^{-1} \, ! \, \Delta_1}{T^{-1} \, ! \, \Delta_2})^{-1}
  \end{array}$
\end{minipage}

\end{source-desc}
\caption{\sourceLang{} syntax and types (repeated)}
\label{fig:source-syntax-types-repeated}
\end{figure}

\begin{figure}
\begin{source-desc}
  \scriptsize
  {\normalsize\textbf{Typing Rules}}
  \\ \textit{Level annotations on types mostly omitted} \\
  \vspace{2mm}\\
  \fbox{$\Gamma \vdash_{\compilemode{} \mid \quotemode{}} v: \effectType{T^{0}}$}\\
  \begin{center}
  \begin{minipage}[t]{0.3\textwidth}
    \centering
    $\inferrule[(Nat)]{ \\ }{\cqtypejudge{m}{\runtimecomptype{\mathbb{N}}{\Delta}}}$
  \end{minipage}%
  \begin{minipage}[t]{0.3\textwidth}
    \centering
    $\inferrule[(Var)]{\Gamma(x) = T^0}{\cqtypejudge{x}{\runtimecomptype{T^0}{\Delta}}}$
  \end{minipage}%
  \begin{minipage}[t]{0.4\textwidth}
    \centering
$\inferrule[(Lambda)]{\cqtypejudge[\Gamma, x: S]{e}{\runtimecomptype{T}{\Delta;\xi}}}{\cqtypejudge{\lambda x.e}{\runtimecomptype{(\functionType[\xi]{S}{T})}{\Delta}}}$
\end{minipage}\\
\end{center}

\vspace{3mm}

\fbox{$\Gamma \vdash_{\compilemode{} \mid \quotemode{}} e: \effectType[\Delta; \xi]{T^{0}}$}\\
\begin{center}
\begin{minipage}[t]{0.25\textwidth}
\centering
$\inferrule[(App)]{\cqtypejudge[\Gamma]{v_1}{\runtimecomptype{(\functionType[\xi]{S}{T})}{\Delta}} \\\\ \cqtypejudge[\Gamma]{v_2}{\runtimecomptype{S}{\Delta}}}{\cqtypejudge{v_1 v_2}{\runtimecomptype{T}{\Delta;\xi}}}$
\end{minipage}%
\begin{minipage}[t]{0.25\textwidth}
  \centering
  $\inferrule[(Continue)]{\cqtypejudge[\Gamma]{v_1}{\runtimecomptype{(\continuationType[\xi]{S}{T})}{\Delta}} \\ \cqtypejudge[\Gamma]{v_2}{\runtimecomptype{S}{\Delta}}}{\cqtypejudge{\continue{v_1}{v_2}}{\runtimecomptype{T}{\Delta;\xi}}}$
  \end{minipage}%
  \begin{minipage}[t]{0.24\textwidth}
    \centering
    $\inferrule[(Return)]{  \\\\ \cqtypejudge[\Gamma]{v}{\runtimecomptype{T}{\Delta}}}{\cqtypejudge{\return{v}}{\runtimecomptype{T}{\Delta;\xi}}}$
  \end{minipage}%
  \begin{minipage}[t]{0.26\textwidth}
    \centering
    $\inferrule[(Do)]{\cqtypejudge[\Gamma]{e_1}{\runtimecomptype{S}{\Delta;\xi}} \\ \cqtypejudge[\Gamma, x:S]{e_2}{\runtimecomptype{T}{\Delta;\xi}}}{\cqtypejudge{\bind{x}{e_1}{e_2}}{\runtimecomptype{T}{\Delta;\xi}}}$
  \end{minipage}

    \vspace{3mm}

  \begin{minipage}[t]{0.25\textwidth}
  \centering
  $\inferrule[(Op)]{\cqtypejudge[\Gamma]{v}{\runtimecomptype{S}{\Delta}} \\ \texttt{op}: S \to T \in \Sigma \\ \texttt{op} \in \xi}{\cqtypejudge{\op{v}}{\runtimecomptype{T}{\Delta;\xi}}}$
\end{minipage}%
\begin{minipage}[t]{0.5\textwidth}
  \centering
  $\inferrule[(Handle)]{\cqtypejudge[\Gamma]{e}{\runtimecomptype{S}{\Delta;\xi_1}} \\ \cqtypejudge{h}{\handlerType{\runtimecomptype{S}{\xi_1}}{\runtimecomptype{T}{\xi_2}}\,!\,\Delta} \\ \forall \texttt{op} \in \xi_1 \setminus \xi_2 \, . \, \texttt{op} \in \textsf{dom}(h)}{\cqtypejudge{\handleWith{e}{h}}{\runtimecomptype{T}{\Delta;\xi_2}}}$
\end{minipage}%
\begin{minipage}[t]{0.25\textwidth}
  \centering
  $\inferrule[(Splice)]{ \\\\ \stypejudge[\Gamma]{e}{\textsf{Code}(T^0 \, ! \, \xi)^{-1} \, ! \, \Delta}}{\cqtypejudge{\splice}{\runtimecomptype{T^0}{\Delta ; \xi}}}$
\end{minipage}
\vspace{3mm}

\end{center}

\fbox{$\Gamma \vdash_{\compilemode{} \mid \quotemode{}} h: (\handlerType{\effectType[\xi_1]{S^0}}{\effectType[\xi_2]{T^{0}}})^{0} \, ! \, \Delta$}\\
\begin{center}
  
\begin{minipage}[t]{0.4\textwidth}
  \centering
$\inferrule[(Ret-Handler)]{  \\\\\\ \cqtypejudge[\Gamma, x: S]{e}{\runtimecomptype{T}{\Delta;\xi_2}}}{\cqtypejudge{\returnHandler{x}{e}}{(\handlerType{\runtimecomptype{S}{\xi_1}}{\runtimecomptype{T}{\xi_2}})\,!\,\Delta}}$
\end{minipage}%
\begin{minipage}[t]{0.6\textwidth}
  \centering
$\inferrule[(Op-Handler)]{\texttt{op}: A \to B \in \Sigma 
\\ \cqtypejudge{h}{\handlerType{\runtimecomptype{S}{\xi}}{\runtimecomptype{T}{\xi_2}}\,!\,\Delta} \\ \cqtypejudge[\Gamma, x: A, k: {\continuationType[\xi_2]{B}{T}} ]{e}{\runtimecomptype{T}{\Delta;\xi_2}} \\ \xi_1 \subseteq \xi_2 \cup \{ \texttt{op} \} \\ \opHandler{x'}{k'}{e'} \notin h} {\cqtypejudge{h;\opHandler{x}{k}{e}}{(\handlerType{\runtimecomptype{S}{\xi_1}}{\runtimecomptype{T}{\xi_2}})\,!\,\Delta}}$
\end{minipage}
\end{center}

\vspace{3mm}

    \fbox{$\Gamma \vdash_{\splicemode{}} v: {T^{-1}}$}\\
    \begin{center} 
    \begin{minipage}[t]{0.2\textwidth}
      \centering
      $\inferrule[(\splicemode{}-Nat)]
      { \\ }
      {\stypejudge{m}{\compiletimetype{\mathbb{N}}}}$
      \end{minipage}%
  \begin{minipage}[t]{0.2\textwidth}
    \centering
  $\inferrule[(\splicemode{}-Var)]
  {\Gamma(x) = \compiletimetype{T^{-1}}}
  {\stypejudge{x}{\compiletimetype{T^{-1}}}}$
  \end{minipage}%
  \begin{minipage}[t]{0.3\textwidth}
    \centering
  $\inferrule[(\splicemode{}-Lambda)]
  {\stypejudge[\Gamma, x:\compiletimetype{S}]{e}{\effectType{\compiletimetype{T}}}}
  {\stypejudge{\function{x}{e}}{\compiletimetype{(\functionType{\compiletimetype{S}}{\compiletimetype{T}})}}}$
  \end{minipage}%
  \begin{minipage}[t]{0.3\textwidth}
  \centering
$\inferrule[(\splicemode{}-Continuation)]
  {\stypejudge[\Gamma, x:\compiletimetype{S}]{e}{\effectType{\compiletimetype{T}}}}
  {\stypejudge{\continuation{x}{e}}{\compiletimetype{(\continuationType{\compiletimetype{S}}{\compiletimetype{T}})}}}$
\end{minipage}
  
  \vspace{3mm}
\end{center}

\fbox{$\Gamma \vdash_{\splicemode{}} e: \effectType{T^{-1}}$}\\
\begin{center}
    
  \begin{minipage}[t]{0.25\textwidth}
    \centering
  $\inferrule[(\splicemode{}-App)]
    {\stypejudge{v_1}{\compiletimetype{(\functionType{\compiletimetype{S}}{\compiletimetype{T}})}} \\ \stypejudge{v_2}{\compiletimetype{S}}}
    {\stypejudge{v_1 \, v_2}{\effectType{\compiletimetype{T}}}}$
  \end{minipage}%
  \begin{minipage}[t]{0.25\textwidth}
    \centering
  $\inferrule[(\splicemode{}-Continue)]
    {\stypejudge{v_1}{\compiletimetype{(\continuationType{\compiletimetype{S}}{\compiletimetype{T}})}} \\ \stypejudge{v_2}{\compiletimetype{S}}}
    {\stypejudge{\continue{v_1}{v_2}}{\effectType{\compiletimetype{T}}}}$
  \end{minipage}%
  \begin{minipage}[t]{0.25\textwidth}
    \centering
  $\inferrule[(\splicemode{}-Return)]
    { \\\\ \stypejudge{v}{\compiletimetype{T}}}
    {\stypejudge{\return{v}}{\effectType{\compiletimetype{T}}}}$
  \end{minipage}%
  \begin{minipage}[t]{0.25\textwidth}
    \centering
  $\inferrule[(\splicemode{}-Do)]
    {\stypejudge{e_1}{\effectType{\compiletimetype{S}}} \\ \stypejudge[\Gamma, x: S]{e_2}{\effectType{\compiletimetype{T}}}}
    {\stypejudge{\bind{x}{e_1}{e_2}}{\effectType{\compiletimetype{T}}}}$
  \end{minipage}
  
  \vspace{3mm}
 
   \begin{minipage}[t]{0.25\textwidth}
    \centering
  $\inferrule[(\splicemode{}-Op)]
    {\stypejudge{v}{\compiletimetype{S}} \\\\ \texttt{op}: \compiletimetype{S} \rightarrow \compiletimetype{T} \in \Sigma \\ \texttt{op} \in \Delta}
    {\stypejudge{\op{v}}{\effectType{\compiletimetype{T}}}}$
  \end{minipage}%
  \begin{minipage}[t]{0.5\textwidth}
    \centering
  $\inferrule[(\splicemode{}-Handle)]
    {\stypejudge{e}{\effectType[\Delta_1]{\compiletimetype{S}}} \\ \stypejudge{h}{\compiletimetype{(\handlerType{\effectType[\Delta_1]{\compiletimetype{S}}}{\effectType[\Delta_2]{\compiletimetype{T}}})}} \\\\ \forall \textsf{op} \in \Delta_1 \setminus \Delta_2. \, \textsf{op} \in \textsf{dom}(h)}
    {\stypejudge{\handleWith{e}{h}}{\effectType[\Delta_2]{\compiletimetype{T}}}}$
  \end{minipage}%
  \begin{minipage}[t]{0.25\textwidth}
    \centering
    $\inferrule[(\splicemode{}-Quote)]{  \\\\ \qtypejudge{e}{\runtimecomptype{T^0}{\Delta ; \xi}}}{\stypejudge{\equote}{\effectType{\compiletimetype{\textsf{Code}(\runtimecomptype{T^0}{\xi})}^{-1}}}}$
  \end{minipage}\\
  \vspace{3mm}

\end{center}
\fbox{$\Gamma \vdash_{\splicemode{}} h: (\handlerType{\effectType[\Delta_1]{S^{-1}}}{\effectType[\Delta_2]{T^{-1}}})^{-1}$}\\ 
\begin{center}
  \begin{minipage}[t]{0.4\textwidth}
    \centering
  $\inferrule[(\splicemode{}-Ret-Handler)]
    { \\\\\\ \stypejudge[\Gamma, x:\compiletimetype{S}]{e}{\effectType[\Delta_2]{\compiletimetype{T}}}}
    {\stypejudge{\returnHandler{x}{e}}{\compiletimetype{\handlerType{\effectType[\Delta_1]{\compiletimetype{S}}}{\effectType[\Delta_2]{\compiletimetype{T}}}}}}$
  \end{minipage}%
  \begin{minipage}[t]{0.6\textwidth}
    \centering
  $\inferrule[(\splicemode{}-Op-Handler)]
    { \texttt{op}: {A} \to B \in \Sigma \\ 
      \stypejudge{h}{\compiletimetype{\handlerType{\effectType[\Delta_1]{\compiletimetype{S}}}{\effectType[\Delta_2]{\compiletimetype{T}}}}}\\
      \stypejudge[\Gamma, x:\compiletimetype{A}, k:{\compiletimetype{(\continuationType[\Delta_2]{\compiletimetype{B}}{\compiletimetype{T}})}} ]{e}{\effectType[\Delta_2]{\compiletimetype{T}}}\\
      \Delta_1 \subseteq \Delta_2 \cup \{ \texttt{op} \} \\
             \opHandler{x'}{k'}{e'} \notin h}
    {\stypejudge{h ; \opHandler{x}{k}{e}}{\compiletimetype{\handlerType{\effectType[\Delta_1]{\compiletimetype{S}}}{\effectType[\Delta_2]{\compiletimetype{T}}}}}}$
  \end{minipage}

\end{center}
  \end{source-desc}
\caption{The typing rules for \sourceLang{}}%
\label{fig:source-s-typing-rules}
\end{figure}

For convenience, \Cref{fig:source-syntax-types-repeated} repeats the syntax and types for \sourceLang{} presented in \Cref{fig:source-syntax-types}.
\Cref{fig:source-s-typing-rules} presents the full typing rules for \sourceLang{}.
Selected rules were presented in \Cref{fig:source-cq-typing-rules} (\Cref{subsection:sourcelang-type-system}).

\section{Full \coreLang{} rules}\label[appendix]{appendix:full-corelang-rules}
\begin{figure}
\begin{core-desc}
  \footnotesize
  {\normalsize\textbf{Syntax}} \\
  $\begin{array}{@{}llll}
    \textbf{Formal Params} & \alpha_R \\
    \textbf{Normal Forms} & n & ::= & x \mid m \in \mathbb{N} \mid \lambda x. t \mid \kappa x. t \mid  \Nat{m} \mid \alpha_R \mid \Var{\alpha}{R} \mid \Lam{n_1}{n_2} \mid \App{n_1}{n_2} \\
    &&& \mid \Continue{n_1}{n_2} \mid \Ret{n} \mid \Do{n_1}{n_2}{n_3} \mid \Op{n} \mid \Hwith{n_1}{n_2}   \\
  &&&\mid \Hret{n_1}{n_2}(n_1, n_2) \mid \Hop{n_1}{n_2}{n_3}{n_4} \\
  \textbf{Terms} & t & := & n_1\;n_2 \mid \return{n}  \mid \bind{x}{t_1}{t_2} \mid \op{t} \mid \handleWith{n}{h} \mid \continue{n_1}{n_2} \\
  &&& \mid \checkfv{n} \mid \checkm{n} \mid \gensym{R} \mid \dlet{n}{t} \mid \tls{t} \mid \err \\
  \textbf{Handlers} & h & := & \returnHandler{x}{t} \mid \opHandler{x}{k}{t} \\
  \end{array}$

\vspace{3mm}

{{\textbf{\normalsize {Typing contexts}}}\\
  $\Gamma ::= \cdot \mid \Gamma, x:T$}

\vspace{3mm}

    {{\normalsize\textbf{Types}}}\\
    \begin{minipage}[t]{0.4\textwidth}
    \textbf{Run-time Pre-types}\\
    $\begin{array}{@{}lllr}
    \text{Effects set } \xi ::= & \emptyset \mid \xi \cup \{ \texttt{op}_i \} \\
    \text{Value type } Q,R ::= & \mathbb{N} \mid  \functionType[\xi]{Q}{R}  \\
    & \mid \continuationType[\xi]{Q}{R} \\
    \text{Computation type} & \effectType[\xi]{R} \\
    \text{Handler type} & \handlerType{\effectType[\xi_1]{Q}}{\effectType[\xi_2]{R}}
    \end{array}$
  \end{minipage}%
  \begin{minipage}[t]{0.6\textwidth}
    \textbf{Types}\\
  $\begin{array}{@{}lllr}
    \text{Effects set } & \Delta ::= \emptyset \mid \Delta \cup \{ \texttt{op}_i \}  \\
    \text{Value type }  S, T ::= & \ldots \mid \textsf{FParam}(R) \mid \textsf{AST}(R) \\
    & \mid \textsf{AST}(\effectType[\xi]{R})  \mid \textsf{AST}(\handlerType{\effectType[\xi_1]{Q}}{\effectType[\xi_2]{R}})\\
    \text{Computation type} & \effectType{T} \\
    \text{Handler type} & \handlerType{\effectType{S}}{\effectType[\Delta_2]{T}}
  \end{array}$
\end{minipage}
\end{core-desc}
\caption{\coreLang{}: syntax and types (repeated)}
\label{fig:core-syntax-repeated}
\end{figure}

\begin{figure}
\begin{core-desc}
{\normalsize \textbf{Operational Semantics}}\\
  {
    \scriptsize

    {\textbf{Auxiliary Definitions}}\\
  {\[\begin{array}{lrcl}
    \text{Evaluation Frame } & F & ::= & \bind{x}{[-]\,}{t_2} \mid \handleWith{[-]}{h} \mid \dlet{\Binder{\alpha}{R}}{[-]} \mid \tls{[-]} \\
    \text{Evaluation Context } & E & ::= & [-] \mid E[F] \\ \vspace{1mm} \\
    \text{Domain of Handler} & \textsf{dom}(h) & \triangleq & \textsf{dom}(\returnHandler{x}{t}) = \emptyset, \\
    &&&\textsf{dom}(h;\opHandler{x}{k}{t}) = \textsf{dom}(h) \cup \{ \textbf{\textsf{op}} \} \\     \vspace{1mm}
    \text{Handled Effects} & \textsf{handled}(E) & \triangleq & \textsf{handled}([-]) = \emptyset, \\
    &&& \textsf{handled}(E[\bind{x}{[-]\,}{t_2}]) = \textsf{handled}(E), \\
    &&& \textsf{handled}(E[\handleWith{[-]}{h}]) = \textsf{handled}(E) \cup \textsf{dom}(h), \\
    &&& \textsf{handled}(E[\dlet{\Binder{\alpha}{R}}{[-]}]) = \textsf{handled}(E), \\
    &&& \textsf{handled}(E[\tls{[-]}]) = \textsf{handled}(E)
  \end{array}
  \]}

 {\textbf{Reduction Rules}}\\
 \textit{Mechanisms related to muting and unmuting are \textbf{\textcolor{coreHighlight}{highlighted}}}\\[2mm]
 $\begin{array}{@{}lcl}
    \reductionRule{App}\\ 
  \transition{\coreConfiguration{(\function{x}{t})n}{E}{U}{M}{I}}{\coreConfiguration{t[n/x]}{E}{U}{M}{I}}\\
  \reductionRule{Seq}\\ 
  \transition{\coreConfiguration{\bind{x}{\return{n}}{t}}{E}{U}{M}{I}}{\coreConfiguration{t[n/x]}{E}{U}{M}{I}}\\
\reductionRule{Hdl}\\
\transition{\coreConfiguration{\handleWith{\return{n}}{h}}{E}{U}{M}{I}}{\coreConfiguration{t[n/x]}{E}{U}{M}{I}}
  \end{array}$\\
  \textit{\textcolor{comment}{where $\returnHandler{x}{t} \in h$}}\\[1mm]
  $\begin{array}{@{}lcl}
  \congruenceRule{Psh} \\
  \transition{\coreConfiguration{F[t]}{E}{U}{M}{I}}{\coreConfiguration{t}{E[F]}{U}{M}{I}} \\
  \congruenceRule{Pop} \\
  \transition{\coreConfiguration{\return{n}}{E[F]}{U}{M}{I}}{\coreConfiguration{F[\return{n}]}{E}{U}{M}{I}}
  \end{array}$\\[1mm]
\begin{minipage}[t]{0.5\textwidth}
  $\begin{array}{@{}lcl}
    \astRule{Gen}\\ 
  \transition{\coreConfiguration{\gensym{R}}{E}{U}{M}{I}}{\coreConfiguration{\return{\Binder{\alpha}{R}}}{E}{U\cup\{\alpha\}}{M}{I}}
  \end{array}$\\
  \textit{\textcolor{comment}{where $\alpha = \textsf{next}(U), \textsf{next}(U) \notin U \text{\textsf{next} deterministic}$}}\\[1mm]
  $\begin{array}{@{}lcl}
  \secRule{Chs}\\
   \transition{\coreConfiguration{\checkfv{n}}{E}{U}{M}{I}}{\coreConfiguration{\return{n}}{E}{U}{M}{I}}
  \end{array}$\\
  \textit{\textcolor{comment}{if $\freevars{n} \not\subseteq \projfvs{E}$}}

  $\begin{array}{@{}lcl}
  \secRule{Chf}\\
   \transition{\coreConfiguration{\checkfv{n}}{E}{U}{M}{I}}{\coreConfiguration{\err}{E}{U}{M}{I}}
  \end{array}$\\
  \textit{\textcolor{comment}{if $\freevars{n} \not\subseteq \projfvs{E}$}}
\end{minipage}%
\begin{minipage}[t]{0.5\textwidth}%
    $\begin{array}{@{}lcl}
      \secRule{Tls}\\
       \transition{\coreConfiguration{\tls{\return{n}}}{E}{U}{M}{I}}{\coreConfiguration{\return{n}}{E}{U}{\textcolor{coreHighlight}{\emptyset}}{\textcolor{coreHighlight}{\top}}}
    \end{array}$\\[\lineskip]
    \\[2.5mm]
  $\begin{array}{@{}lcl}
    \secRule{Cms}\\
   \transition{\coreConfiguration{\checkm{n}}{E}{U}{M}{I}}{\coreConfiguration{\return{n}}{E}{U}{M}{I}}\end{array}$\\
    \textit{\textcolor{comment}{if $\freevars{n} \setminus M \subseteq \projfvs{E}$}}\\
  $\begin{array}{@{}lcl}
    \secRule{Cmf}\\
   \transition{\coreConfiguration{\checkm{n}}{E}{U}{M}{I}}{\coreConfiguration{\err}{E}{U}{M}{I}}
  \end{array}$\\
  \textit{\textcolor{comment}{if $\freevars{n} \setminus M \not\subseteq \projfvs{E}$}}
\end{minipage}
\\[1mm] 
  $\begin{array}{@{}lcl}
  \secRule{Dlt} \\
  \transition{\coreConfiguration{\dlet{\Binder{\alpha}{R}}{\return{n}}}{E}{U}{M}{I}}{\coreConfiguration{\return{n}}{E}{U}{\textcolor{coreHighlight}{M'}}{\textcolor{coreHighlight}{I'}}}\end{array}$\\
\textit{\textcolor{coreHighlight}{if $\textsf{len}(E) > I$ then $M' = M, I' = I$, else $M' = \emptyset, I' = \top$}}\\[1.4mm]
$\begin{array}{@{}lcl}
   \effectRule{Op}\\
   \transition{\coreConfiguration{\op{v}}
                                                   {E_1[\handleWith{E_2}{h}]}
                                                   {U}
                                                   {M}
                                                   {I}}
                                {\coreConfiguration{c[v/x, \text{cont}/ k]}
                                                   {E_1}
                                                   {U}
                                                   {\textcolor{coreHighlight}{M \cup \projfvs{E_2}}}
                                                   {\textcolor{coreHighlight}{I'}}}\end{array}$\\
  \textit{\textcolor{comment}{where cont $=\kappa x. \, \handleWith{E_2[\return{x}]}{h}$ and $\opHandler{x}{k}{c} \in h$ and $\textbf{\textsf{op}} \notin \textsf{handled}(E_2)$ and \textcolor{coreHighlight}{$I' = \textsf{min}(\textsf{len}(E_1), I)$}}}\\[1mm]
  $\begin{array}{@{}lcl}
  \effectRule{Cnt}\\
   \transition{\coreConfiguration{\continue{(\kappa x. E_2[\return{x}])}{n}}{E_1}{U}{M}{I}}{\coreConfiguration{\return{n}}{E_1[E_2]}}{U}{M}{I}
  \end{array}
  $
}

\end{core-desc}
\caption{The operational semantics of \coreLang{}}
\label{fig:core-opsem}
\end{figure}

\begin{figure}
  \begin{core-desc}
    {\large\textbf{Typing Rules}}\\
    \vspace{1mm}\\
    \scriptsize
    \fbox{$\Gamma \vdash n : T$}\\
    \begin{center}
    \begin{minipage}[t]{0.2\textwidth}
      \centering
      $\inferrule[(Nat)]
      { \\ }
      {\type{m}{\mathbb{N}}}$
      \end{minipage}%
  \begin{minipage}[t]{0.2\textwidth}
    \centering
  $\inferrule[(Var)]
  {\Gamma(x) = T}
  {\type{x}{T}}$
  \end{minipage}%
  \begin{minipage}[t]{0.3\textwidth}
    \centering
  $\inferrule[(Lambda)]
    {\type[, x:S]{t}{\effectType{T}}}
    {\type{\function{x}{t}}{\functionType{S}{T}}}$
  \end{minipage}%
  \begin{minipage}[t]{0.3\textwidth}
  \centering
$\inferrule[(Continuation)]
  {\type[, x:S]{t}{\effectType{T}}}
  {\type{\continuation{x}{t}}{\continuationType{S}{T}}}$
\end{minipage} 

  \vspace{3mm}

  \begin{minipage}[t]{0.25\textwidth}
      \centering
    $\inferrule[(FParam)]{ \\\\ }{\type{\Binder{\alpha}{R}}{\textsf{FParam}(R)}}$
    \end{minipage}%
    \begin{minipage}[t]{0.20\textwidth}
      \centering
    $\inferrule[(Nat-AST)]{ \\\\ }{\type{\texttt{Nat}({m})}{\textsf{AST}(\mathbb{N})}}$
    \end{minipage}%
    \begin{minipage}[t]{0.20\textwidth}
      \centering
    $\inferrule[(Var-AST)]{ \\\\ \type{n}{\textsf{FParam}(R)} }{\type{\texttt{Var}({n})}{\textsf{AST}(R)}}$
    \end{minipage}%
    \begin{minipage}[t]{0.35\textwidth}
      \centering
    $\inferrule[(Lambda-AST)]{\type{n_1}{\textsf{FParam}(Q)}\\{\type{n_2}{\textsf{AST}(\effectType[\xi]{R})}}}{\type{\Lam{n_1}{n_2}}{\textsf{AST}(\functionType[\xi]{Q}{R})}}$
    \end{minipage}
    
\end{center}
  
      \vspace{3mm}

\begin{center}
  \begin{minipage}[t]{0.33\textwidth}
      \centering
    $\inferrule[(App-AST)]{\type{n_1}{\textsf{AST}(\functionType[\xi]{Q}{R})}\\{\type{n_2}{\textsf{AST}(Q)}}}{\type{\App{n_1}{n_2}}{\textsf{AST}(\effectType[\xi]{R})}}$
    \end{minipage}%
    \begin{minipage}[t]{0.33\textwidth}
      \centering
    $\inferrule[(Continue-AST)]{\type{n_1}{\textsf{AST}(\continuationType[\xi]{Q}{R})}\\{\type{n_2}{\textsf{AST}(Q)}}}{\type{\Continue{n_1}{n_2}}{\textsf{AST}(\effectType[\xi]{R})}}$
    \end{minipage}%
    \begin{minipage}[t]{0.33\textwidth}
      \centering
    $\inferrule[(Return-AST)]{ \\\\ {\type{n}{\textsf{AST}(R)}}}{\type{\Ret{n}}{\textsf{AST}(\effectType[\xi]{R})}}$
    \end{minipage}
\end{center}

\vspace{3mm}

\begin{center}
  \begin{minipage}[t]{0.33\textwidth}
      \centering
    $\inferrule[(Do-AST)]{ \\\\ {\type{n_1}{\textsf{AST}(\effectType[\xi]{Q})}} \\ \type{n_2}{\textsf{FParam}({Q})} \\{\type{n_3}{\textsf{AST}(\effectType[\xi]{R})}}}{\type{\Do{n_1}{n_2}{n_3}}{\textsf{AST}(\effectType[\xi]{R})}}$
    \end{minipage}%
  \begin{minipage}[t]{0.33\textwidth}
      \centering
    $\inferrule[(Op-AST)]{ \\\\\\ \texttt{op}: Q \rightarrow R \in \Sigma \\ \texttt{op} \in \xi \\{\type{n}{\textsf{AST}(Q)}}}{\type{\Op{n_1}}{\textsf{AST}(\effectType[\xi]{R})}}$
    \end{minipage}%
    \begin{minipage}[t]{0.33\textwidth}
      \centering
    $\inferrule[(Handle-AST)]{\type{n_1}{\textsf{AST}(\effectType[\xi_1]{Q})} \\ \type{h}{\textsf{AST}(\handlerType{\effectType[\xi_1]{Q}}{\effectType[\xi_2]{R}})} \\{\type{n}{\textsf{AST}(Q)}} \\ \forall \textsf{op} \in \xi_1 \setminus \xi_2. \, \textsf{op} \in \textsf{dom}(n_2)}{\type{\Hwith{n_1}{n_2}}{\textsf{AST}(\effectType[\xi_2]{R})}}$
    \end{minipage}
\end{center}
\vspace{3mm}
\begin{center}
  \begin{minipage}[t]{0.5\textwidth}
      \centering
    $\inferrule[(Ret-Handler-AST)]{ \\\\\\ \type{n_1}{\textsf{FParam}(Q)} \\ \type{n_2}{\textsf{AST}(\effectType[\xi_2]{R})}}{\type{\Hret{n_1}{n_2}}{\textsf{AST}(\handlerType{\effectType[\xi_1]{Q}}{\effectType[\xi_2]{R}})}}$
    \end{minipage}%
  \begin{minipage}[t]{0.5\textwidth}
      \centering
    $\inferrule[(Op-Handler-AST)]{
      \textsf{op}: A \to B \in \Sigma \\
      \type{n_1}{{\textsf{AST}(\handlerType{\effectType[\xi_1]{Q}}{\effectType[\xi_2]{R}})}}\\ 
    \type{n_2}{\textsf{FParam}(A)}\\
    \type{n_3}{\textsf{FParam}(\continuationType[\xi_2]{B}{R})}\\ \type{n_4}{\textsf{AST}(\effectType[\xi_2]{R})} \\ \xi_1 \subseteq \xi_2 \cup \{ \texttt{op} \}}{\type{\Hop{n_1}{n_2}{n_3}{n_4}}{\textsf{AST}(\handlerType{\effectType[\xi_1]{Q}}{\effectType[\xi_2]{R}})}}$
    \end{minipage}
\end{center}

  \fbox{$\Gamma \vdash t : \effectType{T}$}\\
  \begin{center}
    
  \begin{minipage}[t]{0.25\textwidth}
    \centering
  $\inferrule[(App)]
    {\type{v_1}{S \oset{\text{\tiny{$\Delta$}}}\longrightarrow T} \\ \type{v_2}{S}}
    {\type{v_1 \, v_2}{\effectType{T}}}$
  \end{minipage}%
  \begin{minipage}[t]{0.25\textwidth}
    \centering
  $\inferrule[(Continue)]
    {\type{v_1}{\continuationType{S}{T}} \\ \type{v_2}{S}}
    {\type{\continue{v_1}{v_2}}{\effectType{T}}}$
  \end{minipage}%
  \begin{minipage}[t]{0.25\textwidth}
    \centering
  $\inferrule[(Return)]
    { \\\\ \type{v}{T}}
    {\type{\return{v}}{\effectType{T}}}$
  \end{minipage}%
  \begin{minipage}[t]{0.25\textwidth}
    \centering
  $\inferrule[(Do)]
    {\type{c_1}{\effectType[\Delta]{S}} \\ \type[, x: S]{c_2}{\effectType{T}}}
    {\type{\bind{x}{c_1}{c_2}}{\effectType{T}}}$
  \end{minipage}
  
  \vspace{3mm}
  \begin{minipage}[t]{0.3\textwidth}
    \centering
  $\inferrule[(Op)]
    {  \\\\ \type{v}{S} \\ \texttt{op}: S \rightarrow T \in \Sigma \\ \texttt{op} \in \Delta}
    {\type{\op{v}}{\effectType{T}}}$
  \end{minipage}%
  \begin{minipage}[t]{0.3\textwidth}
    \centering
  $\inferrule[(Handle)]
    {\type{c}{\effectType[\Delta_1]{S}} \\ \type{h}{\handlerType{\effectType[\Delta_1]{S}}{\effectType[\Delta_2]{T}}} \\ \forall \textsf{op} \in \Delta_1 \setminus \Delta_2. \, \textsf{op} \in \textsf{dom}(h)}
    {\type{\handleWith{c}{h}}{\effectType[\Delta_2]{T}}}$
  \end{minipage}%
  \begin{minipage}[t]{0.2\textwidth}
      \centering
    $\inferrule[(Tls)]{ \\\\\\ \type{t}{\effectType{T}}}{\type{\tls{t}}{\effectType{T}}}$
    \end{minipage}%
    \begin{minipage}[t]{0.2\textwidth}
      \centering
    $\inferrule[(DLet)]{ \\\\ \type{n}{\textsf{FParam}(R)} \\ \type{t}{\effectType{T}}}{\type{\dlet{n}{t}}{\effectType{T}}}$
    \end{minipage} 
    
\vspace{3mm}
\begin{minipage}[t]{0.2\textwidth}
      \centering
    $\inferrule[(Err)]{ \\ }{\type{\err{}}{\effectType{T}}}$
    \end{minipage}%
\begin{minipage}[t]{0.3\textwidth}
      \centering
    $\inferrule[(Mkvar)]{ \\ }{\type{\gensym{R}}{\effectType{\textsf{FParam}(R)}}}$
    \end{minipage}%
    \begin{minipage}[t]{0.25\textwidth}
      \centering
    $\inferrule[(Check)]{\type{n}{T} \\ T \text{ of } \textsf{AST } \text{type} }{\type{\checkfv{n}}{\effectType{T}}}$
    \end{minipage}%
    \begin{minipage}[t]{0.25\textwidth}
      \centering
    $\inferrule[(Check-M)]{\type{n}{T} \\ T \text{ of } \textsf{AST } \text{type} }{\type{\checkm{n}}{\effectType{T}}}$
    \end{minipage}

  \vspace{3mm}

\end{center}

\fbox{$\Gamma \vdash h: \handlerType{\effectType[\Delta_1]{S}}{\effectType[\Delta_2]{T}}$}\\
\begin{center}

  \begin{minipage}[t]{0.5\textwidth}
    \centering
  $\inferrule[(Ret-Handler)]
    {  \\\\  \\\\   {\type[, x:S]{c}{\effectType[\Delta_2]{T}}}}
    {\type{\returnHandler{x}{c}}{\handlerType{\effectType[\Delta_1]{S}}{\effectType[\Delta_2]{T}}}}$
  \end{minipage}%
  \begin{minipage}[t]{0.5\textwidth}
    \centering
  $\inferrule[(Op-Handler)]
    { \texttt{op}: A \to B \in \Sigma \\ 
      \type{h}{\handlerType{\effectType[\Delta_1]{S}}{\effectType[\Delta_2]{T}}}\\
      \type[, x:A, k:{\continuationType[\Delta_2]{B}{T}} ]{c}{\effectType[\Delta_2]{T}}\\
      \Delta_1 \subseteq \Delta_2 \cup \{ \texttt{op} \} \\
             \opHandler{x'}{k'}{c'} \notin h}
    {\type{h ; \opHandler{x}{k}{c}}{\handlerType{\effectType[\Delta_1]{S}}{\effectType[\Delta_2]{T}}}}$
  \end{minipage}

\end{center}
  \end{core-desc}
  \caption{\coreLang{} typing rules}%
  \label{fig:core-lang-typing-rules-full}
\end{figure}

For convenience, \Cref{fig:core-syntax-repeated} repeats the syntax and types of \Cref{fig:core}. \Cref{fig:core-opsem,fig:core-lang-typing-rules-full} give the full
operational semantics and full typing rules for \coreLang{},
completing the subset presented in \Cref{fig:core}
(\Cref{section:core-lang}).

\section{Full elaboration rules}\label[appendix]{appendix:full-elaboration-rules}
\begin{figure}
  \begin{source-desc}
    \scriptsize
  \renewcommand{\cqmode}{\compilemode{} \mid \quotemode{}}
    {\normalsize\textbf{Term Elaboration}}\\[1mm]
    $\arraycolsep=1.4pt
    \begin{array}{@{}lll}
      \elaborate{m}_{\cqmode{}} & = & \return{\Nat{m}}\\[4pt]
      \elaborate{x}_{\cqmode{}} & = & \varToAST{x}\\[4pt]
      \elaborate{\lambda x: T^0. \, e}_{\cqmode} & = & \bind{x}{\gensym{\erase{T^0}}}{\bind{\texttt{body}}{\elaborate{e}_{\cqmode}}{\return{\Lam{x}{\texttt{body}}}}}\\[4pt]
      \elaborate{v_1 v_2}_{\cqmode} & = & \bind{f}{\elaborate{v_1}_{\cqmode}}{\bind{\texttt{a}}{\elaborate{v_2}_{\cqmode}}{\return{\App{f}{a}}}}\\[4pt]
      \elaborate{\return{v}}_{\cqmode} & = & \bind{a}{\elaborate{v}_{\cqmode}}{\return{\Ret{a}}}\\[4pt]
      \elaborate{\bind{x: T^0}{e_1}{e_2}}_{\cqmode} & = & \bind{a}{\elaborate{e_1}_{\cqmode}}{\bind{x}{\gensym{\erase{T^0}}}{\bind{b}{\elaborate{e_2}}{\return{\Do{x}{a}{b}}}}}\\[4pt]
      \elaborate{\op{v}}_{\cqmode} & = & \bind{a}{\elaborate{v}_{\cqmode}}{\return{\Op{a}}}\\[4pt]
      \elaborate{\handleWith{e}{h}}_{\cqmode} & = & \bind{a}{\elaborate{e}_{\cqmode}}{\bind{b}{\elaborate{h}_{\cqmode}}{\return{\Hwith{a}{b}}}}\\[4pt]
      \elaborate{\returnHandler{x: T^0}{e}}_{\cqmode} & = & \bind{x}{\gensym{\erase{T^0}}}{\bind{a}{\elaborate{e}_{\cqmode}}{\return{\Hret{x}{a}}}}\\[4pt]
      \elaborate{h ; \opHandler{x: T^0}{k: \continuationType[\xi]{A^0}{B^0}}{e}}_{\cqmode} & = & \bind{a}{\elaborate{h}_{\cqmode}}{\bind{x}{\gensym{\erase{T^0}}}{}}\\
      && \bind{k}{\gensym{\erase{\continuationType[\xi]{A^0}{B^0}}}}{\bind{b}{\elaborate{e}_{\cqmode}}{\return{\Hop{a}{x}{k}{b}}}}\\[4pt]
      \elaborate{\splice}_{\compilemode{}} & = & \tls{\elaborate{e}_{\compilemode{}}} \\[4pt]
      \elaborate{\splice}_{\quotemode{}} & = & \elaborate{e}_{\quotemode{}} \\[16pt]
      \elaborate{x}_{\splicemode{}} & = & x\\[4pt]
        \elaborate{\lambda x: T^0. \, e}_{\splicemode{}} & = & \lambda x. \, \elaborate{e}_{\splicemode{}}\\[4pt]
      \elaborate{v_1 v_2}_{\splicemode{}} & = & \elaborate{v_1}_{\splicemode{}} \, \elaborate{v_2}_{\splicemode{}}\\[4pt]
      \elaborate{\return{v}}_{\splicemode{}} & = & \return{\elaborate{v}_{\splicemode{}}}\\[4pt]
      \elaborate{\bind{x}{e_1}{e_2}}_{\splicemode{}} & = & \bind{x}{\elaborate{e_1}_{\splicemode{}}}{\elaborate{e_2}_{\splicemode{}}}\\[4pt]
      \elaborate{\op{v}}_{\splicemode} & = & \op{\elaborate{v}_{\splicemode}}\\[4pt]
      \elaborate{\handleWith{e}{h}}_{\splicemode} & = & \handleWith{\elaborate{e}_{\splicemode}}{\elaborate{h}_{\splicemode}}\\[4pt]
      \elaborate{\returnHandler{x}{e}}_{\splicemode} & = & \returnHandler{x}{\elaborate{e}_{\splicemode}}\\[4pt]
      \elaborate{h ; \opHandler{x}{k}{e}}_{\splicemode} & = & {\elaborate{h}_{\splicemode} ; \opHandler{x}{k}{\elaborate{e}_{\splicemode}}}\\[4pt]
      \elaborate{\equote}_{\splicemode} & = & \elaborate{e}_{\quotemode{}}
      \end{array}$
  \end{source-desc}
  \caption{Full term elaboration from \sourceLang{} to \coreLang{}}%
  \label{fig:full-term-elaboration}
\end{figure}

\Cref{fig:full-term-elaboration} presents the full term elaboration from \sourceLang{} to \coreLang{}.  Selected rules
were presented in \Cref{fig:elaboration} (\Cref{section:elaboration}).

\section{Full rules for refined environment classifiers}\label[appendix]{appendix:full-refined-env-classifiers}

\begin{figure}
\begin{rec-desc}
  \renewcommand{\cqtypejudge}[4][\gamma]{#2 \vdash^{#1}_{\compilemode{} \mid \quotemode{}} #3: #4}
  \scriptsize
  {\normalsize\textbf{\compilemode{}$\mid$ \quotemode{}-Typing Rules}}
  \\ \textit{Level annotations on types mostly omitted} \\
  \vspace{2mm}\\
  \fbox{$\Gamma \vdash^{\gamma}_{\compilemode{} \mid \quotemode{}} v: \effectType{T^{0}}$}\\
  \begin{center}
  \begin{minipage}[t]{0.2\textwidth}
    \centering
    $\inferrule[(Nat)]{ \\ }{\cqtypejudge{\Gamma}{m}{\runtimecomptype{\mathbb{N}}{\Delta}}}$
  \end{minipage}%
  \begin{minipage}[t]{0.2\textwidth}
    \centering
    $\inferrule[(Var)]{(x: T^0)^\gamma \in \Gamma}{\cqtypejudge{\Gamma}{x}{\runtimecomptype{T^0}{\Delta}}}$
  \end{minipage}%
  \begin{minipage}[t]{0.3\textwidth}
    \centering
$\inferrule[(Lambda)]{\cqtypejudge[\gamma']{\Gamma, \gamma', \gamma \sqsubseteq \gamma', (x: S)^{\gamma'}}{e}{\runtimecomptype{T}{\Delta;\xi}}}{\cqtypejudge[\gamma]{\Gamma}{\lambda x.e}{\runtimecomptype{(\functionType[\xi]{S}{T})}{\Delta}}}$
\end{minipage}%
\begin{minipage}[t]{0.3\textwidth}
    \centering
$\inferrule[(Sub-Val)]{\cqtypejudge[\gamma']{\Gamma}{v}{\runtimecomptype{T}{\Delta}} \\ \Gamma \vDash \gamma' \sqsubseteq \gamma }{\cqtypejudge{\Gamma}{v}{{\runtimecomptype{T}{\Delta}}}}$
\end{minipage}\\
\end{center}

\vspace{3mm}

\fbox{$\Gamma \vdash^{\gamma}_{\compilemode{} \mid \quotemode{}} e: \effectType[\Delta; \xi]{T^{0}}$}\\
\begin{center}
\begin{minipage}[t]{0.37\textwidth}
\centering
$\inferrule[(App)]{\cqtypejudge{\Gamma}{v_1}{\runtimecomptype{(\functionType[\xi]{S}{T})}{\Delta}} \\ \cqtypejudge{\Gamma}{v_2}{\runtimecomptype{S}{\Delta}}}{\cqtypejudge{\Gamma}{v_1 v_2}{\runtimecomptype{T}{\Delta;\xi}}}$
\end{minipage}%
\begin{minipage}[t]{0.4\textwidth}
  \centering
  $\inferrule[(Continue)]{\cqtypejudge{\Gamma}{v_1}{\runtimecomptype{(\continuationType[\xi]{S}{T})}{\Delta}} \\ \cqtypejudge{\Gamma}{v_2}{\runtimecomptype{S}{\Delta}}}{\cqtypejudge{\Gamma}{\continue{v_1}{v_2}}{\runtimecomptype{T}{\Delta;\xi}}}$
  \end{minipage}%
  \begin{minipage}[t]{0.23\textwidth}
    \centering
    $\inferrule[(Return)]{\cqtypejudge{\Gamma}{v}{\runtimecomptype{T}{\Delta}}}{\cqtypejudge{\Gamma}{\return{v}}{\runtimecomptype{T}{\Delta;\xi}}}$
  \end{minipage}

      \vspace{3mm}

  \begin{minipage}[t]{0.55\textwidth}
    \centering
    $\inferrule[(Do)]{\cqtypejudge{\Gamma}{e_1}{\runtimecomptype{S}{\Delta;\xi}} \\ \cqtypejudge[\gamma']{\Gamma, \gamma', \gamma \sqsubseteq \gamma', (x:S)^{\gamma'}}{e_2}{\runtimecomptype{T}{\Delta;\xi}}}{\cqtypejudge{\Gamma}{\bind{x}{e_1}{e_2}}{\runtimecomptype{T}{\Delta;\xi}}}$
  \end{minipage}%
  \begin{minipage}[t]{0.4\textwidth}
  \centering
  $\inferrule[(Op)]{\cqtypejudge{\Gamma}{v}{\runtimecomptype{S}{\Delta}} \\ \texttt{op}: S \to T \in \Sigma \\ \texttt{op} \in \xi}{\cqtypejudge{\Gamma}{\op{v}}{\runtimecomptype{T}{\Delta;\xi}}}$
\end{minipage}

\vspace{3mm}

\begin{minipage}[t]{0.7\textwidth}
  \centering
  $\inferrule[(Handle)]{\cqtypejudge{\Gamma}{e}{\runtimecomptype{S}{\Delta;\xi_1}} \\ \cqtypejudge{\Gamma}{h}{\handlerType{\runtimecomptype{S}{\xi_1}}{\runtimecomptype{T}{\xi_2}}\,!\,\Delta} \\ \forall \texttt{op} \in \xi_1 \setminus \xi_2 \, . \, \texttt{op} \in \textsf{dom}(h)}{\cqtypejudge{\Gamma}{\handleWith{e}{h}}{\runtimecomptype{T}{\Delta;\xi_2}}}$
\end{minipage}

\vspace{3mm}

\begin{minipage}[t]{0.3\textwidth}
  \centering
  $\inferrule[(Splice)]{\stypejudge[\Gamma]{e}{(\textsf{Code}(T^0 \, ! \, \xi)^{\gamma})^{-1} \, ! \, \Delta}}{\cqtypejudge{\Gamma}{\splice}{\runtimecomptype{T^0}{\Delta ; \xi}}}$
\end{minipage}%
\begin{minipage}[t]{0.3\textwidth}
    \centering
$\inferrule[(Sub-Expr)]{\cqtypejudge[\gamma']{\Gamma}{e}{\runtimecomptype{T}{\Delta ; \xi}} \\ \Gamma \vDash \gamma' \sqsubseteq \gamma }{\cqtypejudge{\Gamma}{e}{{\runtimecomptype{T}{\Delta  ; \xi}}}}$
\end{minipage}\\

\end{center}

\vspace{3mm}

\fbox{$\Gamma \vdash_{\compilemode{} \mid \quotemode{}} h: (\handlerType{\effectType[\xi_1]{S^0}}{\effectType[\xi_2]{T^{0}}})^{0} \, ! \, \Delta$}\\
\begin{center}
  
\begin{minipage}[t]{0.4\textwidth}
  \centering
$\inferrule[(Ret-Handler)]{  \\\\\\ \cqtypejudge[\gamma']{\Gamma, \gamma', \gamma \sqsubseteq \gamma', (x: S)^{\gamma'}}{e}{\runtimecomptype{T}{\Delta;\xi_2}}}{\cqtypejudge{\Gamma}{\returnHandler{x}{e}}{(\handlerType{\runtimecomptype{S}{\xi_1}}{\runtimecomptype{T}{\xi_2}})\,!\,\Delta}}$
\end{minipage}%
\begin{minipage}[t]{0.55\textwidth}
  \centering
$\inferrule[(Op-Handler)]{\texttt{op}: A \to B \in \Sigma 
\\ \cqtypejudge{\Gamma}{h}{\handlerType{\Gamma}{\runtimecomptype{S}{\xi}}{\runtimecomptype{T}{\xi_2}}\,!\,\Delta} \\ \cqtypejudge[\gamma']{\Gamma, \gamma', \gamma \sqsubseteq \gamma', (x: A)^{\gamma'}, (k: {\continuationType[\xi_2]{B}{T}})^{\gamma'}}{e}{\runtimecomptype{T}{\Delta;\xi_2}} \\ \xi_1 \subseteq \xi_2 \cup \{ \texttt{op} \} \\ \opHandler{x'}{k'}{e'} \notin h} {\cqtypejudge{\Gamma}{h;\opHandler{x}{k}{e}}{(\handlerType{\runtimecomptype{S}{\xi_1}}{\runtimecomptype{T}{\xi_2}})\,!\,\Delta}}$
\end{minipage}

\vspace{3mm}

\begin{minipage}[t]{\textwidth}
    \centering
$\inferrule[(Sub-Hdlr)]{\cqtypejudge[\gamma']{\Gamma}{h}{\handlerType{\effectType[\xi_1]{S}}{\effectType[\xi_2]{T}}} \\ \Gamma \vDash \gamma' \sqsubseteq \gamma }{\cqtypejudge{\Gamma}{h}{\handlerType{\effectType[\xi_1]{S}}{\effectType[\xi_2]{T}}} }$
\end{minipage}\\

\end{center}
  \end{rec-desc}
\caption{The \compilemode{}$\mid$\quotemode{}-typing rules for \recLang{}}%
\label{fig:rec-typing-rules-full-cqmode}
\end{figure}

\begin{figure}
  \begin{rec-desc}
      \scriptsize
        {\normalsize\textbf{\splicemode{}-Typing Rules}}
  \\ \textit{Level annotations on types mostly omitted} \\
  \vspace{2mm}\\

     \fbox{$\Gamma \vdash_{\splicemode{}} v: {T^{-1}}$}\\
    \begin{center} 
    \begin{minipage}[t]{0.2\textwidth}
      \centering
      $\inferrule[(\splicemode{}-Nat)]
      { \\ }
      {\stypejudge{m}{\compiletimetype{\mathbb{N}}}}$
      \end{minipage}%
  \begin{minipage}[t]{0.2\textwidth}
    \centering
  $\inferrule[(\splicemode{}-Var)]
  {\Gamma(x) = \compiletimetype{T^{-1}}}
  {\stypejudge{x}{\compiletimetype{T^{-1}}}}$
  \end{minipage}%
  \begin{minipage}[t]{0.3\textwidth}
    \centering
  $\inferrule[(\splicemode{}-Lambda)]
  {\stypejudge[\Gamma, x:\compiletimetype{S}]{e}{\effectType{\compiletimetype{T}}}}
  {\stypejudge{\function{x}{e}}{\compiletimetype{(\functionType{\compiletimetype{S}}{\compiletimetype{T}})}}}$
  \end{minipage}%
  \begin{minipage}[t]{0.3\textwidth}
  \centering
$\inferrule[(\splicemode{}-Continuation)]
  {\stypejudge[\Gamma, x:\compiletimetype{S}]{e}{\effectType{\compiletimetype{T}}}}
  {\stypejudge{\continuation{x}{e}}{\compiletimetype{(\continuationType{\compiletimetype{S}}{\compiletimetype{T}})}}}$
\end{minipage}
  
  \vspace{3mm}
\end{center}

\fbox{$\Gamma \vdash_{\splicemode{}} e: \effectType{T^{-1}}$}\\
\begin{center}
    
  \begin{minipage}[t]{0.33\textwidth}
    \centering
  $\inferrule[(\splicemode{}-App)]
    {\stypejudge{v_1}{\compiletimetype{(\functionType{\compiletimetype{S}}{\compiletimetype{T}})}} \\ \stypejudge{v_2}{\compiletimetype{S}}}
    {\stypejudge{v_1 \, v_2}{\effectType{\compiletimetype{T}}}}$
  \end{minipage}%
  \begin{minipage}[t]{0.67\textwidth}
    \centering
  $\inferrule[(\splicemode{}-Continue)]
    {\stypejudge{v_1}{\compiletimetype{\continuationType{\compiletimetype{\textsf{Code}(\effectType[\xi_1]{S})^{\gamma}}}{\compiletimetype{\textsf{Code}(\effectType[\xi_2]{T})^{\gamma'}}}}} \\ \stypejudge{v_2}{\compiletimetype{\textsf{Code}(\effectType[\xi_1]{S})^{\gamma}}}}
    {\stypejudge{\continue{v_1}{v_2}}{\effectType{\compiletimetype{\textsf{Code}(\effectType[\xi_2]{T})^{\gamma'}}}}}$
  \end{minipage}

  \vspace{3mm}

  \begin{minipage}[t]{0.23\textwidth}
    \centering
  $\inferrule[(\splicemode{}-Return)]
    {\stypejudge{v}{\compiletimetype{T}}}
    {\stypejudge{\return{v}}{\effectType{\compiletimetype{T}}}}$
  \end{minipage}%
  \begin{minipage}[t]{0.32\textwidth}
    \centering
  $\inferrule[(\splicemode{}-Do)]
    {\stypejudge{e_1}{\effectType{\compiletimetype{S}}} \\ \stypejudge[\Gamma, x: S]{e_2}{\effectType{\compiletimetype{T}}}}
    {\stypejudge{\bind{x}{e_1}{e_2}}{\effectType{\compiletimetype{T}}}}$
  \end{minipage}%
  \begin{minipage}[t]{0.45\textwidth}
    \centering
  $\inferrule[(\splicemode{}-Op)]
    {\strut \stypejudge{v}{\compiletimetype{S}} \\ \texttt{op} \in \Delta  \\ \texttt{op}: \compiletimetype{S} \rightarrow \compiletimetype{\textsf{Code}(\effectType[\xi]{T})^{\gamma}} \in \Sigma }
    {\stypejudge{\op{v}}{\effectType{\compiletimetype{\textsf{Code}(\effectType[\xi]{T})^{\gamma}}}}}$
  \end{minipage}

  \vspace{3mm}

  \begin{minipage}[t]{\textwidth}
    \centering
  $\inferrule[(\splicemode{}-Handle)]
    {\stypejudge{e}{\effectType{\compiletimetype{\textsf{Code}(\effectType[\xi_1]{S})^{\gamma}}}} \\ \stypejudge{h}{\compiletimetype{\handlerType{\effectType[\Delta_1]{\compiletimetype{(\textsf{Code}(\effectType[\xi_1]{S})^{\gamma})}}}{\effectType[\Delta_2]{\compiletimetype{(\textsf{Code}(\effectType[\xi_2]{T})^{\gamma})}}}}} \\ \forall \textsf{op} \in \Delta_1 \setminus \Delta_2. \, \textsf{op} \in \textsf{dom}(h)}
    {\stypejudge{\handleWith{e}{h}}{\effectType[\Delta_2]{\compiletimetype{\textsf{Code}(\effectType[\xi_2]{T})^{\gamma}}}}}$
  \end{minipage}

  \vspace{3mm}

  \begin{minipage}[t]{0.25\textwidth}
    \centering
    $\inferrule[(\splicemode{}-Quote)]{ \rqtypejudge{\gamma}{e}{\runtimecomptype{T}{\Delta ; \xi}}}{\stypejudge[\Gamma]{\equote}{\textsf{Code}(T \, ! \, \xi)^{\gamma} \, ! \, \Delta}}$
  \end{minipage}%
  \begin{minipage}[t]{0.4\textwidth}
    \centering
    $\inferrule[(\splicemode{}-Sub)]{\Gamma \vDash \gamma' \sqsubseteq \gamma \\ \stypejudge[\Gamma]{e}{\textsf{Code}(T \, ! \, \xi)^{\gamma'} \, ! \, \Delta}}{\stypejudge[\Gamma]{e}{\textsf{Code}(T \, ! \, \xi)^{\gamma} \, ! \, \Delta}}$
  \end{minipage}\\
  \vspace{3mm}

\end{center}
\fbox{$\Gamma \vdash_{\splicemode{}} h: (\handlerType{\effectType[\Delta_1]{S^{-1}}}{\effectType[\Delta_2]{T^{-1}}})^{-1}$}\\ 
\begin{center}
  \begin{minipage}[t]{\textwidth}
    \centering
  $\inferrule[(\splicemode{}-Ret-Handler)]
    { \\\\\\ \stypejudge[\Gamma, x: \textsf{Code}{(\effectType[\xi_1]{S})}^{\gamma}]{e}{\effectType[\Delta_2]{\textsf{Code}(\effectType[\xi_2]{T})^{\gamma}}}}
    {\stypejudge{\returnHandler{x}{e}}{\handlerType{\effectType[\Delta_1]{\compiletimetype{(\textsf{Code}(\effectType[\xi_1]{S})^{\gamma})}}}{\effectType[\Delta_2]{\compiletimetype{(\textsf{Code}(\effectType[\xi_2]{T})^{\gamma})}}}}}$
  \end{minipage}

  \vspace{3mm}

  \begin{minipage}[t]{\textwidth}
    \centering
  $\inferrule[(\splicemode{}-Op-Handler)]
    { \texttt{op}: {A} \to \textsf{Code}(\effectType[\xi]{B})^{\gamma} \in \Sigma \\ 
      \stypejudge{h}{\handlerType{\effectType[\Delta_1]{\compiletimetype{(\textsf{Code}(\effectType[\xi_1]{S})^{\gamma})}}}{\effectType[\Delta_2]{(\compiletimetype{\textsf{Code}(\effectType[\xi_2]{T})^{\gamma}})}}}\\
      \stypejudge[\Gamma, x:\compiletimetype{A}, k:{\compiletimetype{(\continuationType[\Delta_2]{\textsf{Code}(\effectType[\xi]{B})^{\gamma}}{{\textsf{Code}(\effectType[\xi_2]{T})^{\gamma}}})}} ]{e}{\effectType[\Delta_2]{\compiletimetype{(\textsf{Code}(\effectType[\xi_2]{T})^{\gamma})}}}\\
      \Delta_1 \subseteq \Delta_2 \cup \{ \texttt{op} \} \\
             \opHandler{x'}{k'}{e'} \notin h}
    {\stypejudge{h ; \opHandler{x}{k}{e}}{\compiletimetype{\handlerType{\effectType[\Delta_1]{\compiletimetype{(\textsf{Code}(\effectType[\xi_1]{S})^{\gamma})}}}{\effectType[\Delta_2]{\compiletimetype{(\textsf{Code}(\effectType[\xi_2]{T})^{\gamma})}}}}}}$
  \end{minipage}
\end{center}
\end{rec-desc}
\caption{The \splicemode{}-typing rules for \recLang{}}%
\label{fig:rec-typing-rules-full-smode}
\end{figure}

\Cref{fig:rec-typing-rules-full-cqmode,fig:rec-typing-rules-full-smode} present the \compilemode{}$\mid$\quotemode{}-typing rules and the  \splicemode{}-typing rules for \recLang{}.
Selected rules were presented in \Cref{fig:refined-env-classifiers-types} (\Cref{section:refined-environment-classifier-calculus}).

\section{The \textsf{erase} function}\label[appendix]{appendix:auxiliary-erase}
The \textsf{erase} function takes a level $0$ type and erases all level annotations, and elaborates effect rows. It is defined by straightforward induction on \sourceLang{} types.

  \[\begin{array}{rclr}
      \erase{\mathbb{N}^0} & = & \mathbb{N}\\
      \erase{(\functionType[\xi]{S^{0}}{T^{0}})^0} & = & \functionType[\elaborate{\xi}]{\erase{S^{0}}}{\erase{T^{0}}} \\
      \erase{(\continuationType[\xi]{S^{0}}{T^{0}})^0} & = & \continuationType[\elaborate{\xi}]{\erase{S^{0}}}{\erase{T^{0}}} \\[4mm]
      \erase{\effectType[\xi]{T^0}} & = & \effectType[\elaborate{\xi}]{\erase{T^0}}\\
      \erase{\effectType{T^0}} & = & \effectType[\elaborate{\Delta}]{\erase{T^0}}\\\
      \erase{\effectType[\Delta; \xi]{T^0}} & = & \effectType[\elaborate{\Delta}; \elaborate{\xi}]{\erase{T^0}}\\
      \erase{\effectType{(\handlerType{\effectType[\xi_1]{S^{0}}}{\effectType[\xi_2]{T^{0}}})^0}} & = & \effectType[\elaborate{\Delta}]{(\handlerType{\effectType[\elaborate{\xi_1}]{\erase{S^{0}}}}{\effectType[\elaborate{\xi_2}]{\erase{T^{0}}}})}\\[4mm]
      \erase{(\handlerType{\effectType[\xi_1]{S^{0}}}{\effectType[\xi_2]{T^{0}}})^0} & = & {\handlerType{\effectType[\elaborate{\xi_1}]{\erase{S^{0}}}}{\effectType[\elaborate{\xi_2}]{\erase{T^{0}}}}}

  \end{array}
  \]
  Notice that erasing the level annotations on level-0 types that are values at compile-time (no compile-time effects set $\Delta$) produces a run-time pre-type. 

\clearpage
\newpage

\section{The implementation of the C4C check in Macocaml}\label[appendix]{appendix:implementation}
\Cref{code:c4c-implementation} lists the implementation of the C4C check in the MacoCaml compiler.

\begin{enumerate}
\item The $\checkfv{}$ primitive corresponds to the \lstinline{check} function on line 6. The $\checkm{}$ primitive relies on the \lstinline{Mute} effect handled on lines 15--17 and performed in the case on lines 18--21, where effects besides \lstinline{Mute} and \lstinline{FreeVar} are handled.
\item The \textbf{\texttt{dlet}} primitive corresponds to lines 11-14. Specifically, it computes the free variables (line 19). Since \lstinline{check} is a no-op if the set of free variables is empty, lines 13--14 either: 
  \begin{enumerate}
    \item resumes the continuation if the set of free variables is empty
    \item performs another \lstinline{FreeVar} effect to check that the remaining free variables are safe. If this check returns successfully, the continuation is resumed. 
  \end{enumerate}
\item err is implemented as an unhandled \lstinline{FreeVar} effect (no corresponding line number).
\end{enumerate}

\begin{figure}[h]
\begin{ocaml}
type _ Effect.t += FreeVar: IdentSet.t -> unit Effect.t
type _ Effect.t += Mute: IdentSet.t -> unit Effect.t

(* Some code omitted *)

let check vars = if not (IdentSet.is_empty vars) then perform (FreeVar vars)

let exec_in_current_scope scope f =
  let muted_vars = ref IdentSet.empty in
  match f () with
   effect FreeVar fvs, k->
     let free = IdentSet.diff fvs (IdentSet.union (!muted_vars) scope) in
     check free;
     continue k ()
 | effect Mute (fvs), k ->
     muted_vars := IdentSet.union (!muted_vars) (fvs);
     continue k ()
 | effect op, k  ->
     perform (Mute (IdentSet.union scope (!muted_vars)));
     muted_vars := IdentSet.empty ;
     (match perform op with v -> continue k v)
 | c, fvs ->
     let free = IdentSet.diff fvs scope in
     check free;
     (c, free)

 let new_scope alphas f =
    let scope = IdentSet.of_list alphas in
    let c, fvs = exec_in_current_scope scope f in
    (c, fvs)

let empty = IdentSet.empty
let free_var var = IdentSet.singleton var
let merge_free_vars fvs1 fvs2 = IdentSet.union fvs1 fvs2
\end{ocaml}
  \captionof{listing}{The implementation of the C4C check in Macocaml}%
\label[listing]{code:c4c-implementation}
\end{figure}

\end{extendedonly}

\end{document}
\endinput